\documentclass[12pt,preprint]{aastex}

\lefthead{Antoniou et al.}  \righthead{Optical spectroscopy of 20 Be-XRBs in the SMC}

\def\ergcm2s{~erg cm$^{-2}$ s$^{-1}$ }

\def\lunit{~erg s$^{-1}$}

\def\lx{~$\rm{L_{X}}$}
\def\etal{et al.~}              

\def\lunit{~erg s$^{-1}$}

\def\Zsun{Z$_{\odot}$}

\def\deg{$^{\circ}$}

\def\chandra{{\it Chandra~}}

\def\xmm{{\it XMM-Newton~}}
\def\xmmsimple{{\it XMM-Newton}}
\def\x2{$\chi^{2}$}

\def\Ha{H$\alpha$\,}

\def\E(B-V){{\it E(B-V)}}

\def\HaSii{${\rm H\alpha}$/${\rm [SII]}$~}
\def\Sii{${\rm [SII]}$\,}

\def\Porb{${\rm P_{orb}}$\,}
\def\Porbsimple{${\rm P_{orb}}$}



\begin{document}
\title{Optical spectroscopy of 20 Be/X-ray Binaries in the Small Magellanic Cloud}
\author{V. Antoniou$^{1,2}$, D. Hatzidimitriou$^{1}$, A. Zezas$^{1,2,3}$, P. Reig$^{1,3}$}
\affil{$^{1}$Physics Department, University of Crete, P.O. Box 2208, GR-710 03, Heraklion, Crete, Greece}
\affil{$^{2}$Harvard-Smithsonian Center for Astrophysics, 60 Garden Street, Cambridge, MA 02138, USA; vantoniou@head.cfa.harvard.edu}
\affil{$^{3}$IESL, Foundation for Research and Technology, 71110 Heraklion, Crete, Greece}

\begin{abstract}
We present a large sample (20 in total) of optical spectra of Small
Magellanic Cloud (SMC) High-Mass X-ray Binaries obtained with
the 2dF spectrograph at the Anglo-Australian Telescope. All of these
sources are found to be Be/X-ray binaries (Be-XRBs),
while for 5 sources we present original classifications. Several
statistical tests on this expanded sample support previous findings for similar spectral-type distributions of
Be-XRBs and Be field stars in the SMC, and of Be-XRBs in the Large
Magellanic Cloud and the Milky Way, although this could be the
result of small samples. On the other
hand, we find that Be-XRBs follow a different distribution
than Be stars in the Galaxy, also in agreement with previous
studies. In addition, we find similar Be spectral type
distributions between the Magellanic Clouds samples. These
results reinforce the relation between the
orbital period and the equivalent width of the \Ha line that holds for
Be-XRBs. SMC Be stars have larger \Ha equivalent widths when compared to Be-XRBs,
supporting the notion of circumstellar disk truncation by the
compact object.
 \end{abstract}


\section{Introduction}\label{intro2dF}
We present a detailed optical spectroscopic study of 20 High-Mass X-ray Binaries (HMXBs) in the Small Magellanic Cloud (SMC) using the 2dF (Two-degree Field) multi-object spectrograph on the 3.9m
Anglo-Australian Telescope (AAT). In this study we have used the 2dF
spectrograph to confirm the classification of candidate Be/X-ray binaries (Be-XRBs; the
most numerous subclass of SMC HMXBs) as emission-line stars and
identify their spectral type. Although this is only a pilot study, it
is at the same time a spectroscopic survey of a relatively large number of Be-XRBs and candidates over a large wavelength range. This provides a deep census of the X-ray binary (XRB) population in this galaxy even in low activity levels, which is critical for understanding the characteristics of the overall population.

The SMC is one of the best targets to study a, as complete as possible, population of young XRBs. Similar studies in the Galaxy are hampered by extinction and distance uncertainties. The Large Magellanic Cloud (LMC) is much more extended than the SMC, requiring large area coverage to obtain sufficient numbers of XRBs, while other Local Group galaxies are too far to reach the quiescent population of HMXBs (typical \lx$\sim10^{33}-10^{35}$\lunit; van Paradijs \& McClintock 1995).

In addition, the SMC hosts a large number of HMXBs (e.g. Haberl \&
Pietsch 2004, McBride \etal 2008, Antoniou \etal 2009) when compared to the Milky Way and the LMC. This excess is explained by
the combination of a star-formation burst at $\sim40$ Myr at regions
where the HMXBs are located (Antoniou \etal 2009; consistent with the
age of maximum formation of Be stars, given by McSwain \& Gies 2005), and by the low SMC metallicity (Antoniou \etal 2009, McBride \etal 2008). Simulations show that neither the metallicity nor the star-formation burst alone can produce such a high number of HMXBs in the SMC (Dray 2006). Thus, the SMC is an excellent laboratory to study the Be-XRB populations.

The most representative studies of spectroscopically identified Be-XRBs
are those of McBride \etal (2008), who investigated the spectral
distribution of 37 in total SMC Be-XRB pulsars, and of Negueruela \&
Coe (2002), who presented high-resolution blue-band spectroscopy for
13 LMC Be-XRBs. However, the main limitation of detailed studies
of the HMXB population in the SMC is the lack of spectral
classification of the optical counterparts of a large fraction of the
detected sources, while the vast majority of the studies available in
the literature either do not present the spectra and/or the criteria
used for the spectral classification. This severely hampers the
construction of a large coherent sample of Be-XRBs. Motivated by the
limitations of the current studies, and in order to classify the
different type of XRBs and separate them from the various interlopers
(background/foreground sources), we initiated an optical spectroscopic
campaign of a large number of candidate SMC Be-XRBs, detected either
in our observations or other published X-ray surveys of the SMC. The
goal of this campaign is to confirm the Be classification of each system (often based on photometric data), identify its spectral type, and even identify any possible interlopers.

The structure of this paper is the following. In Section \ref{sample},
we describe the sample of sources we observed with the 2dF
spectrograph; and in Section \ref{obsreduction}, the observations and
data reduction, along with the optical and near-infrared properties of
the sources. In Section \ref{criteriaSpTypes}, we
discuss the criteria used for the classification of the optical spectra, and the comparison of the classification of sources from the
present work with those of previous studies. In Section \ref{2dFdiscussion}, we discuss these results in the context of the Be stars and Be-XRB populations in the Galaxy and the LMC, and we investigate correlations between the \Ha emission and the orbital period and infrared colors of the systems. Finally, we present the main results of this study in Section \ref{conclusions}.

\section{Sample}\label{sample}
Our sample is primarily selected from the shallow \chandra survey of
the SMC (A. Zezas \etal 2009, in preparation; Antoniou \etal 2009), and it also
includes candidate and known Be-XRBs from the census of Haberl \&
Pietsch (2004; based mainly on observations with \xmmsimple). The overall strategy of
this pilot study was to observe the optical counterparts of
confirmed and candidate Be-XRBs. In
total, we have observed 113 and 27 \chandra and
\xmm sources, respectively. We note that even though the 2dF
spectrograph allows the allocation of $\sim$400 fibers, the stellar
density of the selected fields, which lie along the SMC ``Bar'', did not allow us to observe more
objects at a given time due to collision of the fibers. Twenty-nine out
of the 113 \chandra sources do
not have an optical match in the OGLE-II and/or MCPS catalogs
(based on the analysis of Antoniou \etal 2009), we thus obtained blind
spectra, i.e. we placed the fiber at the location of an X-ray source,
in order to obtain a spectrum of any object at that position. For the remaining \chandra sources, the typical
V-band magnitude of the optical counterparts was $19.3\pm2.1$ mag. In
addition to the \chandra sources we also observed 9 confirmed and 18
candidate Be-XRBs from the compilation of Haberl \&
Pietsch (2004).

\section{Observations and data reduction}\label{obsreduction}

The optical spectra for this work were obtained during service time
in 2004 November, with the 2dF multiple-fiber spectrograph at the
prime focus of the 3.9-m Anglo-Australian Telescope (Lewis \etal
2002). The observations were taken at airmass between 1.35 and 1.48,
and the seeing was $\sim1.6\arcsec$. We used the $300{\rm lines/mm}$
grating (300B), which gave a 4400\AA\, wavelength coverage (from 3640\AA\, to 8040\AA\, region) at an instrumental dispersion of
 4.3\AA/pixel. Four 2250s exposures were obtained, providing a total exposure time of 9000s. The on-target exposures
 were preceded and followed by arc calibration exposures (CuAr and CuHe) and fiber-flat-field exposures.
 Approximately 30 ``sky" fibers located on star-free regions (15 for each spectrograph) were assigned,
 to ensure good definition of the sky background. Successful use of the 2dF spectrograph depends on the source
 positions being accurate to better than 0.5 arcsec, in order to avoid significant
 light loss from the 2-arcsec-diameter fibers. Positions were used from OGLE-II (which has a small astrometric
error of $\sim0.7\arcsec$; Udalski \etal 1998) whenever possible to fulfill this requirement. In order to achieve accurate pointing and guide the telescope successfully during the observations we chose 4 sources as fiducial stars evenly spaced through the 2dF field. These stars were chosen from the 2MASS catalog, however they have consistent coordinates in the OGLE-II catalog and they abide by the suggestions listed in the 2dF manual (e.g. brighter than 15.0 magnitude stars which should not cover more than a one magnitude range). Nevertheless, some loss in counts is expected to be caused by the small number of fiducial stars used to guide the field plate and due to the small positional mismatches between fiber and object positions.

The data preparation and spectral extraction were performed using
the {\it 2dfdr} software (Lewis \etal 2002). The main steps include
bias subtraction, extraction of the spectra from the CCD image,
division by a normalized flat-field, wavelength calibration,
calibration of the fiber throughputs, and subtraction of the scaled
median sky spectrum (derived from the sky-fiber spectra). Extraction
of the spectra, spectral line fitting and EW line
measurements were performed with the {\it DIPSO} v3.6-3 and FIGARO
v4.11 packages of STARLINK. In particular, we measured the center of each \Ha line and its Full-Width at Half-Maximum (FWHM) using a Gaussian fit ({\it elf} subroutine), while for the EW we used the {\it ew} subroutine of the {\it DIPSO} package. 
Due to the different throughput
of each fiber (which is also wavelength-dependent), it is difficult
to flux calibrate spectra from a multi-fiber instrument such as 2dF,
as  a flux standard should be -in principle- observed through each
fiber. An average response curve can be applied to give an
approximate relative flux calibration, however, the results for
individual spectra can vary considerably (Lewis \etal 2002).
Therefore, flux calibration was not attempted. This does not affect
our analysis, as our classification criteria are based on the presence or absence of spectral lines rather than their absolute intensity. Whenever comparisons between strengths of
lines are used, care is taken to use lines close in wavelength, in
which case the fiber response is identical, for all practical
purposes.

Fibers that were assigned to objects fainter than $\sim16.5$ mag in V-band, yielded low signal-to-noise (S/N) ratio spectra. We
defined a nominal ${\rm S/N}$ ratio of each spectrum using the counts in
the 5600\AA-6200\AA\, range (measured using the {\it istat} package
of FIGARO v4.11) which can be considered as representative of the
local continuum. Spectra with $<$400 average counts in this
wavelength range were excluded from the present study, as it was
difficult to perform accurate spectral classification for them. This leaves 58 spectra for further consideration. All but one of these 58 sources exhibit \Ha emission.

 The presence of \Ha emission is one of the main classification criteria for Be-XRBs. We thus focus on the
 study of \Ha emission-line objects. However, the inner regions of the SMC show strong
 and variable diffuse emission from HII regions and supernova remnants (SNRs). Due to hardware limitations it is not possible to obtain sky spectra within a few arcseconds from each source\footnotemark,\footnotetext{The absolute minimum is 30 arcsec (2mm), but typically it is 30-40 arcsec depending on the location in the field and the target distribution (http://www.aao.gov.au/AAO/2df/aaomega/aaomega\_faq.html\#fibsep)} which in turn does not allow us to accurately subtract the variable local background. Therefore, some of our object spectra may be
 contaminated by the typical interstellar emission lines: ${\rm [OIII]\, \lambda5007}$, ${\rm [NII]\, \lambda\lambda6548, 6584}$, and ${\rm [SII]\, \lambda\lambda6716, 6731}$. For this reason we adopt a conservative approach and we focus only on objects with minimal interstellar contamination. The selection of such objects is based on the width of
the \Ha line and their \HaSii ratio.

Be stars exhibit broad \Ha emission (e.g. Coe \etal 2005), whereas the
interstellar emission lines are typically narrow. Therefore, selection
of objects with broad \Ha emission allows us to identify objects with
small interstellar contamination. We choose a low limit for the FWHM of the \Ha line $\geq9.5$\AA, based
on the average FWHM of this line from high S/N spectra of known
Be-XRBs and Be-XRB pulsars, which are observed as part of the present
work. In total, we find 22 sources with \Ha FWHM greater than
9.5\AA. Two of these sources have broad but extremely weak \Ha
emission when compared to the other sources, thus they are not
considered for further investigation. Based on the above criteria our
sample is limited to 20 sources. We note that there could be
additional bona-fide Be-XRBs in the remaining sample, however the
absence of a broad \Ha component makes their classification uncertain.

 The selection of these sources as bona-fide Be-XRBs is confirmed by their high \HaSii ratio. Be stars typically do not show forbidden high excitation emission lines (with the only exception of the subclass of B[e] stars; e.g. Lamers \etal 1998). Therefore, the presence of \Sii $\lambda$6716\AA\ emission is a signature of significant contamination from the diffuse interstellar emission. We determine the
average \Ha to \Sii strength ratio of the diffuse emission from the spectra of the sky fibers which uniformly 
probe the observed area. This gives an average local ``sky" value of \Sii to \Ha ratio of $0.15\pm0.03$. All objects which fulfill the \Ha width criterion also have \Sii to \Ha ratio at least
$3\sigma$ below the average sky ratio. This gives us further confidence that the emission-line objects we selected  for further study are indeed Be stars. In addition, although we cannot rule out the possibility that some of the remaining observed sources could be Be-XRBs, none of the discarded sources with good S/N has low enough \Sii to \Ha ratio to fulfill the above criterion. This indicates that we do not miss a significant population of Be-XRBs among the remaining sources. However, because of these criteria, our sample is biased towards objects with stronger \Ha emission.

Based on the above analysis, we found 20 objects with good
quality optical spectra obtained in this run, and known X-ray
properties typical of HMXBs (e.g. hard X-ray spectrum and/or detection
of pulsations). Out of the 20 objects, the spectra of which are
presented here, only 2 have good quality published optical spectra in
the full wavelength range ($\sim{\rm 3800\AA-7000\AA}$), while one source has been observed only in the red portion of the optical
spectrum around H$\alpha$ ($\sim{\rm 6000\AA-7000\AA}$). These 20
sources are presented in Table \ref{otherXrayname}. In Column 1 we
list the X-ray source ID as given in Antoniou \etal (2009; (Field
ID)-(Src ID within the field)) for \chandra sources, while the XMM ID
refers to sources from Haberl \& Pietsch (2004). Other names for the
{\it XMM-Newton} sources and their references (Column 2) are taken
from the latter work. In Column 3 we present the X-ray pulsar ID
(defined as the pulse period in
seconds\footnotemark).\footnotetext{Based on the online census of
Malcolm Coe as of 22 August 2008
(http://www.astro.soton.ac.uk/$\sim$mjc/).} Finally, in Column 4 we
list the associated emission-line object from the catalog of Meyssonnier \& Azzopardi (1993; hereafter [MA93]). With the present observations we will investigate the Be-XRB nature of these 20 objects and we will study their spectral characteristics, in conjunction with their optical and near-infrared properties.

The optical and near-infrared counterparts of the studied X-ray
sources are present in Table \ref{2dFoptIDs}. In particular, the
optical counterparts of the \chandra sources are taken from Antoniou
\etal (2009). Following the same approach for the \xmm sources, we
cross-correlated their positions with the OGLE-II (Udalski \etal 1998)
and MCPS catalogs (Zaritsky \etal 2002) using a search radius of
5\arcsec\, around each X-ray source (Brusa \etal 2007). In Column
1 we give the X-ray source ID (as in Table
\ref{otherXrayname}), and in Columns 2  and 3 the X-ray source
coordinates (Right Ascension and Declination, respectively). In Column 4 we present
the optical counterpart of the X-ray source from the OGLE-II and/or
MCPS catalogs. OGLE-II sources are named as O-F-NNNNNN, where F and NNNNNN are the field and optical source number, respectively (from Udalski \etal 1998), and MCPS sources are named as Z-NNNNNN where NNNNNN is the line number of the source in Table 1 of Zaritsky \etal (2002). The angular separation (in arcseconds) of the counterpart to the X-ray source is given in Column 5. The optical ({\it V,B-V}) data of the sources are presented in Columns [6] and [7] along with their errors (these data are taken directly from the original catalogs without applying any reddening or zero-point correction): apparent magnitude in the {\it V}\, band (Column 6), and {\it B - V}\, color (Column 7). For one source (XMM-47) we found an optical counterpart in slightly larger distance from the X-ray source position (within 5.3\arcsec\, and 4.9\arcsec\, in the OGLE-II and MCPS catalogs, respectively). The near-infrared counterparts of the Be-XRBs studied here are taken from the 2MASS catalog (Skrutskie \etal 2006). Column 8 lists the closest 2MASS counterpart ID within 2\arcsec\, from the X-ray source position, and Columns 9 and 10 the J, and K magnitudes along with their errors. Four sources (4-3, XMM-15, XMM-47 and XMM-5) do not have a 2MASS counterpart within 2\arcsec\, from the X-ray source position.

\begin{figure}
\begin{center}
\rotatebox{270}{\includegraphics[width=6cm]{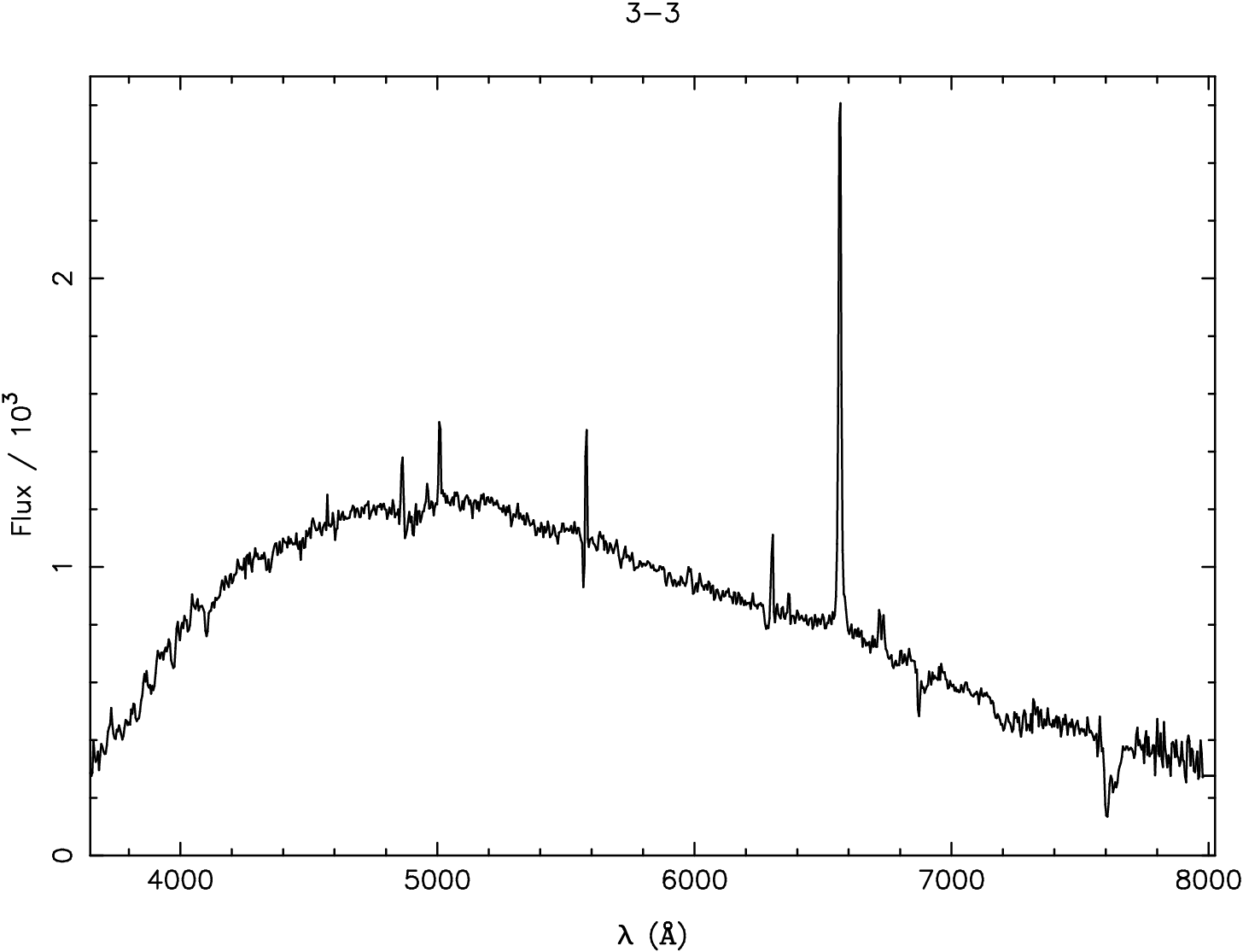}}
\vspace{0.5cm}
\rotatebox{270}{\includegraphics[width=6cm]{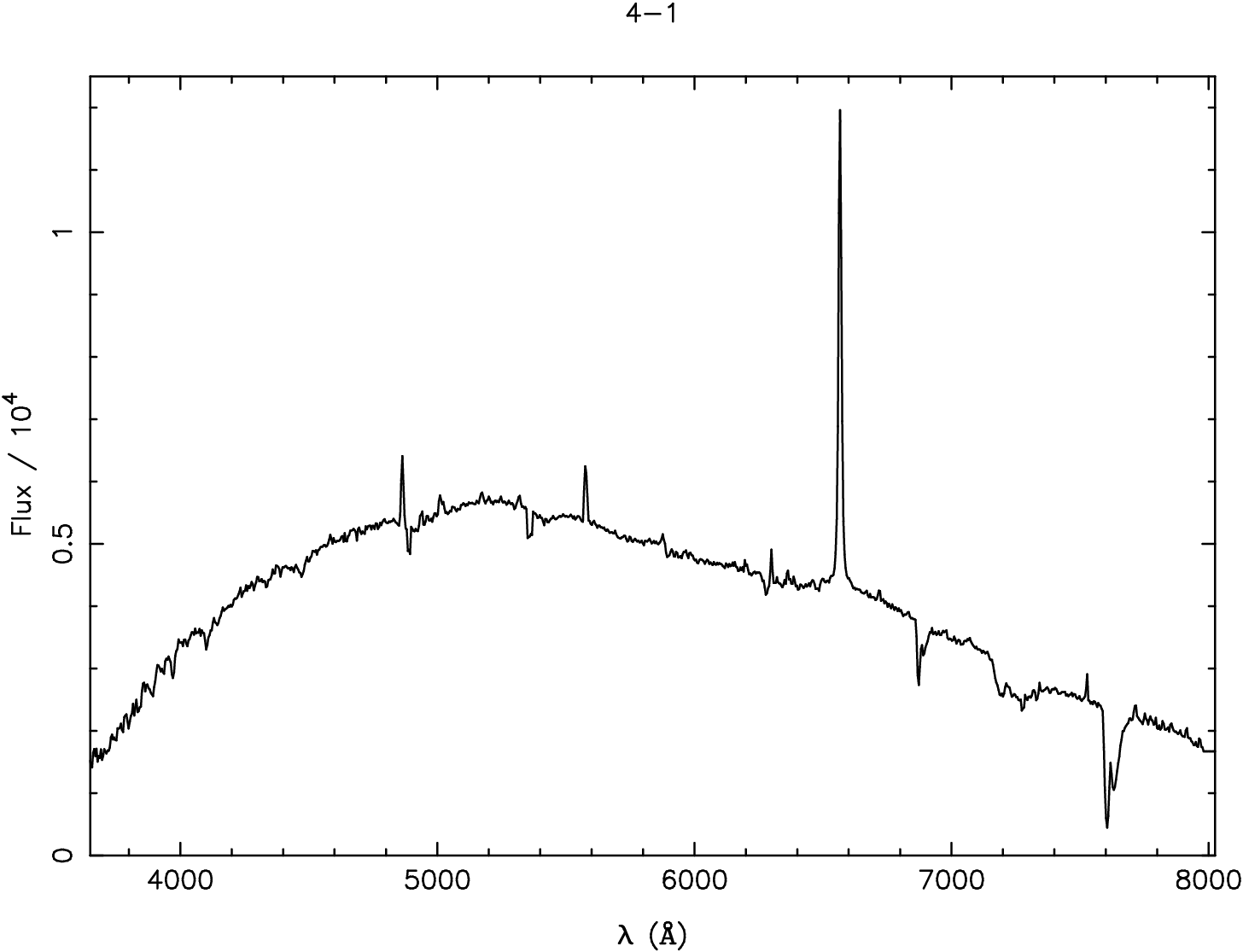}}
\vspace{0.5cm}
\rotatebox{270}{\includegraphics[width=6cm]{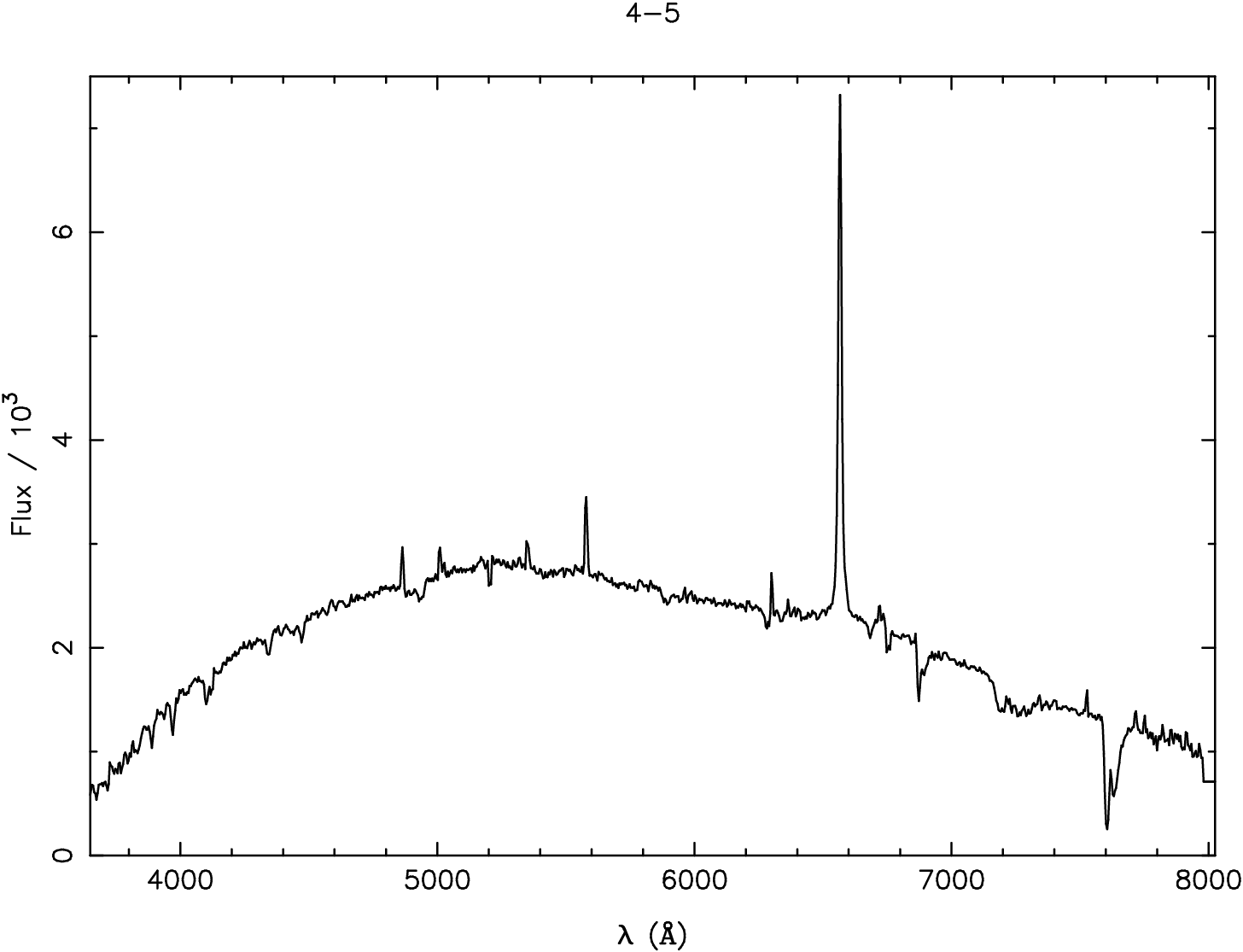}}
\rotatebox{270}{\includegraphics[width=6cm]{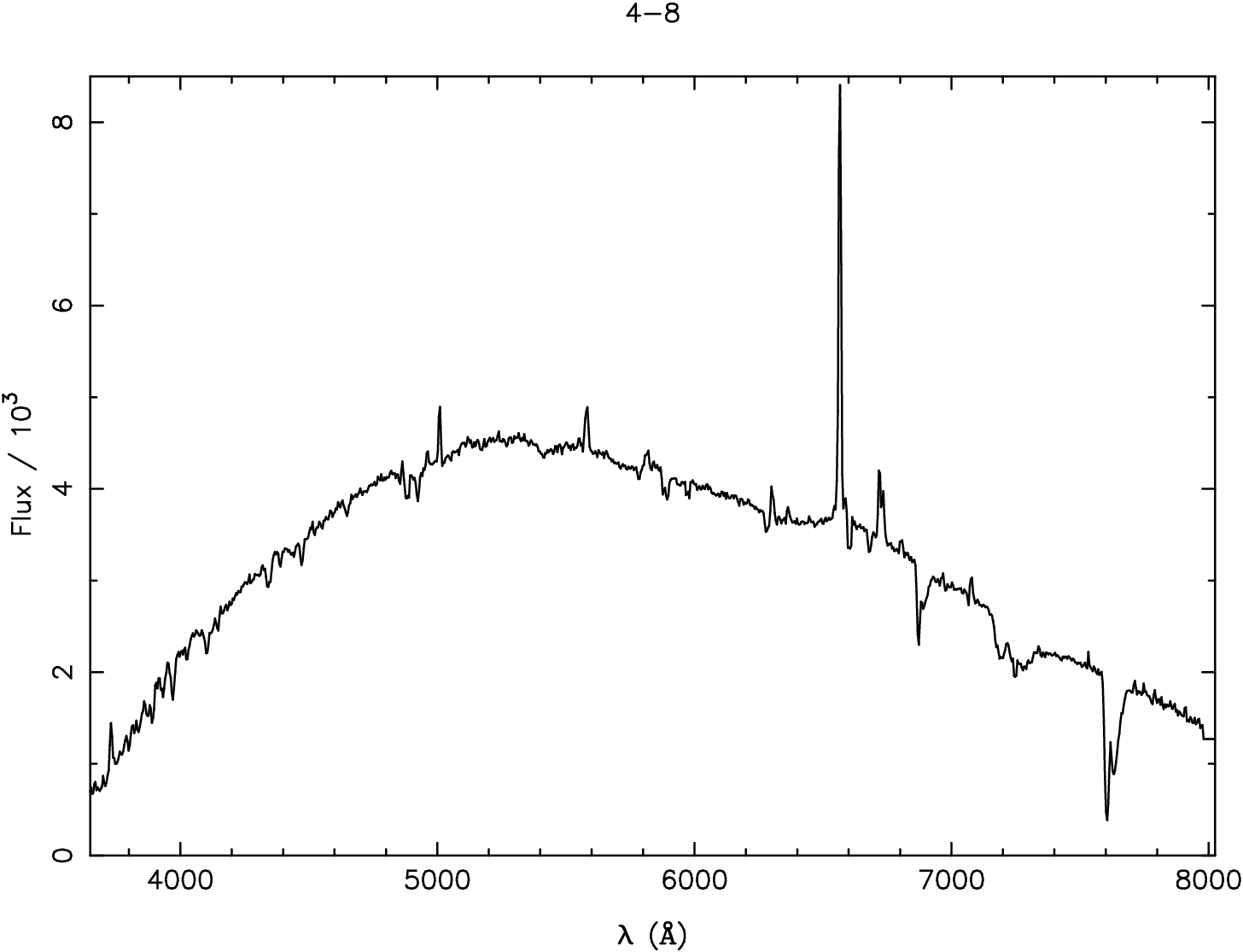}}
\vspace{0.5cm}
\rotatebox{270}{\includegraphics[width=6cm]{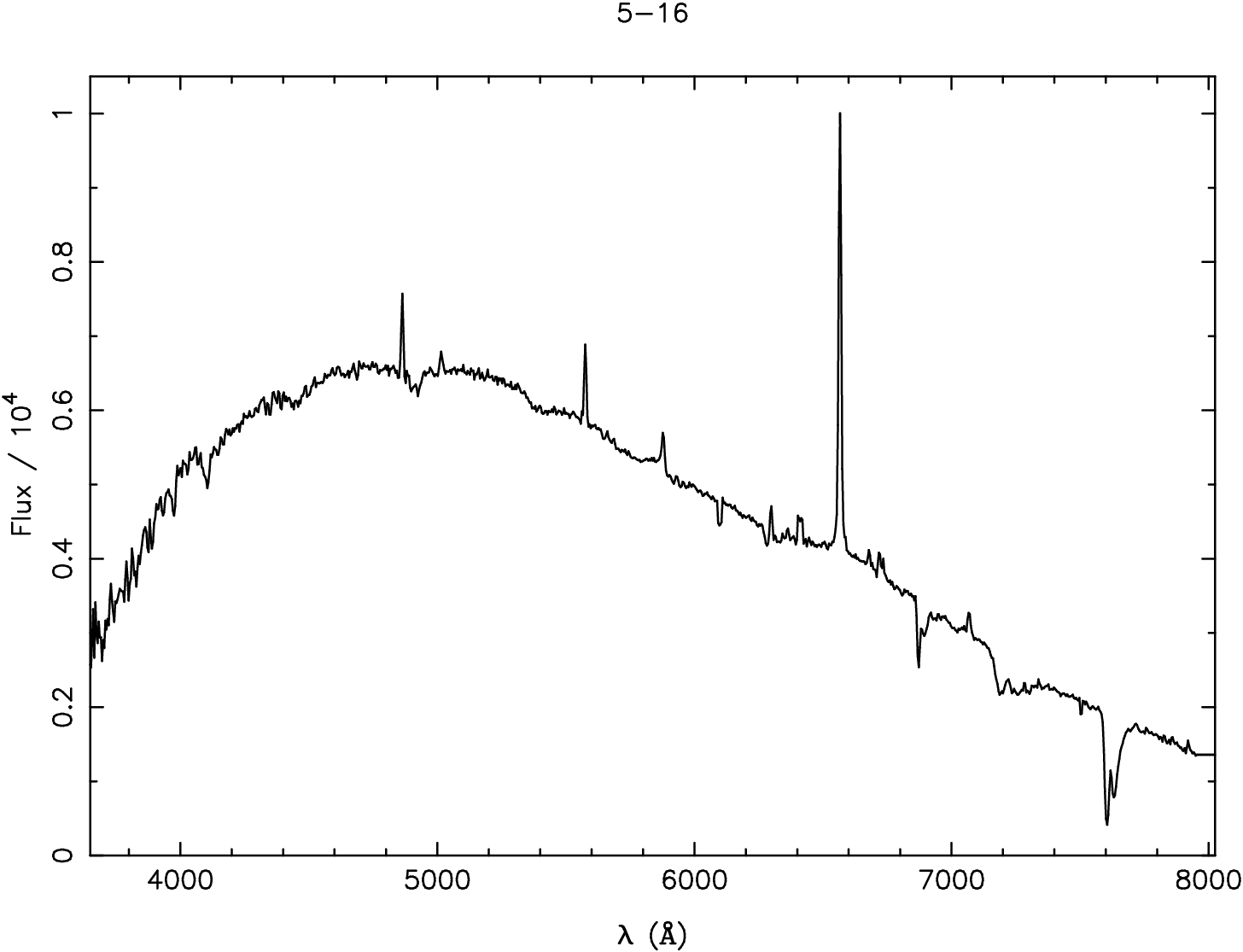}}
\rotatebox{270}{\includegraphics[width=6cm]{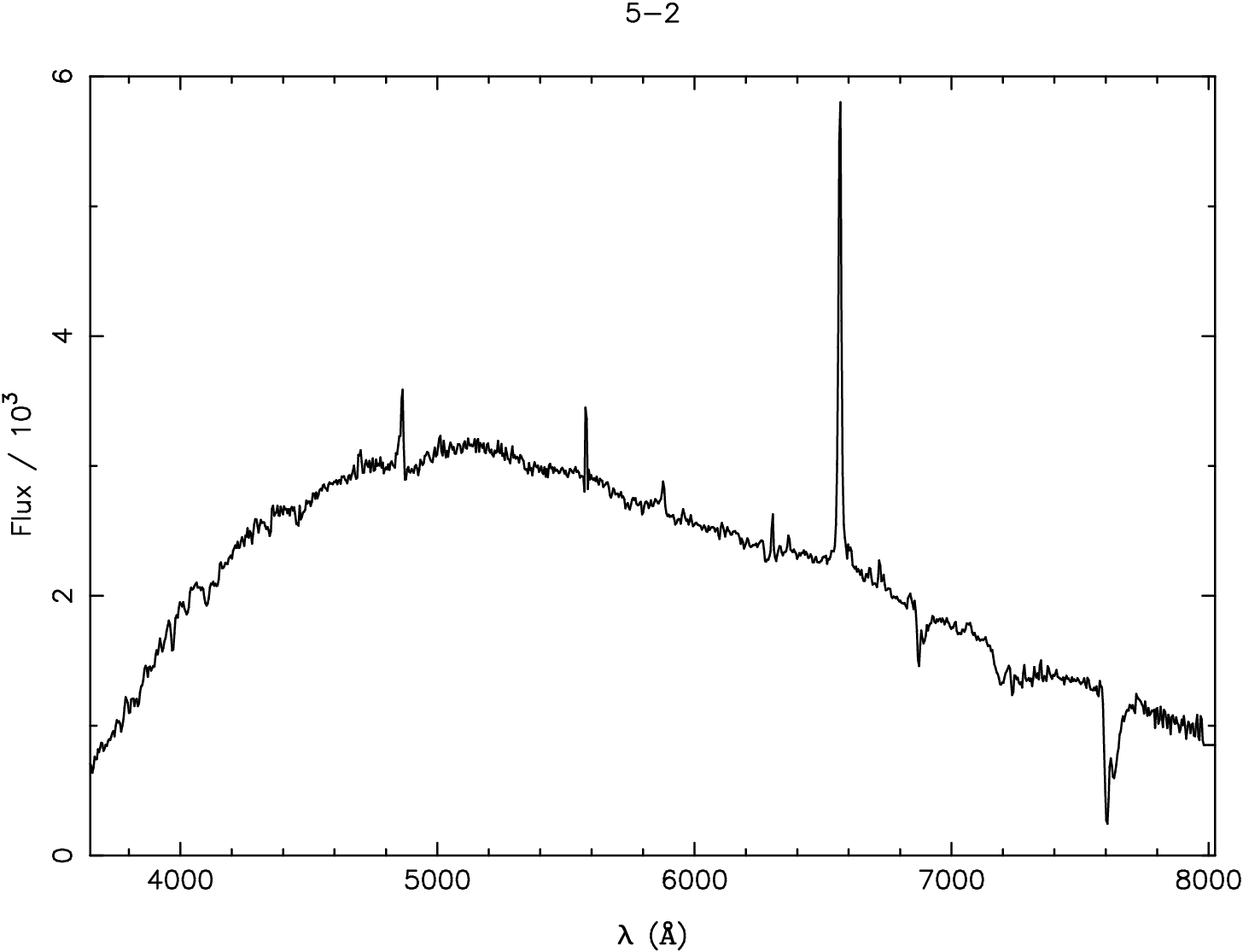}}
\end{center}
\caption{Full wavelength range ($\sim{\rm 3650\AA-8000\AA}$) spectra
of the 20 SMC Be-XRBs obtained with the 2dF multi-object spectrograph
on the AAT. The flux is arbitrarily normalized.}\label{2dFspecBeXRBs}
\end{figure}

 The full (3650\AA-8000\AA) wavelength calibrated spectra of the 20 SMC Be-XRBs are given in Figure \ref{2dFspecBeXRBs}. The early-type nature of the stars is apparent in all spectra, as is their strong \Ha emission. The contribution of telluric emission lines is sometimes not well subtracted (due to the difficulties in sky subtraction in multifiber observations discussed earlier) leaving emission and/or absorption features around the [OI]\,$\lambda\lambda5577,6300$\AA\, lines. Moreover, telluric absorption bands of molecular oxygen are obvious in the red part of our spectra, around 6875\AA, and 7610\AA. Finally, some diffuse interstellar emission lines can also be seen in some of the spectra, including [OIII]\,$\lambda4959$\AA, [OII]\,$\lambda5007$\AA, [NII]\,$\lambda6585$\AA\ and [SII]\,$\lambda\lambda6716,6731$\AA.

In Table \ref{EWFWHM} we present for each selected X-ray source
(Column [1]) the measurements of the center (Column [2]), FWHM (Column
[4]), and EW (Column [6]) of the \Ha emission line, along with their
errors (Columns [3], [5], and [7], respectively) from the interactive
fits. The errors represent the 68\% confidence intervals for one
interesting parameter. We found
published \Ha EW measurements for 8 of our sources. We present these
values in Column [8] along with their source (given in
parenthesis). They are in good agreement, despite the expected
scatter, given the variable nature of Be stars (e.g. McSwain \etal
2008), and the different disk inclination in which they may have been
observed (Dachs \etal 1986). The \Ha emission-line profiles in the
spectra of Table \ref{EWFWHM} are very simple with no signs of double
or split peaks. However, because of the resolution of the used grating, we may not have been able to resolve any asymmetries or double peaks. In addition, in Column [9] we give the orbital period of the HMXBs in this sample from the literature (Liu \etal 2005 and references therein), and we use these values in order to study the properties of the decretion disk in relation to the parameters of the binary system (see discussion in Section \ref{Porbital} below).

In Figure \ref{EWdistr} we show the distribution of the \Ha EWs for
Be-XRBs (solid histograms) and Be stars (open
histograms). We note here that by using the term Be stars throughout this paper
we refer to field and cluster Be stars that do not exhibit X-ray emission characteristic of XRBs. These distributions are shown for the
SMC (top panel), the LMC (middle panel) and the Milky Way (bottom
panel). The data for the SMC Be-XRBs are taken from the present study and are supplemented by the catalog of Coe \etal (2005), while for the SMC Be stars
we used data from the study of the star-cluster NGC330 by Martayan \etal (2007a). The
data for the LMC Be-XRBs are taken from Liu \etal (2005; and
references therein), while the LMC Be sample is from the study of the star-cluster NGC2004 by Martayan \etal (2006a). The data for the Galactic Be-XRBs
are taken from Reig (2007) and Liu \etal
(2006; and references therein), while for the Galactic Be stars we
used data from Ashok \etal (1984), supplemented with data from
Fabregat \& Reglero (1990), Dachs \etal (1986), and Dachs,  Hummel \&
Hanuschik (1992). We note that the above Galactic Be samples contain field stars. In order for the above comparisons to be valid, we
selected Be-XRBs and Be stars within the same spectral-type
range in each of the 3 galaxies. Be-XRBs in the Milky Way  and the LMC
have spectral types up to B2, while the majority of such systems in
the SMC is found to have spectral types B3 or earlier (see discussion
in Section \ref{SpTypediscuss} and Figure \ref{Bspecsubtdistr} below). The
spectral types of Galactic sources with \Ha EW measurements used here are taken
from Andrillat \& Fehrenbach (1982) and Skiff (2007; and references
therein), while for the SMC and LMC Be stars we used data from Martayan \etal (2007b) and (2006b), respectively.

 In total we have used 41 Galactic Be stars, 21 Galactic Be-XRBs, 116
SMC Be stars, 35 SMC Be-XRBs, 5 LMC Be-XRBs and 42 LMC Be stars. From
Figure \ref{EWdistr} we see that the \Ha EW distribution of SMC
Be-XRBs peaks at $\sim{\rm 15-25\AA}$, while that of SMC Be field
stars peaks at $\sim{\rm 30-45\AA}$. This is in agreement with Reig
\etal (1997) and Zamanov \etal (2001) who reported that Be stars in
XRBs have on average a lower \Ha EW when compared to Be
stars, due to the truncation of the disk by the compact object. For
the LMC and the Milky Way, the comparison is hampered by the small
size of the samples. On the other hand, Dachs \etal (1992) found that
Be stars with larger \Ha EW have larger radii of their \Ha emitting
circumstellar disks. Grundstrom \& Gies (2006) used numerical models to show that the predicted radius of the Be disk, based on \Ha EW, is monotonically related to the observed disk radius with long baseline interferometry. From the above figure, we also see that the Be
stars in the Magellanic Clouds (hereafter MCs) have larger \Ha EW
values when compared to the Galactic sample, thus indicating larger disks around
them. Martayan \etal (2007a), who find similar results, suggest that
the higher \Ha EWs (especially in the SMC) indicate the presence of
more extended circumstellar envelopes around a fraction of Be stars. We also note that some Be stars may be in fact in binary systems. However, even if some of those prove to be binaries, we do not expect their \Ha EW distributions to change significantly.

\subsection{Optical and near-infrared photometry of the Be-XRB 2dF sample}\label{2dFoptcparts}
The identification and classification of the optical counterparts of
the X-ray sources allow us to identify the interlopers and determine
the type of XRBs (e.g. Be-XRBs, Supergiant XRBs, Low-Mass XRBs). In
Figure \ref{CMD2dF} we present the extinction corrected absolute {\it V}
magnitude vs. {\it B-V} color (${\rm M_{{V}_{o}}}$ versus $(B-V)_{\rm {o}}$)
color-magnitude diagram (CMD) of the counterparts of the X-ray sources
with 2dF spectra (black circles). The magnitudes, colors and their
errors are taken from the OGLE-II catalog (Udalski \etal 1998), except
for sources with matches only in the MCPS catalog (Zaritsky \etal
2002), based on data from Table \ref{2dFoptIDs}\footnotemark
\footnotetext{We correct the original data presented in this table for
the extinction to the SMC using E(B-V)=0.09 and Av=0.29 (for a more
detailed justification of this choice see Antoniou \etal 2009).}. With
gray squares we also show the NGC330 Be stars (Martayan \etal
2007a), while with gray small dots we present the OGLE-II stars that
lie in our \chandra field 4 and have small errors in the {\it V} and {\it B} band
($<0.2$ mag). The main sequence, the red giant branch and the red
clump loci are clearly seen. In this figure, we overplot the isochrones from the Geneva database (Lejeune \&
Schaerer 2001)
for the metallicity of the young SMC stars ($\rm{[Fe/H]}=-0.68\pm0.13$ dex,
equivalent to $\rm{Z}=0.004$\footnotemark; \footnotetext{For the conversion, we use the relation
$\rm{[Fe/H]}\equiv \log(\rm{Z}/$\Zsun$)$, with \Zsun=$0.02$ for the
solar metallicity of $\rm{[Fe/H]}=0$ (Russell \& Dopita 1992).}Luck \etal 1998), for
ages of 8.7 Myr to 275.4 Myr. The position of all the X-ray sources in the CMD agrees well with the
NGC330 Be stars. We also see that the Be-XRBs from our sample and the
Be stars from Martayan \etal (2007a) are at the right of the main
sequence, as expected given that Be stars are generally redder than
B-type stars (e.g. McSwain \& Gies 2005), while they all lie between
the $\sim15.5$ Myr and $\sim85$ Myr isochrones.

Most Be stars exhibit infrared excess (Gehrz, Hackwell \& Jones 1974)
due to the disk contribution to the continuum emission. Several
authors have observed a similar excess in Be-XRBs (e.g. Coe \etal
2005). In Figure \ref{JKo} we present the distribution of the J-K
color for the Be-XRBs of the present work (black filled histogram). We
also show the J-K color distribution of Be stars of luminosity class
III-V (gray hatched histogram) and of all non-emission B-type stars
(i.e. excluding the above Be stars sample; gray filled histogram)
using data from the 2dF SMC spectroscopic survey of Evans \etal
(2004). For clarity, the number of sources in the Be-XRB and Be stars sample is
normalized by a factor of 2. We find that the J-K distribution of
Be-XRBs peaks at $\sim$0.3 mag, while that of Be stars at
$\sim$0.5-0.9 mag. Non-emission B-type stars show a skewed
distribution, which peaks at ${\rm (J-K)}\sim$-0.1 mag.

\section{Determination of spectral types}\label{criteriaSpTypes}

The primary spectral classification criteria for OB stars are usually based on metal line strengths except for the earliest subtypes which are based on HeI and HeII lines (e.g. Lennon 1997). However, due to the low metallicity of the SMC sources, the metal lines are very weak (see also Evans \etal 2004) hampering the classification based on them. Therefore, we adopt the classification scheme of Evans \etal (2004), although, due to the lower resolution of our spectra, it was not always possible to use all the proposed indicators.

According to this scheme, the spectral classification is based on the
blue part of the spectrum (3800\AA-5000\AA\ wavelength range). The
presence of HeII absorption lines (HeII\,$\lambda\lambda,4200, 4541,
4686$) indicates spectral types earlier than or equal to B0. Strong HeI lines are also
indicative of early spectral types. Their strength increases with
decreasing temperature, while it peaks
around B2. The main characteristic of early B-type stars is the
absence of lines of ionized helium (HeII), which are still present in
late O-type stars. Thus, in B0.5 stars, HeII\,$\lambda4200$ and HeII\,$\lambda4541$ are absent, while HeII\,$\lambda4686$ is weak. In B1 stars, HeII\,$\lambda4686$ disappears, while SiIV\,$\lambda\lambda4088,4116$ lines appear. For spectral types later than B1.5 the OII+CIII\,$\lambda4640-4650$ blend decreases rapidly, and disappears for stars later than B3. For spectral types B2 or later MgII\,$\lambda4481$ and  SiIII\,$\lambda4553$ become stronger. Comparison between the strengths of these two lines is used to distinguish between B2, B2.5, and B3 types. For spectral types B5 or later the classification is based mainly on the absence of SiIII\,$\lambda4553$. In B8 stars HeI\,$\lambda4143$ is stronger than SiII\,$\lambda4128$, which in turn is stronger than HeI\,$\lambda4121$. Finally, for B9 stars MgII\,$\lambda4481$ appears stronger than HeI\,$\lambda4471$. In Table \ref{identifications} we summarize the above line identifications (Column 1) for each B spectral type subclass (Column 2).

The resulting spectral types for the 20 objects of Table
\ref{otherXrayname} are shown in Table \ref{spclass}, while sample
spectra in the 3800\AA-5000\AA\ wavelength range are shown in Figure
\ref{normspectra}, with characteristic spectral features marked. In
Table \ref{spclass} we give the X-ray source ID (Column 1), the
spectral classification derived from this study (Column 2), and any
previous classification of the same sources (Column 3) along with
their reference (given in parenthesis). A ``Be?" (i.e. tentative Be)
input in Column 3 indicates sources for which the classification as
a Be star is based on the positional coincidence of a hard X-ray
source with an emission-line object from the catalog of [MA93] and/or
with a photometrically classified early-type (OB) star (e.g. Antoniou
\etal 2009). In total, 5 out of the 20 sources studied here have been previously identified as candidate Be stars based only on photometric data and/or positional coincidence with an emission-line object from the catalog of [MA93]. In addition, two sources with hard X-ray spectra have been previously spectroscopically classified as B-type stars (i.e. non-Be stars) in the study of Evans \etal (2004). Furthermore, one object was classified as B-type star showing Balmer emission, but no further classification has been provided (Garmany \& Humphreys 1985). The present study examines these classifications, since only spectroscopy can unambiguously identify the nature of the donor with that of a Be star. Due to the absence of flux calibration and the relatively low resolution of our spectra it was not  possible to assign with certainty a spectral luminosity class to our objects. For the same reason, Evans \etal (2004), who have used the same spectrograph, have relied on a photometric/spectroscopic approach to estimate the luminosity class.

Because later-type main-sequence stars are also fainter, and because the present spectroscopic sample is obviously flux limited, there is an obvious selection effect favoring earlier spectral types. The faintest of these 20 sources (which has $\sim950$ counts in the 5600\AA-6200\AA\, wavelength range) has been classified here as a B0-4e star, thus our sample is complete up to mid-B.

\subsection{Comparison with previous studies}

We find good agreement between our and other published spectral-type
classifications for 9 out of the 15 sources with published classification. However, for 6 sources (XMM-3, XMM-5, 5-2, 5-7, XMM-39, and XMM-47) we derive different (or in some cases slightly different) spectral types. In addition, due to the low resolution of the spectra used in this work
and the low S/N ratio of sources 6-1, 5-12, 6-4, and 3-3 we can only assign to them broad spectral types. The blue part of the spectrum (3800\AA-5000\AA) of the last three sources is presented in Figure \ref{normspectra}.  In summary, for all but four sources (5-2, 5-7, XMM-3, and XMM-47 for
which we find 0.5, 1, 0.5 and 1 spectral subtype differences,
respectively) we find spectral classifications in good agreement with
previous studies. We estimate that our spectral types are accurate to better than $\pm1$ subclass in most cases, especially for earlier than B2 spectral types. Next we discuss in detail the spectral classification criteria for all sources in Table \ref{spclass}:\\

\begin{itemize}

\item{We classify source XMM-3 as a B0.5e type star based on the
presence of weak HeII\,$\lambda4686$, absent HeII\,$\lambda4541$, and
very weak (almost absent) HeII\,$\lambda4200$. The spectrum of source
XMM-3 exhibits strong Balmer lines in absorption (with the strongest
one being H$\epsilon\,\lambda3970$), and moderate
H$\beta\,\lambda4861$ in emission. Weak HeI ($\lambda\lambda4144,
4387, 4438, 4471, 4922$) lines and OII+CIII\,$\lambda4640-4650$ blend
are also present. Source XMM-3 has been assigned a slightly later spectral type (B1-1.5Ve) by McBride \etal (2008).} 

\item{Based only on photometric data, source 4-8 has been recently
classified as a later than B3 emission-line star (Haberl \etal
2008). In the present work, we classify it as a B1.5e star, based on
the presence of strong \Ha, weak SiIII\,$\lambda4553$ and
OII+CIII\,$\lambda4640-4650$ blend, and the absence of
SiIV\,$\lambda4116$ and HeII\,$\lambda4686$. Similarly to source
XMM-5, it shows strong HeI+II\,$\lambda4026$ and
HeI\,$\lambda\lambda4144, 4387, 4471, 4922$ lines, while
H$\beta\,\lambda4861$ shows weak emission.}

\item{Source XMM-5 has been classified as B0-5(II) by Evans \etal
(2004). Based on the presence of strong \Ha, weak SiIII\,$\lambda4553$
and OII+CIII\,$\lambda4640-4650$ blend, and the absence of
SiIV\,$\lambda4116$ and HeII\,$\lambda4686$, we classify this source
as B1.5e. Other present helium lines include HeI+II\,$\lambda4026$ and HeI\,$\lambda\lambda4144, 4387, 4471, 4922$. The blue part of the spectrum (3800\AA-5000\AA) of source XMM-5 is shown in Figure \ref{normspectra}.}  

\item{Sources 4-2, 4-5, 5-4, and 6-2 are classified as B1e stars in this
work. SiIV\,$\lambda\lambda4088, 4116$ lines and
OII+CIII $\lambda4640-4650$ blend are present, while HeII\,$\lambda 4686$ is
absent. H$\gamma\,\lambda4340$ is partially
filled-in by emission in source 4-2, and completely filled-in in
sources 5-4 and 6-2. Sources 4-5
and 5-4 are classified here
for the first time, while they are listed as candidate
Be-XRBs associated with objects [MA93]300 (Haberl \& Pietsch 2004) and [MA93]798 (Antoniou
\etal 2009), respectively. In addition, sources 6-2 and 4-2 have been previously
classified as B1 III-Ve and B1-3 IV-Ve stars by McBride \etal (2008).}

\item{We classify sources 4-1 and 5-16 as B0.5e stars, based on
the presence of the HeI\,$\lambda\lambda4144, 4387, 4438, 4922$ and
HeII\,$\lambda\lambda4339, 4686$ lines and the absence of the HeII\,$\lambda\lambda4200, 4541$ lines. 
The OII+CIII $\lambda4640-4650$ blend is weak, while in the case of source 5-16 H$\eta\, 3835$ and H$\gamma\,\lambda4340$ are
partially filled-in by emission. Coe \etal (2002) have classified source 4-1 as
a B0-0.5 Ve star, while source 5-16 has been classified as a O9.5-B1 IV-Ve star
by McBride \etal (2008).}

\item{Our classification of source XMM-15 as a B0.5e star,
based on very weak emission in HeII\,$\lambda\lambda4200, 4541$ and
HeI\,$\lambda\lambda4009, 4121, 4713$, absent HeII\,$\lambda4686$ and
SiIV\,$\lambda\lambda4088, 4116$, and strong HeI\,$\lambda\lambda4387,
4438, 4471$ and OII+CIII\,$\lambda4640-4650$ absorption, is in agreement with
previous studies (McBride \etal 2008). In Figure \ref{normspectra} we
present 
the blue part (3800\AA-5000\AA) of its spectrum. This source shows
H$\zeta\,\lambda3889$, H$\epsilon\,\lambda3970$ and
H$\delta\,\lambda4101$ in absorption, partly filled-in by
emission. H$\gamma\,\lambda4340$ is completely filled-in by emission
(absent from the spectrum), while the strong H$\beta\,\lambda4861$
emission line is the most prominent feature of the blue part of the
spectrum.}

\item{Source 4-3 has not been previously classified. It is only listed as candidate
Be-XRB associated with object [MA93]414 (Haberl \& Pietsch 2004). In
this work, we classify it as a B0.5e star, based on the absence of
HeII\,$\lambda\lambda4200, 4541$ lines, and on the presence of weak HeII\,$\lambda 4686$.}

\item{Source XMM-17 is classified here as a B0e star, based on the presence
of HeII\,$\lambda\lambda4200, 4541, 4686$ lines, in agreement with
McBride \etal (2008), who have classified it as a O9.5-B0 Ve star.}

\item{Source 6-1 is classified here as a B1-3e star. The HeII\,$\lambda 4686$ line is absent, while SiIV\,$\lambda 4116$
is present, indicating a spectral type of B1. However,
MgII\,$\lambda 4481$ is stronger than SiIII\,$\lambda 4553$, which is common in
B3 type stars. Finally, the OII+CIII $\lambda4640-4650$ blend is present but weak, thus indicating a spectral type later than B1.5. Taking into account all the above, we can only assign a broad spectral type of B1-3e to source
6-1, which is identical to the classification of McBride \etal (2008). Furthermore,
H$\gamma\,\lambda4340$ is partially 
filled-in by emission.}

\item{We classify source 5-12 as O9e-B0e, which is the earliest
spectral type found in this sample. We based this classification on
the presence of HeI\,$\lambda\lambda4144, 4387, 4471$ and
HeII\,$\lambda\lambda4200, 4541, 4686$ lines. HeI+II\,$\lambda4026$
and HeI\,$\lambda\lambda4009, 4922$ are also present. The Balmer
series H$\eta\,\lambda3835$, H$\epsilon\,\lambda3970$,
H$\delta\,\lambda4101$ are strongly in absorption. However,
H$\gamma\,\lambda4340$ is party filled-in by emission, while
H$\beta\,\lambda4861$ shows weak emission. Moreover, this spectrum
shows strong absorption of the OII+CIII\,$\lambda4640-4650$ blend,
indicative of an early B-type star.}

\item{Source 6-4 has been broadly classified as a B-type star which
shows Balmer emission (Garmany \& Humphreys 1985). In the present
work, we study this object more extensively. The composite nature of
this spectrum is revealed based on the HeI\,$\lambda4471$ and
HeII\,$\lambda4541$ lines, that indicate early-type (OB)
components. H$\beta\,\lambda4861$ is almost completely filled-in by
emission, while H$\eta\,\lambda3835$, H$\zeta\,\lambda3889$,
H$\epsilon\,\lambda3970$, H$\delta\,\lambda4101$ and
H$\gamma\,\lambda4340$ are in absorption. The strong
NIV\,$\lambda4058$ emission is the most prominent feature in the blue
portion of its optical spectrum (3800\AA-5000\AA; Figure
\ref{normspectra}). The NV\,$\lambda\lambda4604-20$ lines are not
present, while HeII\,$\lambda4686$ line is not in broad, low-intensity
emission, as it is typical for O3 If* stars (Walborn \& Fitzpatrick
1990). We do not detect the weak NIII\,$\lambda\lambda4634-40-42$
triplet emission, which in combination with strong 
HeII\,$\lambda4686$ absorption, has been used as a signature of O((f))
stars (e.g. Ma{\'{\i}}z-Apell{\'a}niz \etal 2004), but this could be due to the
spectral resolution (most probably blended with
NII\,$\lambda4631$). In contrast to the above indicators which suggest
an O spectral-type, HeII\,$\lambda4200$ is absent, indicating a B-type
star. Other helium lines clearly seen in this spectrum, include
HeI+II\,$\lambda4026$ and HeI\,$\lambda4922$. Taking into account the
above spectral features, we conclude that this is a composite spectrum
and we assign an O2((f))+OBe type to this system. Based on the current spectral resolution, a classification is not straightforward and only higher resolution spectroscopy can unambiguously identify the two components and their correct spectral types.}

\item{We classify source 5-2 as a B0.5e star based on the presence of
weak HeII\,$\lambda4686$, and the absence of
HeII\,$\lambda\lambda4200, 4541$ lines. Strong HeI+II\,$\lambda4026$
and HeI\,$\lambda\lambda4121, 4144, 4471$ absorption lines are in
agreement with the above classification. HeI\,$\lambda4922$ appears
weaker, while HeI\,$\lambda5876$ shows strong, broad emission. This
source has been classified as a later type star (B1-2 IV-Ve) by 
McBride \etal (2008).}

\item{In the same work, source 5-7 is classified as an O9 Ve star
(McBride \etal 2008). We find that this is a later (B0e) spectral-type
star based on the presence of HeI and HeII lines in the blue part of
its spectrum (3800\AA-5000\AA; Figure \ref{normspectra}). In
particular, HeII\,$\lambda\lambda4200, 4541$ lines are present, but
they are much more weaker than HeI\,$\lambda4144$. HeII\,$\lambda4686$
is also weak. HeI\,$\lambda\lambda4387, 4471, 4922$ lines are clearly
seen in this spectrum, while the OII+CIII,$\lambda4640-4650$ blend is
also present. The above lines are tell-tale signatures of a B0 star.}

\item{Source 3-3, which has been classified as B0-B2 by McBride \etal (2008), shows several conflicting spectral-type indicators. In our spectrum, the MgII\,$\lambda4481$ line, which appears in later than B2 spectral types, is present. It is also stronger than SiIII\,$\lambda4553$, indicating a B3 spectral type. For spectral types B5 or later SiIII\,$\lambda4553$ is absent, hence the latest spectral type that can be assigned to source 3-3 is B4. In addition, HeII\,$\lambda4686$ and the OII+CIII\,$\lambda4640-4650$ blend are very weak (almost absent), and on that account source 3-3 would be of B0.5-1.5 spectral type. On the other hand, SiIV\,$\lambda4116$ is present, which is typical of B1 stars, but not SiIV\,$\lambda4088$. Furthermore, HeII\,$\lambda4200$ and HeII\,$\lambda4541$ are present, indicating B0 and B0.5 spectral types, respectively. Based on the above spectral features, we only assign to source 3-3 a wide B0-4 spectral type. We also note that Balmer lines H$\zeta\,\lambda3889$ and H$\gamma\,\lambda4340$ are both filled-in by emission. The spectrum of this source also shows emission in NII\,$\lambda4045$ and MgI\,$\lambda4571$.}

\item{Source XMM-39 exhibits the strongest H$\beta\,\lambda4861$
emission of the sources from the current sample. Moreover,
H$\delta\,\lambda4101$ and H$\gamma\,\lambda4340$ are partly and
completely filled-in by emission, respectively. The optical
counterpart of this source has been observed with 2dF by Evans \etal
(2004), and it has been classified as a non-emission line B0-5(II)
star. We classify it as a B1.5e star because we find weak
SiIII\,$\lambda4553$ absorption and moderate
OII+CIII\,$\lambda4640-4650$ blend. This spectrum also shows
unidentified emission around 3951\AA.}

\item{Source XMM-47 has been classified as B0.5(IV)e by Evans \etal
(2004), however, following the above criteria, we classify it as
B1.5e. In particular, absence of the SiIV\,$\lambda4116$ $4650$ line
and OII+CIII\,$\lambda4640-4650$ blend suggests a B1.5 or later
spectral type source. On the other hand, no presence of the
MgII\,$\lambda4481$ line is indicative of an earlier than B2 star,
while a very weak HeII\,$\lambda4686$ line indicates a B0.5 (or
slightly earlier) spectral type (although the presence of this line is
ambiguous. We thus classify source XMM-47 as a B1.5e star.}

\end{itemize}

\section{Discussion }\label{2dFdiscussion}

In the previous sections we presented the spectral classification of
optical counterparts of 20 SMC HMXBs observed with the 2dF
spectrograph. We identify all of these sources as Be-XRBs. We also
compared these results with previous studies and we found that our
classification is in good agreement in all but four cases. Next we
discuss the implications of these results in the context of the
spectral-type distribution of Be-XRBs comparisons in the MCs and
the Galaxy. In addition, we investigate the distribution of the \Ha EW
measurements for Be-XRBs and Be field stars in these 3 galaxies and
their \Ha-orbital period (\Porbsimple) relation. We also investigate the intrinsic reddening values of the systems presented in this work.

\subsection{Distribution of spectral types}\label{SpTypediscuss}

SMC and Galactic Be-XRBs are found to follow the same spectral-type distribution (McBride \etal 2008). 
The same holds for LMC and Galactic Be-XRBs (Negueruela \& Coe 2002). 
Extending the above comparisons to Be stars, Coe \etal (2005)
observed somewhat similar distributions for Be-XRBs and Be stars in
the SMC, however using photometrically derived spectral types. In the
present work we supplement the SMC Be-XRB sample used in McBride \etal
(2008) with spectral-type classifications for 10 additional sources,
and we investigate differences in the spectral-type distributions of
both Be-XRBs and Be stars in the MCs and the Milky Way. For sources
common in the two studies, we have used the classification derived from the present work.

 In Figure \ref{Bspecsubtdistr} we present the B spectral subtype distributions of Be-XRBs (solid histograms) and of Be stars (dashed histograms) in the SMC (top panel), the LMC (middle panel), and the Milky Way (bottom panel). Negative spectral subtypes correspond to O-type stars. Whenever only a broad spectral class was available, we equally divided the contribution in the different subtypes. 

For the SMC, the data for Be stars are taken from Evans \etal
(2004; 168 in total), and for Be-XRBs from our sample supplemented by the recent study of
McBride \etal (2008; 46 in total). The Galactic Be
stars (148 in total) are taken from the study of Slettebak (1982). For
the Galactic Be-XRBs we used 20 sources from the census of Reig
(2007), 4 additional from the work of McBride \etal (2008), and 2 more
from the catalog of Liu \etal (2006). For the 14 LMC Be-XRBs we used
the spectral types from the census of Liu \etal (2005). The LMC Be
stars (103 in total) are taken from Martayan \etal (2006b).

In order to check the null hypothesis that different samples of
Be-XRBs and Be stars are drawn from the same populations, we perform
the two-sample Kolmogorov-Smirnov (KS) test (e.g., Conover 1971). The results of the
p-values are presented in Table \ref{KStest}, a two-way table with the
compared distributions for Be-XRBs and Be stars listed in its
rows and columns. Because of the small size of the LMC Be-XRBs sample
we also check these results with the two-sided Mann-Whitney-Wilcoxon
test (Bauer 1972). In addition, because it is very difficult to obtain
accurate spectral types for stars later than mid B-type in the
MCs, we repeat
the above analysis by using the Peto \& Prentice Generalized Wilcoxon
Test (Lavalley, Isobe \& Feigelson 1992) and assuming lower limits to the spectral subtypes
later than B3 in all samples and later than B1 and B2 in the LMC and Galactic
Be-XRB samples, respectively. The three tests agree only in the case of very small
($\ll$0.1) or very large ($\gg$0.9) p-values, while intermediate values
are not informative. In particular, small p-values ($\ll$0.1) are
indicative of the different populations from where these distributions
are drawn. Based on these results, we cannot definitely say if the SMC
Be-XRBs follow a different spectral-type distribution from Galactic
and LMC Be-XRBs. This is consistent with McBride \etal (2008), who
note that there is indication for similar distributions in the SMC and
the Milky Way. We find similar spectral-type
distributions between LMC and Galactic Be-XRBs (p-value=1.0), in agreement with
Negueruela \& Coe (2002), although we note the small
size of the samples. Regarding the distributions of Be-XRBs and
Be stars, we find that in the SMC the two populations are
consistent with each other, in contrast to the Milky Way populations
(in agreement with Negueruela, 1998). Moreover, we find similar Be spectral type
distributions between the MCs samples, while we do not
find evidence for differences between the MCs and the Milky Way.

Previous studies find the highest fraction of Be stars at B2 spectral
types (e.g. Martayan \etal 2007b). McBride \etal (2008) find a
spectral distribution cutoff for the SMC Be-XRBs around B2 and they
suggest this might be due to significant angular-momentum losses, even
before the Be-XRB evolutionary phase (i.e. during the first stage of
mass transfer). Ekstr{\"o}m \etal (2008) believe that the Be-star
phenomenon may be limited to a given spectral-type range because of
the mass loss and the meridional currents. In particular, they argue
that for spectral types earlier than the favored one, mass loss prevents the star from approaching the critical velocity limit (at which the Be-star phenomenon appears), while for the later spectral types, the meridional currents (which transfer angular momentum to the surface) are not efficient enough for accelerating the envelope. Regarding the effects of metallicity in the fraction of Be stars in a galaxy, it is now widely accepted that the lower the metallicity the higher this fraction, because of their higher rotational velocities (e.g. Maeder \etal 1999, Wisniewski \& Bjorkman 2006, Martayan \etal 2006b, 2007a; after correcting for multi- and single-epoch observations for the Milky Way and the MCs, respectively). Moreover, the Be star phase can last longer in low metallicity environments such as the MCs when compared to the Galaxy (Martayan \etal 2007b).

As pointed out in Zorec \& Briot (1997) the observed populations of Be
stars suffer from many incompleteness effects, especially in faint
magnitude limits. Since the quantification of these selection effects
is still in its infancy and depends on the used samples, we opted to
use the available uncorrected samples and simply treat the number of
stars at late spectral types as lower limits. This approach has lower
discriminating power but it does not rely on modeling assumption about
the selection effects.

Furthermore, the above samples include
both field and cluster Be star populations. In order to check if there
are different spectral type distributions due to their different ages
and rotational velocity characteristics, we examined 2 Magellanic Clouds samples which distinguish between them (Martayan \etal 2007b
and 2006b for the SMC and the LMC, respectively). We do not find any
difference in the spectral type distributions of Be field and cluster
populations, which could be due to the small size of these samples (79 and
52 Be stars for the SMC, and 27 and 20 for the LMC,
respectively). In any case, in the present work we do not distinguish between
the different Be star populations, especially since all Be-XRBs used
in our study are located in the field. Although there is a possibility
that they were formed in clusters and then ejected due to supernova
kicks, there is no way to test it. 

From the above analysis it is obvious that in comparisons of the spectral-type distribution of different populations, one should keep in mind that the Be phenomenon is a transient phase in the life of some B-type stars, greatly depending on the rotational velocity and perhaps binarity (e.g. McSwain et al. 2008). Because of the transient nature of the Be-star disks, we may then get only a fraction of the real underlying Be-star population with single-epoch observations, as is the majority in the SMC.

\subsection{Orbital Period}\label{Porbital}

The orbital period of an XRB system is related to the size of its
orbit, while the \Ha EW is considered an indicator of the size of the
decretion disk (e.g., Grundstrom \& Gies 2006; as observed by long baseline
interferometry of nearby Be stars). Reig \etal (1997) found a strong
correlation between orbital period and \Ha EW, and attributed it to the presence of the neutron star, which appears to act as a barrier which prevents the formation of a extended disk in systems with short orbital period. Figure \ref{EWPorb} is an updated version of the
${\rm P_{orb}}$-EW(H$\alpha$) diagram for Be-XRBs including in
addition to the 15 Galactic Be-XRBs (open circles) used in the work of
Reig (2007), 14 SMC sources (filled squares) derived
from the present work (values are taken from Table \ref{EWFWHM} and
references therein) and 2 recently detected X-ray pulsars (SXP7.78,
SXP455\footnotemark \footnotetext{Based on the online census of
Malcolm Coe as of 22 August 2008
(http://www.astro.soton.ac.uk/$\sim$mjc/).}). The orbital periods of these 2 sources are taken from Liu
\etal (2005; and references therein), while for the \Ha EW
measurements we used the work of Coe \etal (2005). In Figure
\ref{EWPorb} we only included sources with well determined orbital
periods, thus other SMC X-ray pulsars (such as SXP756, SXP8.80,
SXP46.6 and SXP304) with published \Ha EW and \Porb values are not
present. Five X-ray sources (out of the
20 used in Section \ref{SpTypediscuss}) do not have published orbital
periods, thus they do not appear in this figure. Source 4-1 is not included in
 Figure \ref{EWPorb} because of its complicated nature (Coe \etal 2002). We also note that the
\Ha EW measurements used here are the maximum found in the literature
(thus these measurements can be considered upper limits of the disk
size; e.g Dachs \etal 1992).

 This plot strongly supports the
correlation of these two quantities despite the large scatter observed
for the shorter period systems and extends it to extragalactic systems. As Reig (2007) points out, the neutron star does not allow the companion star to develop a large decretion disk in cases of small orbital period systems, leading to the tidal truncation of the disk, and in turn smaller \Ha EW values. The values of the linear regression
correlation coefficient for the SMC, Milky Way and the combined sample are
0.71, 0.89, and 0.81, respectively. Thus, the SMC
sample (solid line) is less correlated when compared to the Galactic
sample (dashed line), while the combined data sample (dotted line)
shows a stronger correlation than the SMC data alone because of the
increased sample size.

From this
plot, we also see that the SMC sources tend to have systematically larger \Ha EW than
Galactic sources with similar orbital periods. This could be due to
their faster rotational velocities (a result of their lower metallicity, as observed for example by Martayan \etal 2007b) and hence larger decretion disks. However, given the truncation effect discussed earlier,
growing a larger disk would require a wider orbit which for the same
orbital period would indicate a lower ellipticity. Alternatively, a
high mass-loss rate would allow a faster growth of the disk after the truncation.

\subsection{Intrinsic reddening}\label{IntrinsicReddening}

In addition to the \Ha EW, the ${\rm (J-K)_{o}}$ color is also an
indicator of the size of the circumstellar disk. By plotting these
quantities, Coe \etal (2005)
find a similar range in the values of Galactic Be stars and SMC
Be-XRBs, but not significant evidence for a correlation between the
\Ha EW and the ${\rm (J-K)_{o}}$ color for the Be-XRBs. Here we
revisit this study using our extended census of Be-XRBs and comparing
with SMC Be stars from the study of NGC330 by Martayan \etal
(2007a). The \Ha EWs for the Be-XRBs are taken from Table \ref{EWFWHM} and the J and K magnitudes from Table
\ref{2dFoptIDs}, corrected for reddening assuming E(J-K)=0.56E(B-V) (Bessell \& Brett
1988), and E(B-V)=0.09 mag (following Antoniou \etal 2009). The \Ha EW
measurements for the NGC330 Be stars are taken from Martayan \etal
(2007a), while the ${\rm (J-K)_{o}}$ color values are the result of the
cross-correlation of the above census with the 2MASS
catalog\footnotemark. \footnotetext{We searched for the nearest 2MASS
counterpart of the NGC330 Be stars within 1.5\arcsec\, from their
position. For the 131 NGC330 stars, we found 88 matches but we kept
only 48, because the remaining do not have constrained J and/or K
magnitude errors listed in the 2MASS catalog.}

A comparison of these two quantities does not show any evidence for
correlation, which we attribute to the fact that the IR and \Ha observations are not
simultaneous, and thus the comparison between the two quantities is
not direct. Therefore, we do not pursue this comparison
further. Instead we focus on the individual distributions of the \Ha
EW and (J-K) color. As shown in Figures \ref{EWdistr} and \ref{JKo}, respectively, the majority of the Be-XRBs presented in this work have
\Ha EWs extending up to the mid range of the corresponding values for Be stars, while both the Be stars and the
Be-XRBs in the SMC have similar values of the ${\rm (J-K)_{o}}$
color. When we compare the results for the Be-XRBs presented in this
work with the study of Coe \etal
(2005), we find compatible values for the \Ha EWs, however
most of the sources in the present work have larger values of the ${\rm (J-K)_{o}}$ color
($\sim0.2-0.4$ mag; Figure \ref{JKo})
than those presented by Coe \etal (2005; ${\rm (J-K)_{o}}<0.2$ mag). This can be the result of the different extinction correction in the two studies. We have used E(J-K)=0.05, while Coe \etal (2005) corrected for all reddening effects by using the same color shift used for determining the spectral class of their objects (shift by (B-V)=-0.13 from their study).

We then investigate the correlation of the \Ha EW measurements to the
intrinsic reddening caused by the circumstellar disks around Be
stars. We define this ``local" reddening as the difference in the {\it
B-V}
color of the observed (extinction corrected) and theoretical
values. The observed {\it B-V} color is taken from the MCPS catalog of
Zaritsky \etal 2002 (since all the Be-XRBs from the present study have
an optical counterpart in this catalog) after correcting for the
extinction to the SMC (E(B-V)=0.09; see \S \ref{2dFoptcparts}). In
order to derive the theoretical {\it B-V} color, we first assign an
effective temperature to each B subtype using the temperature scale
for main-sequence B-type stars at the SMC metallicity (Figure 33(d) of
Hunter \etal 2007). We then derived the theoretical {\it B-V} color using
the Geneva isochrones database (Lejeune \& Schaerer 2001) assuming
Z=0.004 (for the SMC metallicity) and the nearest corresponding
available effective temperature to the one derived above. The \Ha EW
versus intrinsic reddening plot is presented in Figure
\ref{LocalReddening}, while the values of these quantities are given
in Table \ref{deltaBV}. In Column 1 we give the source ID, in Column
2 the spectral classification from this study and in Columns 3 and
4 the extinction corrected $(B-V)_{\rm {o}}$ color and the corresponding error
(using data listed in Table \ref{2dFoptIDs}), respectively. In Column
5 we give the nearest B subtype to the classified sources used in
order to derive the effective temperature. In Column 6 we give the
{\it B-V} color from the isochrones, and in Column 7 the intrinsic
reddening (extinction corrected photometric minus theoretical {\it B-V}
values). In Table \ref{deltaBV} and Figure \ref{LocalReddening} we did
not include source 6-4 because its composite spectrum indicates that
there might be more than one stars contributing to its emission. From this
figure, we do not find any correlation between these two quantities, though
they are both related to the decretion disk of the Be star. The large
scatter in the plot is mainly due to the fact that the theoretical
values for the {\it (B-V)} color are in some degree uncertain. First, we had
to assume a mean subtype for sources with uncertain spectral
class (e.g. X-ray sources 3-3, 6-1, 5-12). Then, we
had to extrapolate the temperature scale of Hunter \etal
(2007) to later than B1 spectral subtypes. And finally, we had to
choose the nearest effective temperature derived above to the one used
in the Geneva database. All the above, in addition to disk inclination
effects, can account, at least in part, for the large scatter of this
plot, and may explain the fact that we do not find any correlation of
the \Ha EW and the intrinsic reddening, though they are both related
to the size of the circumstellar disk around Be stars. We also note
that even if we use the mean {\it B-V} color for each B subtype using data
from Evans \etal (2004), we do not find any correlation either. The
fact that the spectroscopic and photometric observations are not
simultaneous may also result in increased scatter.

\section{Summary}\label{conclusions}

We have presented the results of a detailed optical spectroscopic analysis of 20 HMXBs in the SMC. These sources have been detected with \chandra (A. Zezas \etal 2009, in preparation) and {\it XMM-Newton} (Haberl \& Pietsch 2004). In the present work, we confirm the Be nature of all these systems. In addition, we provide their full wavelength range ($\sim{\rm 3650\AA-8000\AA}$) spectra.

The classification of these 20 systems is accurate to better than
$\pm1$ subclass and is proved to be in good agreement with previous
studies. In a few cases, only wide range
spectral-types could be assigned, and for these systems higher
resolution spectroscopy is needed in order to derive more accurate
classifications. We also find similar spectral-type distributions for
Be-XRBs and Be field stars in the SMC. With the present available data we cannot conclude that the
Be-XRBs in the SMC and the Milky Way or the LMC follow a different
spectral-type distribution. In addition, we find similar Be spectral type
distributions between the MCs samples, while we do not
find evidence for differences between the MCs and the Milky
Way.

Moreover, our results reinforce the relation between the
orbital period and the equivalent width of the \Ha line that holds for
Be-XRBs. This
relation is the result of truncated decretion disks in smaller period
systems, due to the interaction of the compact object with the circumstellar disk of
the Be star. For similar orbital periods, SMC sources tend to have systematically larger \Ha EW than
Galactic sources. This could be due to
their faster rotational velocities (a result of their lower metallicity, as observed for example by Martayan \etal 2007b) and hence larger decretion disks. 

 We also
investigate the near-infrared properties of the 20 Be-XRBs by
cross-correlating their X-ray positions with sources in the 2MASS
catalog. For sources with a 2MASS counterpart we do not find evidence
for 
correlation of the \Ha EW values and the J-K color, though they are
both related to the size of the decretion disk. The same holds for the
\Ha EW values when compared to the intrinsic reddening of the sources
(defined as the observed minus the theoretical {\it B-V} color).

\acknowledgements
We thank the referee, Virginia McSwain, for useful comments which have improved this
paper. We would also like to thank Rob Sharp for performing the 2dF
observations during service time and Nolan Walborn for fruitful
discussions on the spectral classification. VA acknowledges support
from Marie Curie grant no. 39965 to the Foundation for Research and
Technology - Hellas, NASA LTSA grant NAG5-13056, and NASA grants
GO2-3117X and GO8-9089C. AZ acknowledges support by FP-7 RERPOT grant 206469. This publication makes use of data products from the Two Micron All Sky Survey, which is a joint project of the University of Massachusetts and the Infrared Processing and Analysis Center/California Institute of Technology, funded by the National Aeronautics and Space Administration and the National Science Foundation.

{}

\makeatletter
\def\jnl@aj{AJ}
\ifx\revtex@jnl\jnl@aj\let\tablebreak=\nl\fi
\makeatother
\begin{deluxetable}{cccc}
\tabletypesize{\scriptsize}
\tablecolumns{4}
\tablewidth{0pt}
\tablecaption{The Be-XRB sample observed with 2dF\label{otherXrayname}}
\tablehead{\colhead{X-ray Src ID\tablenotemark{\dag}} & \colhead{Other Source Names} & \colhead{Pulsar ID\tablenotemark{\ddag}} & \colhead{[MA93]\tablenotemark{\star}}\\[0.2cm]
\colhead{[1]} & \colhead{[2]} & \colhead{[3]} & \colhead{[4]}}
\startdata
 XMM-3                   & XMMU J004723.7-731226 (1) = (RX J0047.3-7312 AX J0047.3-7312) (2)                                                    & SXP264      & 172           \\[0.1cm]
   4-8                        & CXOU J004814.2-731004 (3) = XMMU J004814.1-731003 (4)                                                                            & SXP25.5    & \nodata  \\[0.1cm]
 XMM-5                    & XMMU J004834.5-730230 (1) = (RX J0048.5-7302) (5)                                                                               & \nodata  & 238           \\[0.1cm]
 4-2 (XMM-7)          & CXOU J004913.6-731138 = XMMU J004913.8-731136 (1) = (RX J0049.2-7311) (6)                             & SXP9.13         & \nodata    \\[0.1cm]
4-5 (XMM-9)       & CXOU J004929.7-731059 (3) = XMMU J004929.9-731058 (1) & SXP893   & 300         \\
                  & = (RX J0049.5-7310) (6)                                &         &              \\[0.1cm]
4-1 (XMM-12)      & CXOU J005044.6-731605 (3) = XMMU J005045.2-731602 (1) & SXP323  & 387           \\
                  & = (RX J0050.8-7316 = AX J0051-733) (2,7)              &         &               \\[0.1cm]
XMM-15                  & AX J0051-722 (8)                                                                                                                                                          & SXP91.1      & 413::       \\[0.1cm]
4-3 (XMM-14)        & CXOU J005057.2-731008 = XMMU J005057.6-731007 (1)    & \nodata         &  414        \\
                    &  = (RX J0050.9-7310, AX J0050.8-7310) (2)             &                 &             \\[0.1cm]
XMM-17                  & XMMU J005152.2-731033 (1) = (RX J0051.9-7311 = AX J0051.6-7311) (2,7)                                            & SXP172       & 504          \\[0.1cm]
   6-1                        & CXOU J005209.0-723804 (3)                                                                                                                                    & SXP82.4       & \nodata   \\[0.1cm]
5-12                         & CXOU J005245.0-722844 (3)              & \nodata    & \nodata\\[0.1cm]
   6-4                        & CXOU J005252.2-724830 (3)                                                                                                                                      & \nodata   & 618         \\[0.1cm]
   5-2                        & CXOU J005323.9-722716 (3)                                                                                                                                       & SXP138       & 667         \\[0.1cm]
  5-16                       & CXOU J005355.3-722646 (3) = 1WGA 0053.8-7226,XTE J0053-724                                                          & SXP46.6    & \nodata  \\[0.1cm]
   5-4                        & CXOU J005446.2-722523 (3)                                                                                                                                    & \nodata    & 798         \\[0.1cm]
   6-2                        & CXOU J005455.8-724511 (3)                                                                                                                                   & SXP504         & 809         \\[0.1cm]
5-7 (XMM-30)      & CXOU J005456.3-722648 (3) =                                         & SXP59.0   & 810          \\
                  & (XTE J0055-724 = 1SAX J0054.9-7226 = 1WGA J0054.9-7226) (9,10,11)   &           &              \\[0.1cm]
3-3 (XMM-33)      & CXOU J005736.0-721934 (3)  = CXOU J005736.2-721934 (11,12)                                                                          & SXP565   & 1020        \\[0.1cm]
XMM-39                  & XMMU J010030.2-722035 (11)                                                                                                                            & \nodata   & 1208        \\[0.1cm]
XMM-47                   & RX J0104.5-7221 (5)                                                                                                                                         & \nodata  & 1470    \\    
\tablenotetext{\dag}{The X-ray source ID corresponds to \chandra
sources as in Antoniou \etal (2009; (Field ID)-(Src ID within the
field)). The XMM ID and their association with sources from other works are from Haberl \& Pietsch (2004).}
\tablenotetext{\ddag}{Source names related to their X-ray pulse periods (in seconds) and based on the online census of Malcolm Coe as of 22 August 2008 (http://www.astro.soton.ac.uk/$\sim$mjc/).}
\tablenotetext{\star}{Emission-line object ID from the catalog of Meyssonnier \& Azzopardi (1993).}
\tablenotetext{::}{The \Ha emission-line nature of the object is very doubtful, mainly due to its severe faintness or because it is a late-type star (Meyssonnier \& Azzopardi 1993).}
\tablerefs{(1) Haberl \& Pietsch (2004), (2) Yokogawa \etal (2003), (3) A. Zezas \etal (2009; in preparation), (4) Haberl \etal (2008), (5) Haberl \& Sasaki (2000), (6) Filipovic \etal (2000), (7) Schmidtke \etal (1999), (8) Corbet \etal (1998), (9) Marshall \etal (1998), (10) Santangelo \etal (1998), (11) Sasaki, Pietsch \& Haberl (2003), (12) Macomb \etal (2003)}
\enddata
\end{deluxetable}

\makeatletter
\def\jnl@aj{AJ}
\ifx\revtex@jnl\jnl@aj\let\tablebreak=\nl\fi
\makeatother
\begin{deluxetable}{cccccccccc}
\tabletypesize{\scriptsize}
\tablecolumns{10}
\tablewidth{0pt}
\tablecaption{Optical and near-infrared counterparts of selected X-ray sources with 2dF spectra from the present work\label{2dFoptIDs}}
\tablehead{\multicolumn{3}{c}{X-ray} & \colhead{Optical} & \colhead{Off.\tablenotemark{\ddag}}  & \colhead{$V$} & \colhead{$(B-V)$} & \colhead{2MASS ID\tablenotemark{\star}}  & \colhead{$J$} & \colhead{$K$} \\
\colhead{Src ID\tablenotemark{\dag}} & \colhead{R.A.} & \colhead{Decl.} & \colhead{Src ID} & \colhead{X-O} & \colhead{} & \colhead{} & \colhead{} & \colhead{}  & \colhead{} \\
\colhead{} & \colhead{(h m s)} & \colhead{(\deg\, \arcmin\, \arcsec\,)} & \colhead{} & \colhead{(\arcsec)} & \multicolumn{2}{c}{(mag)} & \colhead{} & \multicolumn{2}{c}{(mag)} \\[0.2cm]
\colhead{[1]} & \colhead{[2]} & \colhead{[3]} & \colhead{[4]} & \colhead{[5]} & \colhead{[6]} & \colhead{[7]} & \colhead{[8]} & \colhead{[9]} & \colhead{[10]}}
\startdata
XMM-3        & 00 47 23.7     & -73 12 27    & O-4-116979   & 1.52   & $16.14\pm0.01$ & $-0.04\pm0.01$ & J00472330-7312275 & $15.88\pm0.10$ & $14.84\pm$\nodata \\
                     &                         &                      & Z-1713720     & 1.25   & $16.03\pm0.03$ &  $0.08\pm0.04$  &                                        &                                &                                   \\[0.15cm]
4-8         &  00 48 14.15   & -73 10  4.1  & O-4-171264   & 0.63   & $15.74\pm0.04$ &  $0.00\pm0.05$   & J00481410-7310045 & $15.02\pm0.06$ & $14.69\pm0.12$ \\
                 &                           &                      & Z-1816472    & 0.30   & $15.30\pm0.05$ &  $0.26\pm0.06$  &                                        &                                &                                 \\[0.15cm]
XMM-5        & 00 48 34.5     & -73 02 30  & O-4-178950     & 1.85   & $14.95\pm0.04$ & $-0.09\pm0.06$  & \nodata & \nodata & \nodata \\
                     &                         &                    & =O-5-43566     & 1.89   & $15.11\pm0.12$ & $-0.23\pm0.13$   &                                        &                                &                                   \\
                     &                         &                    & Z-1857369       & 1.64   & $14.78\pm0.03$ &  $0.00\pm0.04$    &                                        &                                &                                   \\[0.15cm]
4-2       & 00 49 13.8     & -73 11 37  & O-5-111490     & 0.84   & $16.52\pm0.02$ & $0.10\pm0.04$   & J00491360-7311378  & $16.11\pm0.09$ & $15.88\pm$\nodata \\
 (XMM-7)                     &                        &                     & Z-1938257       & 0.75   & $16.44\pm0.04$ & $0.19\pm0.05$   &                                        &                                &                                   \\[0.15cm]
4-5         &  00 49 29.74   & -73 10 58.5  & O-5-111500   & 0.61   & $16.30\pm0.01$ &  $0.09\pm0.02$  & J00492984-7310583 & $15.59\pm0.11$ & $15.01\pm0.16$ \\
(XMM-9)&                         &                        & Z-1971979     & 0.65   & $16.15\pm0.03$ &  $0.20\pm0.05$  &                                        &                                &                                \\[0.15cm]
4-1         &  00 50 44.61   & -73 16  5.3  & O-5-180026   & 0.55   & $15.44\pm0.04$ & $-0.04\pm0.05$   & J00504470-7316054 & $15.30\pm0.05$ & $14.81\pm0.11$ \\
(XMM-12)          &                           &                       & Z-2131651     & 0.60   & $15.48\pm0.04$ & $-0.11\pm0.05$ &                                        &                                &                                \\[0.15cm]
XMM-15       & 00 50 56.9     & -72 13 31    & Z-2158744     & 3.22   & $15.06\pm0.06$ & $-0.08\pm0.06$ & \nodata & \nodata & \nodata \\[0.15cm]
 4-3     & 00 50 57.6     & -73 10 08    & O-5-271074   & 2.10   & $14.54\pm0.01$ & $-0.06\pm0.01$ & \nodata & \nodata & \nodata \\
 (XMM-14)                      &                        &                       & Z-2159045     & 1.93   & $14.35\pm0.05$ &  $0.08\pm0.06$ &                                        &                                &                                \\[0.15cm]
XMM-17       & 00 51 52.3     & -73 10 33    & O-6-22749    & 1.36   & $14.48\pm0.02$ & $-0.08\pm0.02$ & J00515203-7310340 & $14.43\pm0.03$ & $14.17\pm0.07$ \\
                       &                        &                      & Z-2280409     & 1.32   & $14.45\pm0.05$ & $-0.07\pm0.06$ &                                        &                                &                                \\[0.15cm]
6-1         & 00 52  8.95    & -72 38  3.5  & O-6-77228    & 0.58   & $15.03\pm0.02$ &  $0.14\pm0.03$       & J00520896-7238032 & $14.63\pm0.03$ & $14.21\pm0.06$ \\
                 &                         &                      & Z-2319498    & 0.64   & $15.23\pm0.03$ & $-0.08\pm0.03$     &                                        &                                &                                  \\[0.15cm]
 5-12            & 00 52 45.04   & -72 28 43.6 & Z-2406014       & 0.35  & $14.92\pm0.08$ & $0.00\pm0.09$ & J00524508-7228437 & $14.97\pm0.05$ & $14.90\pm0.12$\\[0.15cm]
6-4         & 00 52 52.22    & -72 48 29.8  & O-6-147662   & 0.27   & $14.42\pm0.05$ & $-0.10\pm0.05$   & J00525230-7248301 & $14.24\pm0.05$ & $13.98\pm0.08$ \\
              &                            &                       & Z-2423181     & 0.56   & $14.36\pm0.03$ & $-0.05\pm0.04$    &                                        &                                &                                \\[0.15cm]
5-2         &  00 53 23.86   & -72 27 15.5  & Z-2498173     & 0.24   & $16.19\pm0.12$ & $-0.09\pm0.12$  & J00532381-7227152 & $16.27\pm0.11$ & $16.51\pm$\nodata \\[0.15cm]
5-16        &  00 53 55.25   & -72 26 45.8  & Z-2573354     & 0.83   & $14.72\pm0.03$ & $-0.07\pm0.03$ & J00535518-7226448 & $14.41\pm0.04$ & $14.00\pm0.07$ \\[0.15cm]
5-4         & 00 54 46.22    & -72 25 23.0  & O-7-70843    & 0.79   & $15.58\pm0.02$ & $0.14\pm0.06$     & J00544633-7225228 & $15.29\pm0.06$ & $14.84\pm0.12$  \\
                &                            &                       & Z-2707354    & 0.83   & $15.36\pm0.05$ & $0.14\pm0.06$   &                                        &                                &                                \\[0.15cm]
6-2         & 00 54 55.78    & -72 45 10.7  & O-7-47103    & 0.40   & $15.01\pm0.01$ & $-0.02\pm0.01$    & J00545586-7245108 & $14.77\pm0.04$ & $14.40\pm0.07$  \\
                 &                           &                       & Z-2729974    & 0.61   & $15.00\pm0.03$ & $-0.03\pm0.04$  &                                        &                                &                                 \\[0.15cm]
5-7         & 00 54 56.34    & -72 26 48.4  & O-7-70829    & 1.19   & $15.30\pm0.01$ & $-0.04\pm0.02$    & J00545618-7226478 & $15.18\pm0.05$ & $15.01\pm0.13$ \\
(XMM-30)     &                  &                        & Z-2730786    & 1.21   & $15.27\pm0.03$ & $-0.05\pm0.04$    &                                        &                                &                                \\[0.15cm]
3-3         & 00 57 36.00    & -72 19 33.9  & O-8-49531    & 0.14   & $16.01\pm0.02$ & $-0.02\pm0.04$ & J00573602-7219341 & $15.74\pm0.07$ & $15.33\pm0.20$  \\
(XMM-33)     &                   &                       & Z-3103982    & 0.53   & $15.99\pm0.03$ & $0.01\pm0.17$  &                                        &                                &                                \\[0.15cm]
XMM-39       & 01 00 30.2     & -72 20 35  & O-9-35989      & 2.04   & $14.65\pm0.02$ & $-0.06\pm0.03$  & J01003000-7220335 & $14.50\pm0.04$ & $14.13\pm0.07$ \\
                      &                         &                     & Z-3499823      & 2.13   & $14.64\pm0.03$ & $-0.06\pm0.14$  &                                        &                                &                                    \\[0.15cm]
XMM-47       & 01 04 35.7     & -72 21 43  & O-10-61612 &  5.28  & $15.19\pm0.02$ &  $-0.01\pm0.03$ & \nodata & \nodata & \nodata \\
                      &                          &                    & Z-4031467   &  4.91 &  $15.13\pm0.03$ &  $0.05\pm0.03$   &                                        &                                &         \\                          
\tablenotetext{\dag}{Same as in Table \ref{otherXrayname}.}
\tablenotetext{\ddag}{The offset (in arcseconds) between the counterpart and the X-ray source.}
\tablenotetext{\star}{Closest counterpart within 2\arcsec\, from the X-ray source position.}
\enddata
\end{deluxetable}

\makeatletter
\def\jnl@aj{AJ}
\ifx\revtex@jnl\jnl@aj\let\tablebreak=\nl\fi
\makeatother
\begin{deluxetable}{ccccccccc}
\tabletypesize{\scriptsize}
\tablecolumns{9}
\tablewidth{0pt}
\tablecaption{Measurements of the \Ha emission in the Be-XRB sample\label{EWFWHM}}
\tablehead{\colhead{X-ray Src ID\tablenotemark{\dag}} & \colhead{Center}  & \colhead{$\rm{\delta (Center)\tablenotemark{\ddag}}$} & \colhead{FWHM} & \colhead{$\rm{\delta (FWHM)\tablenotemark{\ddag}}$}  & \colhead{EW} & \colhead{$\rm{\delta (EW)\tablenotemark{\ddag}}$} & \colhead{EW (Refer.)} & \colhead{${\rm P_{orb}}$\tablenotemark{\star}}\\
\colhead{} & \colhead{(\AA)} & \colhead{(\AA)} & \colhead{(\AA)} & \colhead{(\AA)} & \colhead{(\AA)} & \colhead{(\AA)} & \colhead{(\AA)} & \colhead{(d)} \\[0.2cm]
\colhead{[1]} & \colhead{[2]} & \colhead{[3]} & \colhead{[4]} & \colhead{[5]} & \colhead{[6]} & \colhead{[7]} & \colhead{[8]} & \colhead{[9]}}
\startdata
 XMM-3        & 6564.57      & 0.06    & 11.32    & 0.14    & -24.40  &   0.48 & $-31.6\pm0.9$ (1) & 48.8 \\[0.1cm]
   4-8        & 6565.61      & 0.12    & 11.08    & 0.26    & -15.68  &   0.33  & \nodata   & \nodata \\[0.1cm]
XMM-5        & 6566.06      & 0.03    & 12.55    & 0.08    & -36.03  &   0.24  & \nodata   & \nodata \\[0.1cm]
  4-2 (XMM-7)        & 6565.49      & 0.03    & 12.13    & 0.08    & -32.41  &   0.40  & $-29.6\pm1.5$ (1)   & 91.5 \\[0.1cm]
4-5 (XMM-9)& 6566.09      & 0.08    & 15.51    & 0.21    & -38.47  &   0.70  & \nodata  & 91.5 \\[0.1cm]
   4-1 (XMM-12) & 6566.27      & 0.08    & 14.11    & 0.21    & -28.30  &   0.82  & $-24.5\pm0.7$ (1) & 0.708 \\[0.1cm]
XMM-15        & 6566.12      & 0.04    & 14.83    & 0.10    & -32.04  &   1.07  & -22 (2)   & 88.4 \\[0.1cm]
4-3 (XMM-14)        & 6564.99      & 0.05    & 10.53    & 0.11    & -20.37  &   0.35  & \nodata   & \nodata  \\[0.1cm]
XMM-17        & 6566.71      & 0.10    & 11.97    & 0.25    & -16.16  &   0.26  & $-13.1\pm0.5$ (1) & 67.0 \\[0.1cm]
   6-1        & 6566.94      & 0.04    & 10.56    & 0.09    & -23.87  &   0.38  & \nodata    & \nodata  \\[0.1cm]
5-12             & 6565.67       & 0.26    & 10.44    & 0.62    &   -5.38  &  0.11  & \nodata   & \nodata \\[0.1cm]
   6-4        & 6565.83      & 0.16    & 13.05    & 0.38    &  -6.59  &   0.18  & \nodata   & \nodata \\[0.1cm]
   5-2        & 6565.92      & 0.09    & 13.26    & 0.24    & -24.02  &   0.63  & \nodata & 125.0 \\[0.1cm]
  5-16        & 6566.00      & 0.05    & 11.76    & 0.12    & -19.48  &   0.46  & $-21.9\pm0.7$ (1)   & \nodata \\[0.1cm]
   5-4        & 6566.84      & 0.09    & 12.83    & 0.24    & -36.19  &   1.06  & \nodata   & \nodata  \\[0.1cm]
   6-2        & 6566.14      & 0.03    & 12.40    & 0.08    & -56.00  &   0.95  & \nodata & 261.0 \\[0.1cm]
5-7 (XMM-30) & 6565.67      & 0.04    & 10.90    & 0.10    & -19.97  &   0.66  & $-25\pm2$ (2)   & 123.0  \\[0.1cm]
3-3 (XMM-33) & 6565.78      & 0.09    & 11.98    & 0.22    & -32.08  &   0.45  & $-28\pm2$ (1) & 95.3 \\[0.1cm]
XMM-39        & 6565.74      & 0.03    & 10.33    & 0.07    & -65.02  &   1.44  & \nodata   & \nodata  \\[0.1cm]
XMM-47        & 6565.82      & 0.06    & 11.94    & 0.16    & -38.64  &   0.60 & \nodata   & \nodata \\
\tablenotetext{\dag}{Same as in Table \ref{otherXrayname}.}
\tablenotetext{\ddag}{All errors are 68\% confidence intervals for one
interesting parameter.}
\tablenotetext{\star}{Orbital period in days from Liu \etal (2005) and references therein.}
\tablerefs{(1) Coe \etal (2005), (2) Stevens, Coe \& Buckley (1999)}
\enddata
\end{deluxetable}

\makeatletter
\def\jnl@aj{AJ}
\ifx\revtex@jnl\jnl@aj\let\tablebreak=\nl\fi
\makeatother
\begin{deluxetable}{cc}
\tabletypesize{\scriptsize}
\tablecolumns{2}
\tablewidth{0pt}
\tablecaption{Line diagnostics for B spectral-type subclasses\label{identifications}}
\tablehead{\colhead{Line Identifications} & \colhead{Spectral Class}\\[0.2cm]
\colhead{[1]} & \colhead{[2]}}
\startdata
HeII\,$\lambda4200$, HeII\,$\lambda4541$, HeII\,$\lambda4686$ present & Earlier than B0 \\[0.1cm]
HeII\,$\lambda4541$, HeII\,$\lambda4686$ present & B0 \\[0.1cm]
HeII\,$\lambda4200$ and HeII\,$\lambda4541$ absent, HeII\,$\lambda4686$ weak      & B0.5 \\[0.1cm]
HeII\,$\lambda4686$ absent, SiIV\,$\lambda\lambda4088,4116$ present                         & B1 \\[0.1cm]
SiIV\,$\lambda4116$ absent, SiIII\,$\lambda4553$ appear                                                  & B1.5 \\[0.1cm]
OII+CIII\,$\lambda4640-4650$ blend decreases rapidly                                                        & Later than B1.5 \\[0.1cm]
MgII\,$\lambda4481$ $<$ SiIII\,$\lambda4553$                                                                       & B2 \\[0.1cm]
MgII\,$\lambda4481$ $\sim$ SiIII\,$\lambda4553$                                                                  & B2.5 \\[0.1cm]
MgII\,$\lambda4481$ $>$ SiIII\,$\lambda4553$                                                                       & B3 \\[0.1cm]
OII+CIII\,$\lambda4640-4650$ blend disappears                                                                     & Later than B3 \\[0.1cm]
SiIII\,$\lambda4553$ absent                                                                                                         & B5 \\[0.1cm]
HeI\,$\lambda4121$ $<$ SiII\,$\lambda4128$  $<$ HeI\,$\lambda4144$                          & B8 \\[0.1cm]
HeI\,$\lambda4471$ $<$ MgII\,$\lambda4481$                                                                       & B9
\enddata
\end{deluxetable}

\makeatletter
\def\jnl@aj{AJ}
\ifx\revtex@jnl\jnl@aj\let\tablebreak=\nl\fi
\makeatother
\begin{deluxetable}{ccc}
\tabletypesize{\scriptsize}
\tablecolumns{3}
\tablewidth{0pt}
\tablecaption{Spectral classification results\label{spclass}}
\tablehead{\colhead{X-ray} & \multicolumn{2}{c}{Classification}\\
\colhead{Src ID\tablenotemark{\dag}} & \colhead{This Study} & \colhead{Previous} \\[0.2cm]
\colhead{[1]} & \colhead{[2]} & \colhead{[3]}}
\startdata
XMM-3             &   B0.5e                                               & B1-1.5 Ve (1)                           \\[0.15cm] 
  4-8                  &   B1.5e                                         & Later than B3e\tablenotemark{\ddag} (2), Be? (3)  \\[0.15cm] 
 XMM-5            &   B1.5e                                           & B0-5 (II) = [2dF]690\tablenotemark{*} (4), Be? (5)    \\[0.15cm] 
4-2 (XMM-7)            &  B1e                                             & B1-3 IV-Ve (1)                                  \\[0.15cm] 
4-5 (XMM-9)    &  B1e                                               & Be? (6)                                       \\[0.15cm] 
  4-1 (XMM-12) &  B0.5e                                         & B0-0.5 Ve (7)                            \\[0.15cm] 
XMM-15           &  B0.5e                                       & B0.5 III-Ve (1)                         \\[0.15cm]
4-3 (XMM-14)           &  B0.5e                                           &  Be? (6)                      \\[0.15cm]
XMM-17           &  B0e                                                & Be ((8), peculiar (9))                 \\
                          &                                                          & O9.5-B0 Ve (1)                        \\[0.15cm]
  6-1                   & B1-3e                                         & B1-5 (II)e = [2dF]5054\tablenotemark{*} (4), Be (8)  \\
                           &                                                      & B1-3 III-Ve (1)                            \\[0.15cm]
 5-12                  &  O9-B0e                                                & Be? (3) \\[0.15cm]  
 6-4                    &  O2((f))+OBe                                 & Be (5), B type with Balmer emission (10) \\[0.15cm]
  5-2                  &  B0.5e                                          & B1-2 IV-Ve (1)                          \\[0.15cm]
 5-16                 &   B0.5e                                         & O9.5-B1 IV-Ve (1)                     \\[0.15cm]
   5-4                  &  B1e                         & Be? (3)                                      \\[0.15cm] 
 6-2                    &  B1e                                      & B1 III-Ve (1)                                \\[0.15cm]
  5-7 (XMM-30) &  B0e                                           &  O9 Ve (1)                                  \\[0.15cm]
3-3 (XMM-33)   &  B0-4e                                       & B0-2 IV-Ve (1)                           \\[0.15cm] 
  XMM-39           &  B1.5e                                              & B0-5 (II) = [2dF]1475\tablenotemark{*} (4), Be? (6) \\[0.15cm]  
XMM-47          &   B1.5e                                           & B0.5 (IV)e = [2dF]1905\tablenotemark{*} (4), Be? (5)  \\
\tablenotetext{\dag}{Same as in Table \ref{otherXrayname}.}
\tablenotetext{\ddag}{Photometric classification.}
\tablenotetext{*}{Sources from Evans \etal (2004) appear as [2dF]ID, where ID comes from the original catalog.}
\tablerefs{(1) McBride \etal (2008), (2) Haberl \etal (2008), (3) Antoniou \etal (2009), (4) Evans \etal (2004), (5) Haberl \& Sasaki (2000), (6) Haberl \& Pietsch (2004), (7) Coe \etal (2002), (8) Coe \etal (2005), (9) Schmidtke \etal (1999), (10) Garmany \& Humphreys (1985)}
\enddata
\end{deluxetable}

\makeatletter
\def\jnl@aj{AJ}
\ifx\revtex@jnl\jnl@aj\let\tablebreak=\nl\fi
\makeatother
\begin{deluxetable}{ccccccc}
\tabletypesize{\scriptsize}
\tablecolumns{7}
\tablewidth{0pt}
\tablecaption{Two-sample Kolmogorov-Smirnov test results (p-values)\label{KStest}}
\tablehead{\colhead{Sample}   &  \colhead{SMC Be-XRBs}  &  \colhead{LMC Be-XRBs}  &  \colhead{MW Be-XRBs}  &  \colhead{SMC Be stars}   &   \colhead{LMC Be stars}  &  \colhead{MW Be stars}}
\startdata
SMC Be-XRBs            &  \nodata    &  0.92        &  0.92      & 1.00     &  1.00   &  0.31            \\[0.1cm]
LMC Be-XRBs            &   0.92      &  \nodata     &  1.00      & 0.66     &  0.82   &  0.08            \\[0.1cm]
MW Be-XRBs             &   0.92      &  1.00        & \nodata    & 0.66     &  0.82   &  0.08            \\[0.1cm]
SMC Be stars           &   1.00      &  0.66        &  0.66      & \nodata  &  1.00   &  0.53            \\[0.1cm]
LMC Be stars           &   1.00      &  0.82        &  0.82      & 1.00     & \nodata &  0.66            \\[0.1cm]
 MW Be stars           &   0.31      &  0.08        &  0.08      & 0.53     &  0.66   &  \nodata         \\
\enddata
\end{deluxetable}

\makeatletter
\def\jnl@aj{AJ}
\ifx\revtex@jnl\jnl@aj\let\tablebreak=\nl\fi
\makeatother
\begin{deluxetable}{ccccccc}
\tabletypesize{\scriptsize}
\tablecolumns{7}
\tablewidth{0pt}
\tablecaption{Intrinsic reddening values\label{deltaBV}}
\tablehead{\colhead{Source ID} & \colhead{Spectral Class} & \colhead{${\rm (B-V)_{o,phot}}$} & \colhead{errBV} & \colhead{Nearest B-subtype} & \colhead{${\rm (B-V)_{theor}}$} & \colhead{${\rm \delta(B-V)}$}\\
\colhead{} & \colhead{} & \colhead{(mag)} & \colhead{(mag)} & \colhead{} & \colhead{(mag)} & \colhead{(mag)}\\[0.2cm]
\colhead{[1]} & \colhead{[2]} & \colhead{[3]} & \colhead{[4]} & \colhead{[5]} & \colhead{[6]} & \colhead{[7]} }
\startdata
  XMM-3        & B0.5e   & -0.01   & 0.04 & B0.5     & -0.28        & 0.27       \\[0.1cm] 
  4-8          & B1.5e   & 0.17    & 0.06 & B1.5     & -0.26        & 0.43       \\[0.1cm] 
  XMM-5        & B1.5e   & -0.09   & 0.04 & B1.5     & -0.26        & 0.17       \\[0.1cm] 
  4-2 (XMM-7)  & B1      & 0.10    & 0.05 & B1       & -0.27        & 0.37       \\[0.1cm] 
  4-5 (XMM-9)  & B1e     & 0.11    & 0.05 & B1       & -0.27        & 0.38       \\[0.1cm] 
  4-1 (XMM-12) & B0.5e   & -0.20   & 0.05 & B0.5     & -0.28        & 0.08       \\[0.1cm] 
  XMM-15       & B0.5e   & -0.17   & 0.06 & B0.5     & -0.28        & 0.11       \\[0.1cm] 
  4-3 (XMM-14) & B0.5e   & -0.01   & 0.06 & B0.5     & -0.28        & 0.27       \\[0.1cm] 
  XMM-17       & B0e     & -0.16   & 0.06 & B0       & -0.28        & 0.12       \\[0.1cm] 
  6-1          & B1-3e   & -0.17   & 0.03 & B2       & -0.25        & 0.08       \\[0.1cm] 
  5-12         & O9e-B0e & -0.09   & 0.09 & O9.5     & -0.28        & 0.19       \\[0.1cm] 
  5-2          & B0.5e   & -0.18   & 0.12 & B0.5     & -0.28        & 0.10       \\[0.1cm] 
  5-16         & B0.5e   & -0.16   & 0.03 & B0.5     & -0.28        & 0.12       \\[0.1cm] 
  5-4          & B1e     & 0.05    & 0.06 & B1       & -0.27        & 0.32       \\[0.1cm] 
  6-2          & B1e     & -0.12   & 0.04 & B1       & -0.27        & 0.15       \\[0.1cm] 
  5-7 (XMM-30) & B0e     & -0.14   & 0.04 & B0       & -0.28        & 0.14       \\[0.1cm] 
  3-3 (XMM-33) & B0-4e   & -0.08   & 0.17 & B2       & -0.25        & 0.17 \\[0.1cm] 
  XMM-39       & B1.5e   & -0.15   & 0.14 & B1.5     & -0.26        & 0.11       \\[0.1cm] 
  XMM-47       & B1.5e   & -0.04   & 0.03 & B1.5     & -0.26        & 0.22       \\[0.1cm] 
\enddata
\end{deluxetable}

\clearpage

\clearpage
\begin{center}
\rotatebox{270}{\includegraphics[width=6cm]{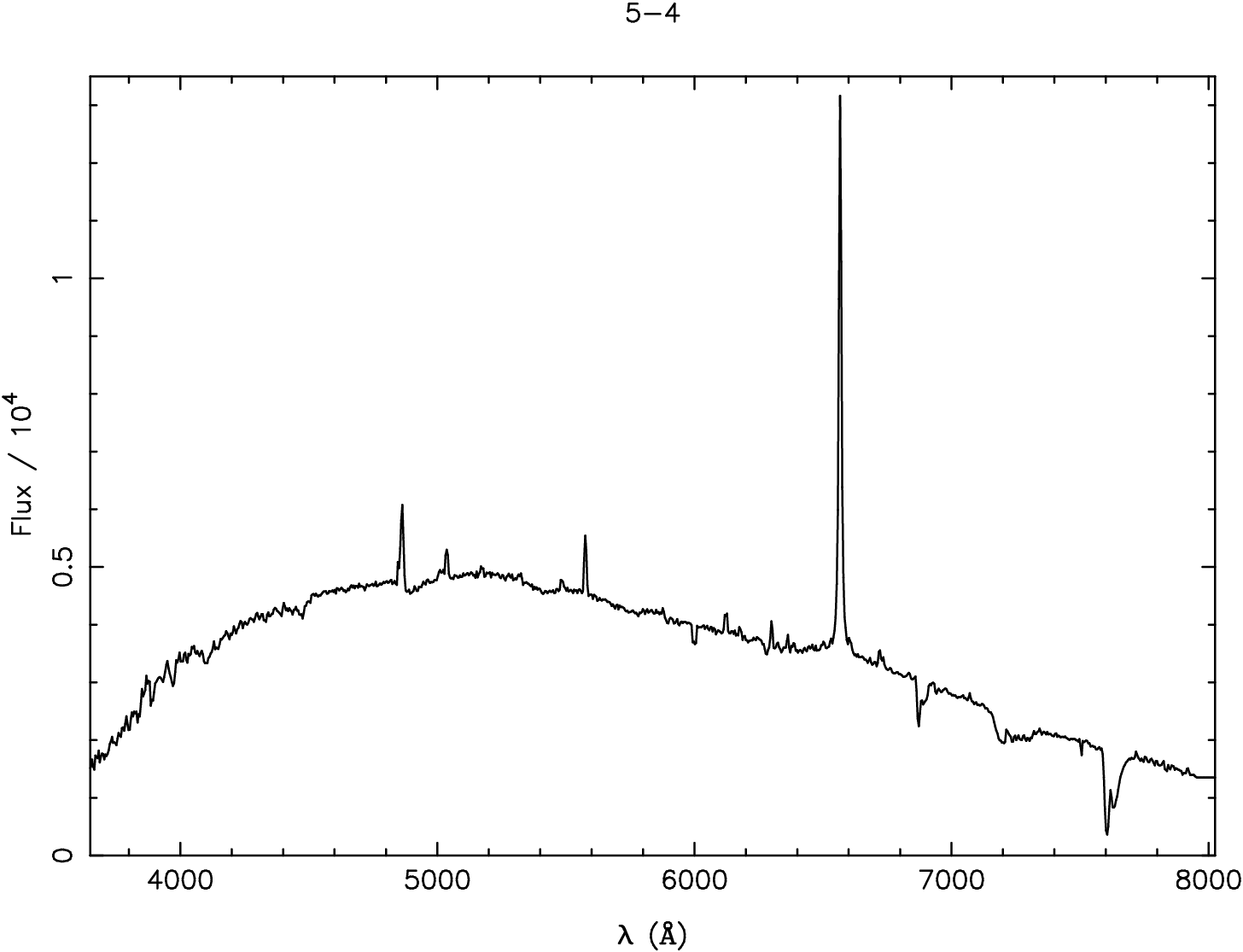}}
\vspace{0.5cm}
\rotatebox{270}{\includegraphics[width=6cm]{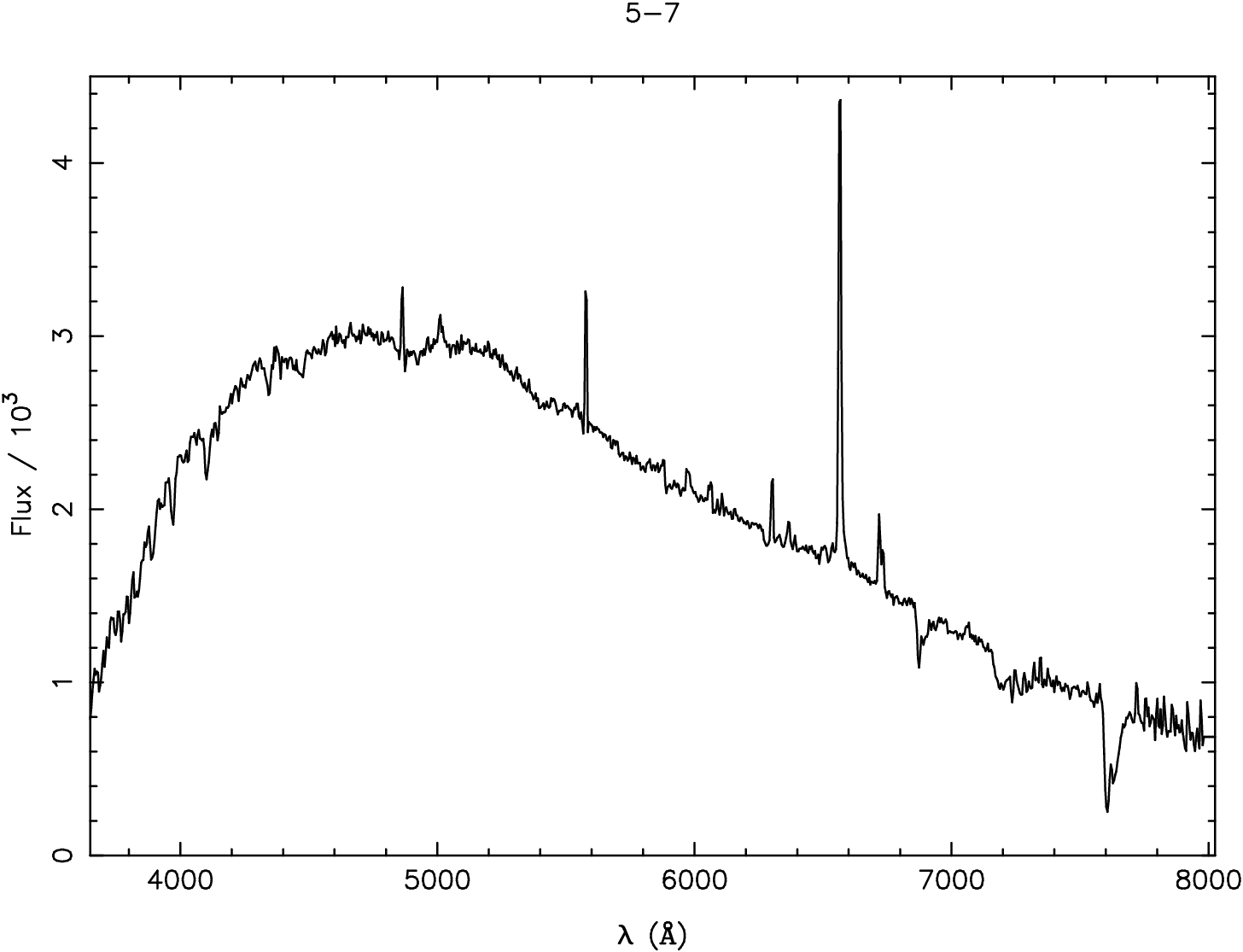}}
\vspace{0.5cm}
\rotatebox{270}{\includegraphics[width=6cm]{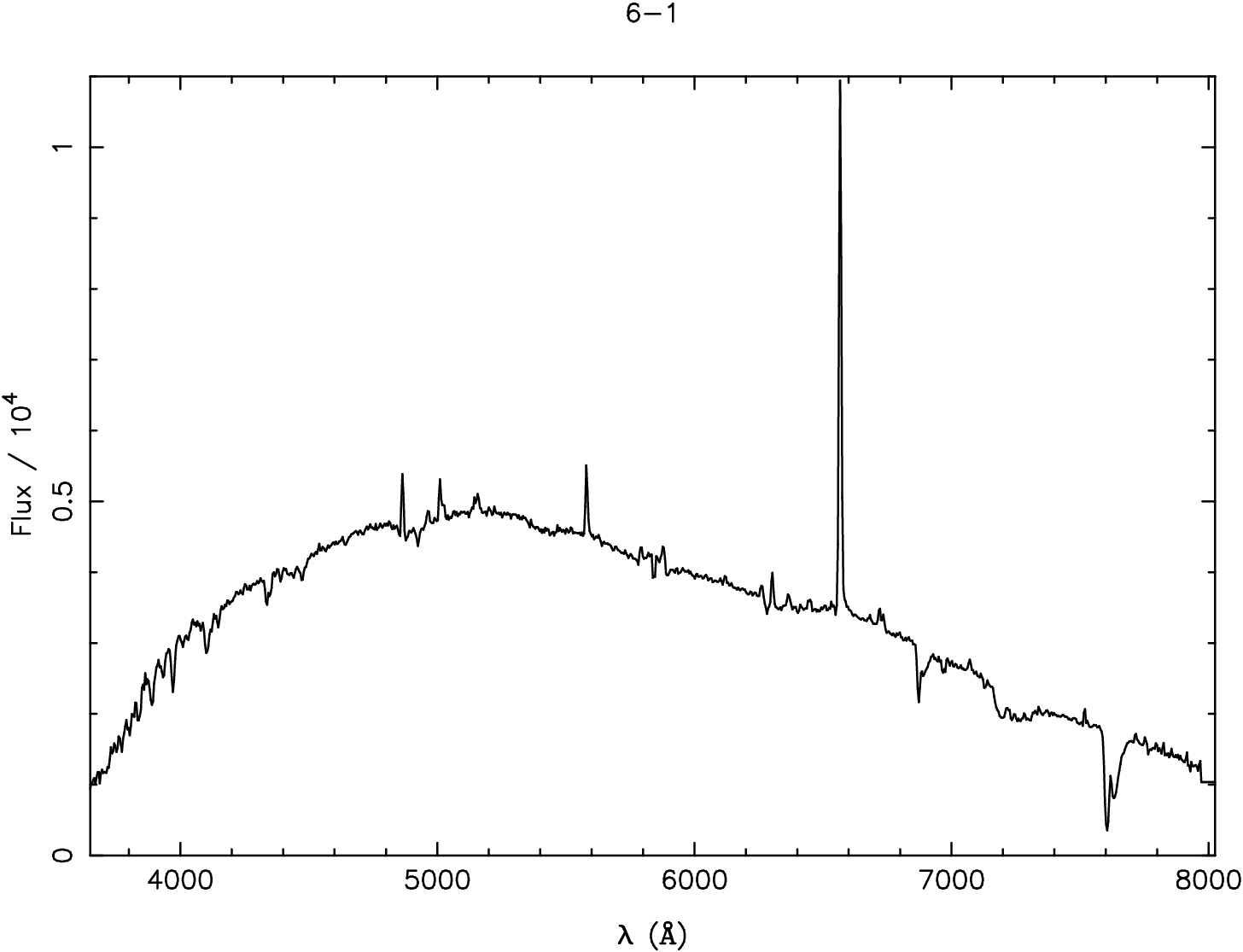}}
\rotatebox{270}{\includegraphics[width=6cm]{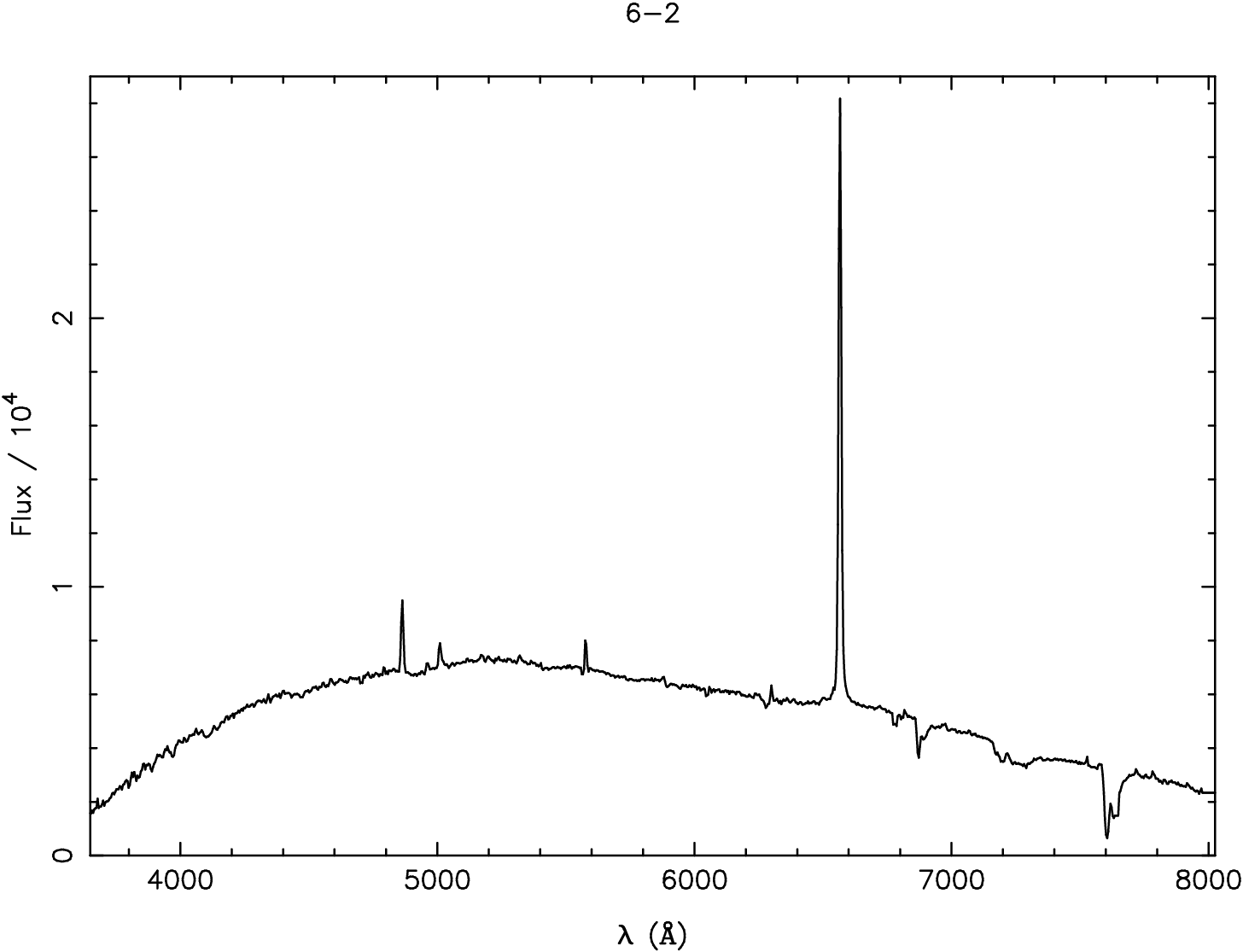}}
\vspace{0.5cm}
\rotatebox{270}{\includegraphics[width=6cm]{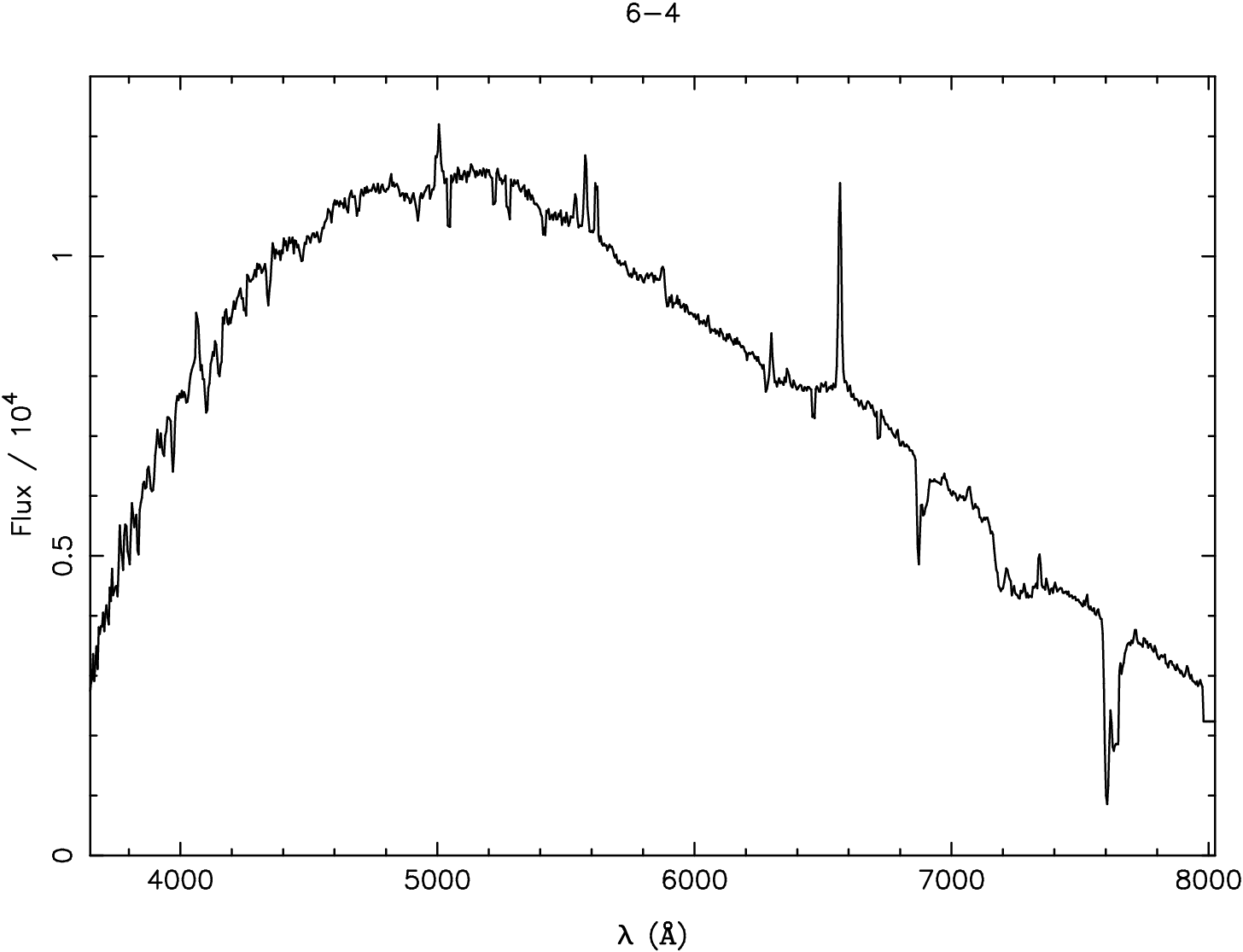}}
\rotatebox{270}{\includegraphics[width=6cm]{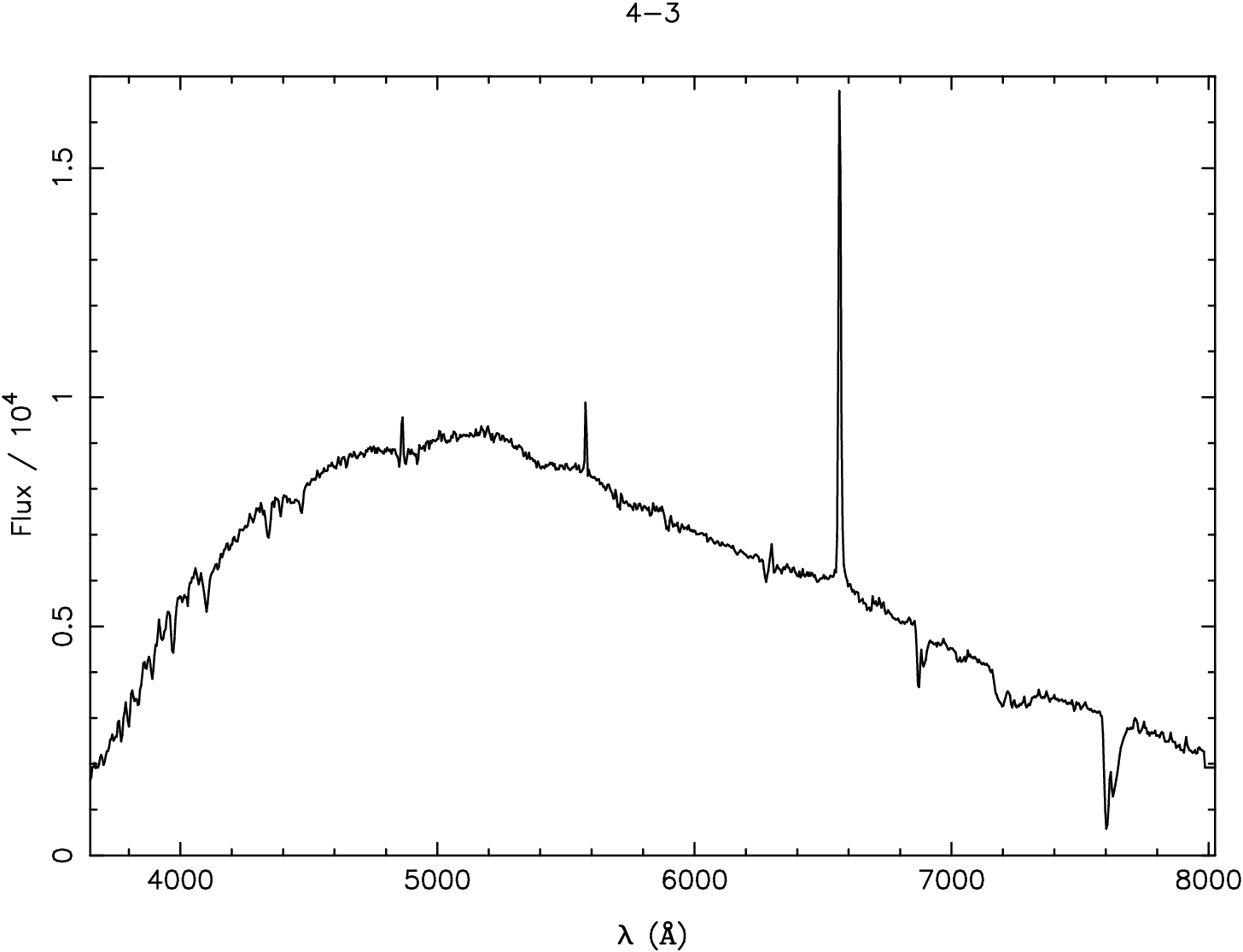}}
\end{center}
\clearpage
\begin{center}
\rotatebox{270}{\includegraphics[width=6cm]{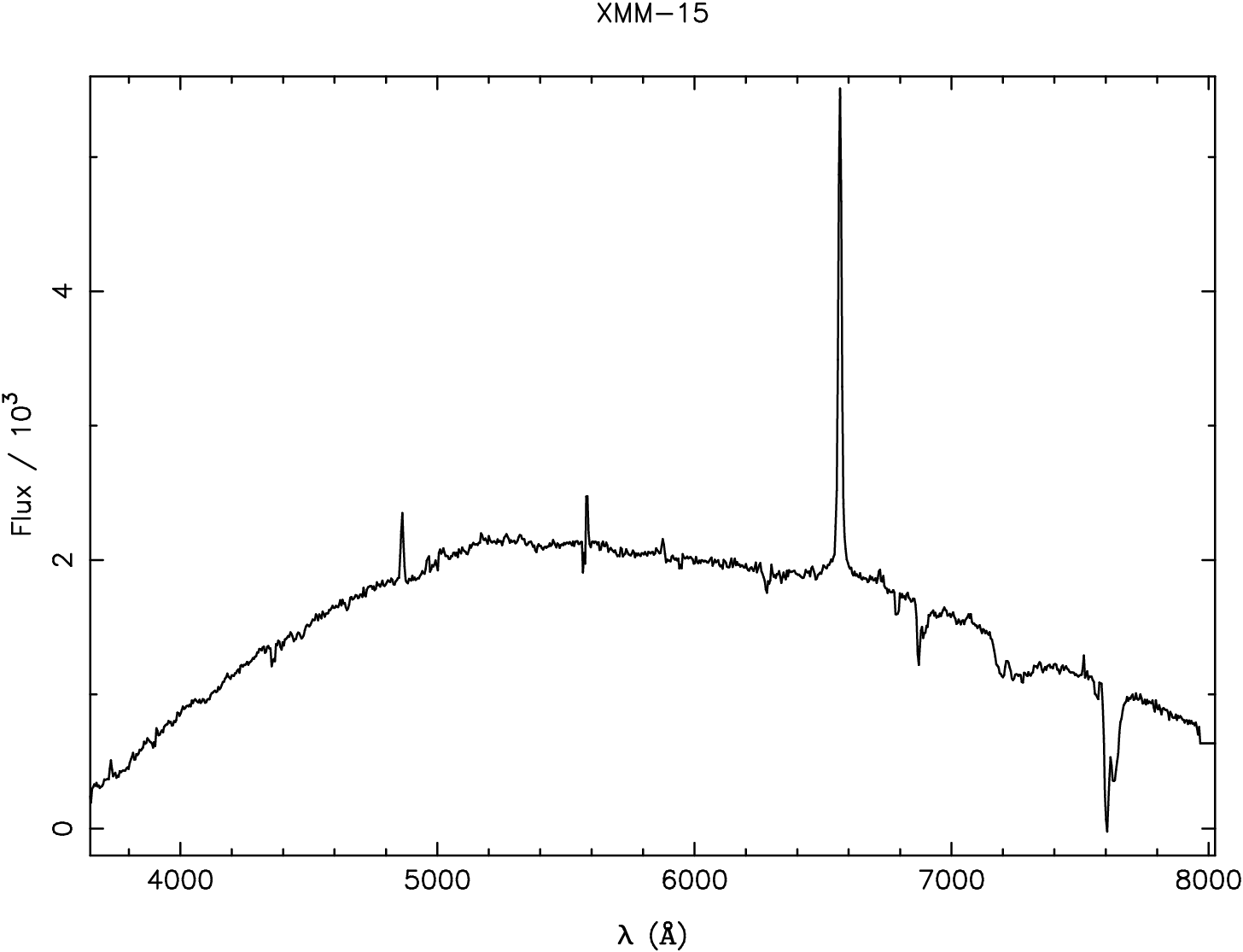}}
\vspace{0.5cm}
\rotatebox{270}{\includegraphics[width=6cm]{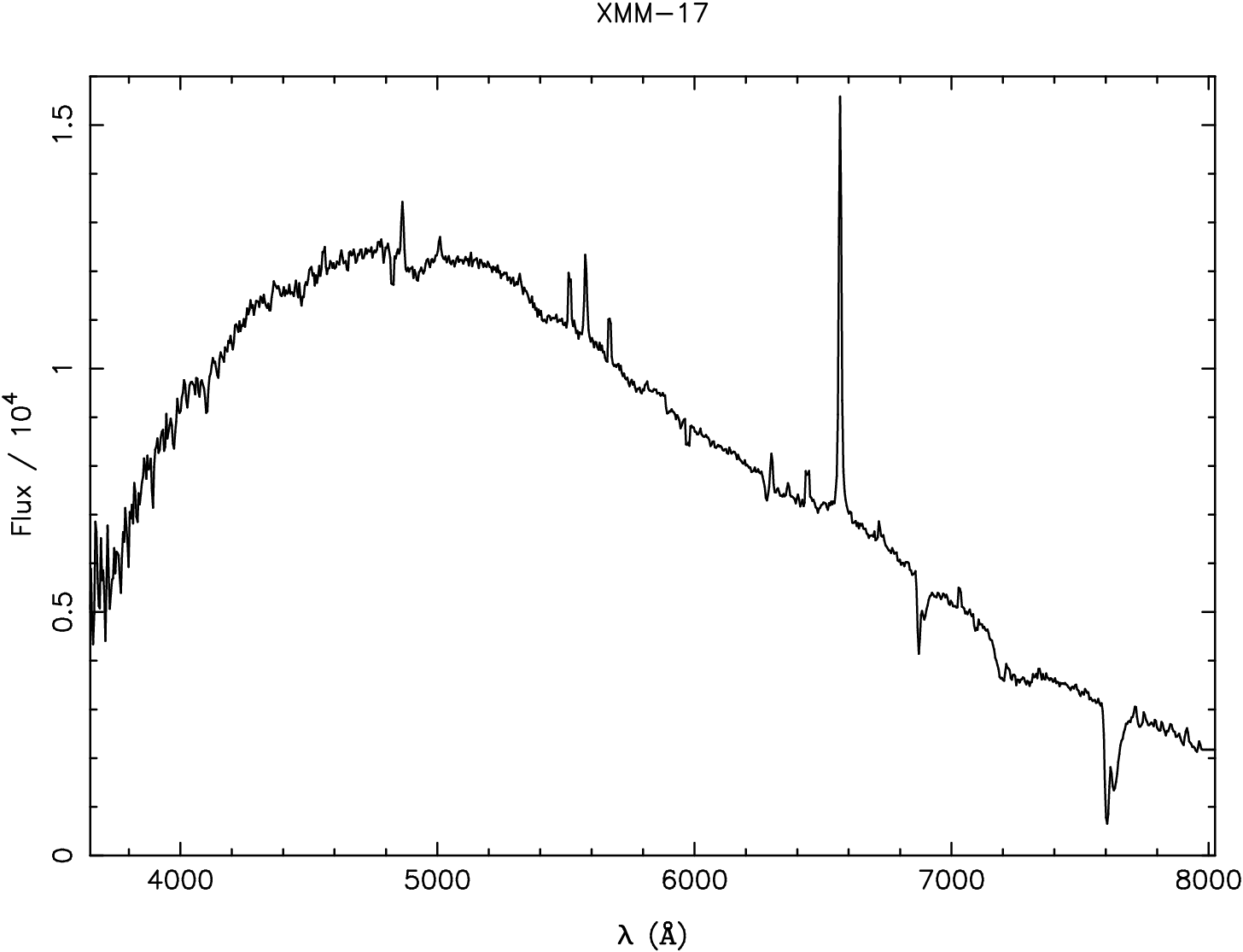}}
\vspace{0.5cm}
\rotatebox{270}{\includegraphics[width=6cm]{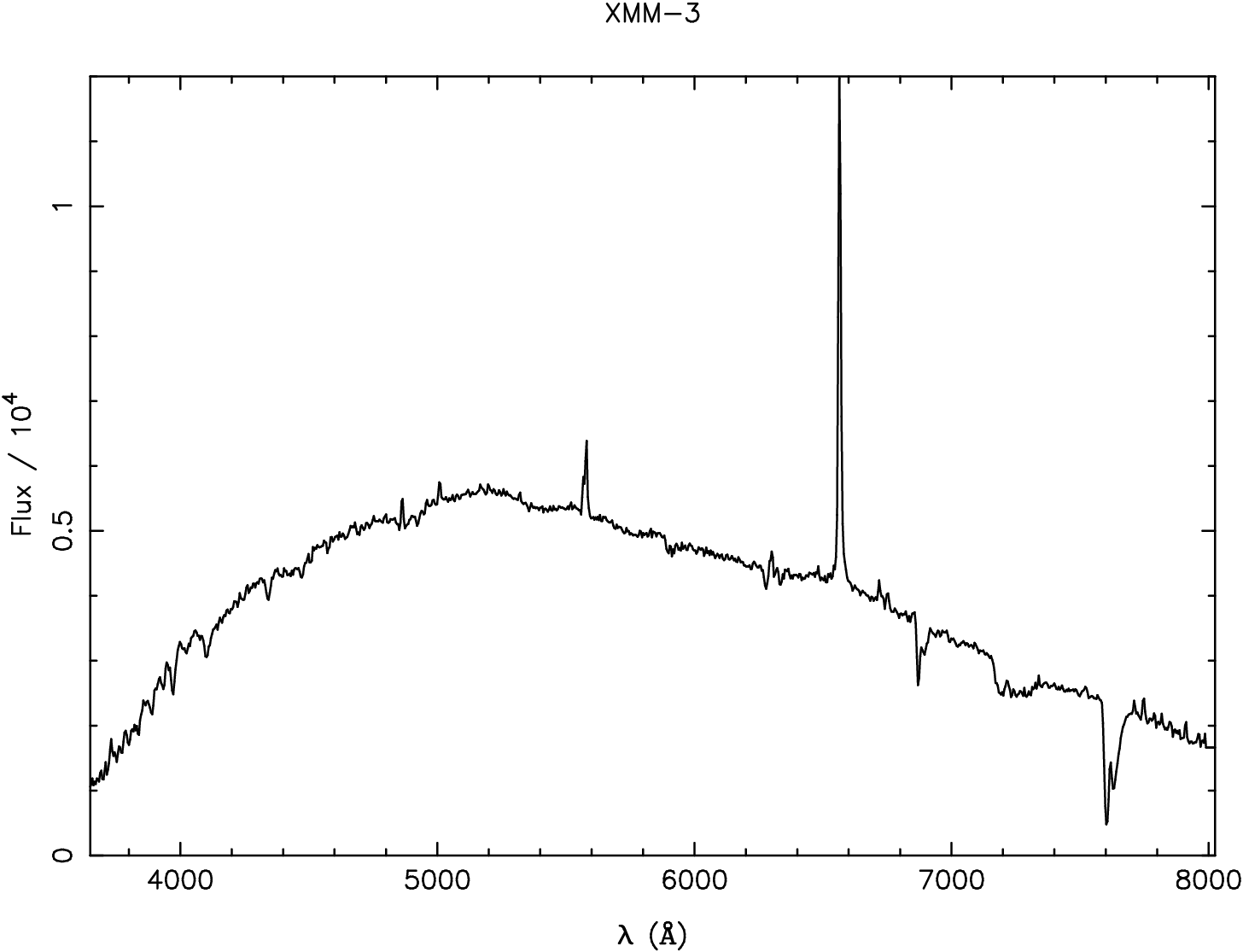}}
\rotatebox{270}{\includegraphics[width=6cm]{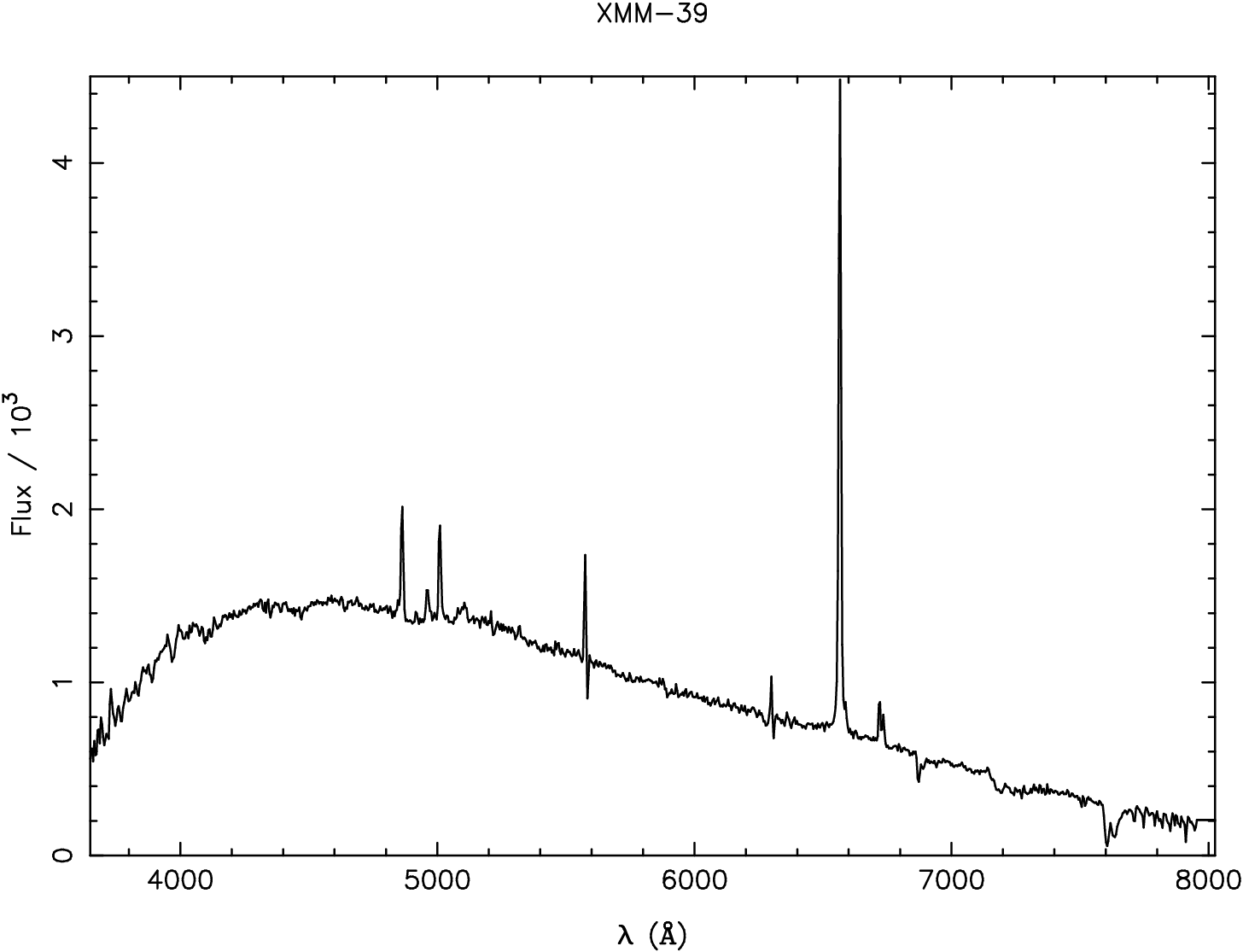}}
\vspace{0.5cm}
\rotatebox{270}{\includegraphics[width=6cm]{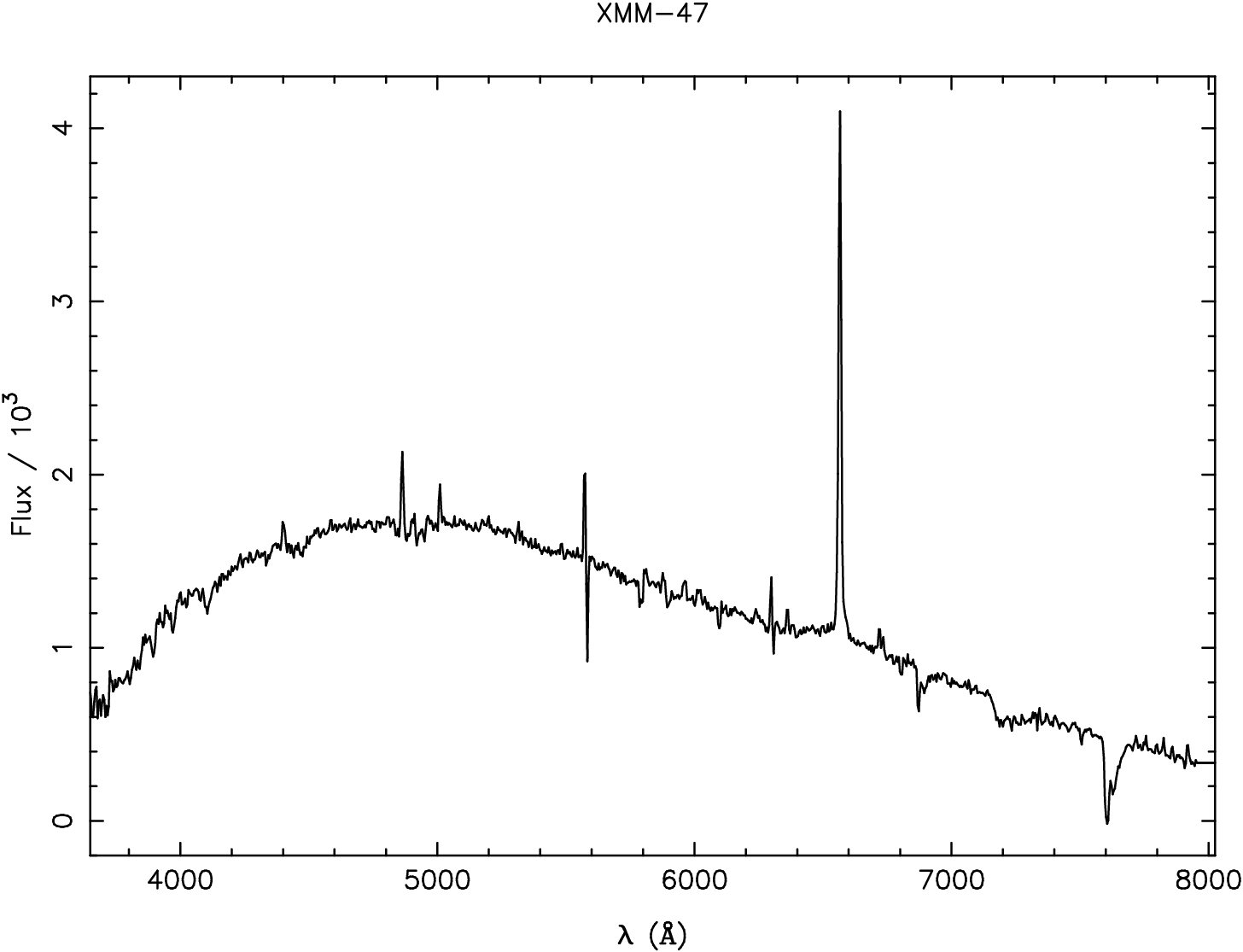}}
\rotatebox{270}{\includegraphics[width=6cm]{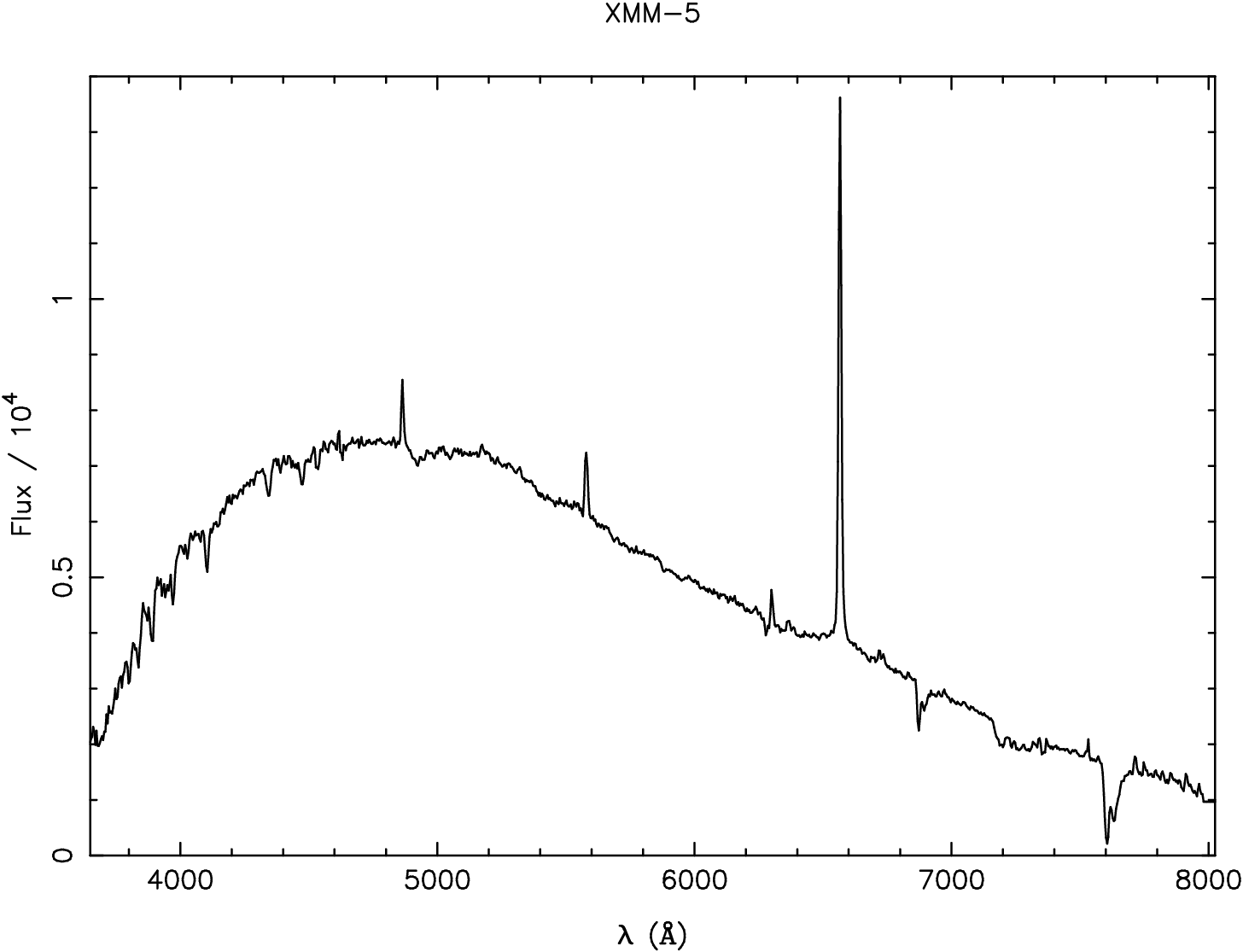}}
\end{center}
\clearpage
\begin{center}
\rotatebox{270}{\includegraphics[width=6cm]{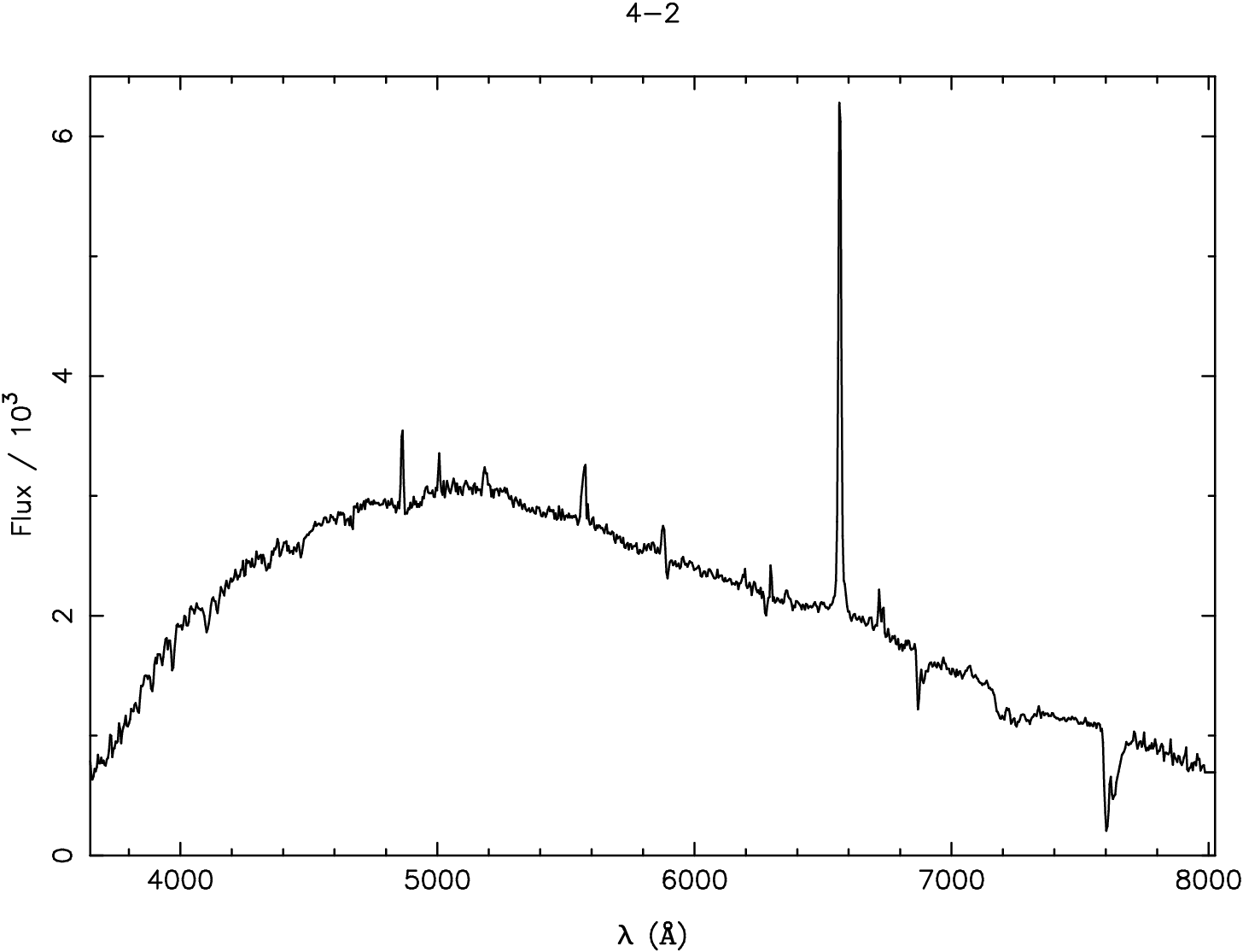}}
\rotatebox{270}{\includegraphics[width=6cm]{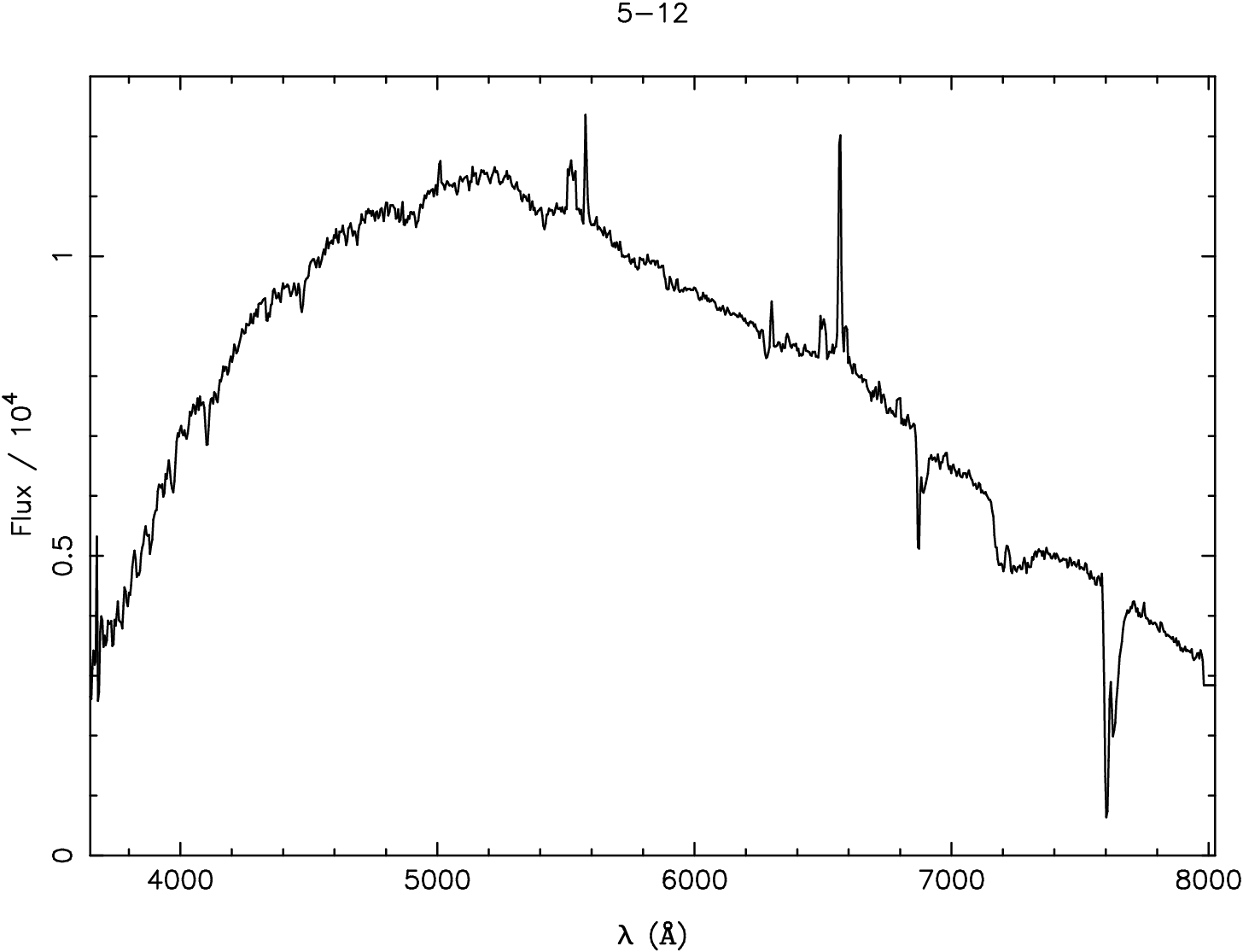}}
\end{center}

\clearpage

\begin{figure}
\centering
\rotatebox{270}{\includegraphics[width=12cm]{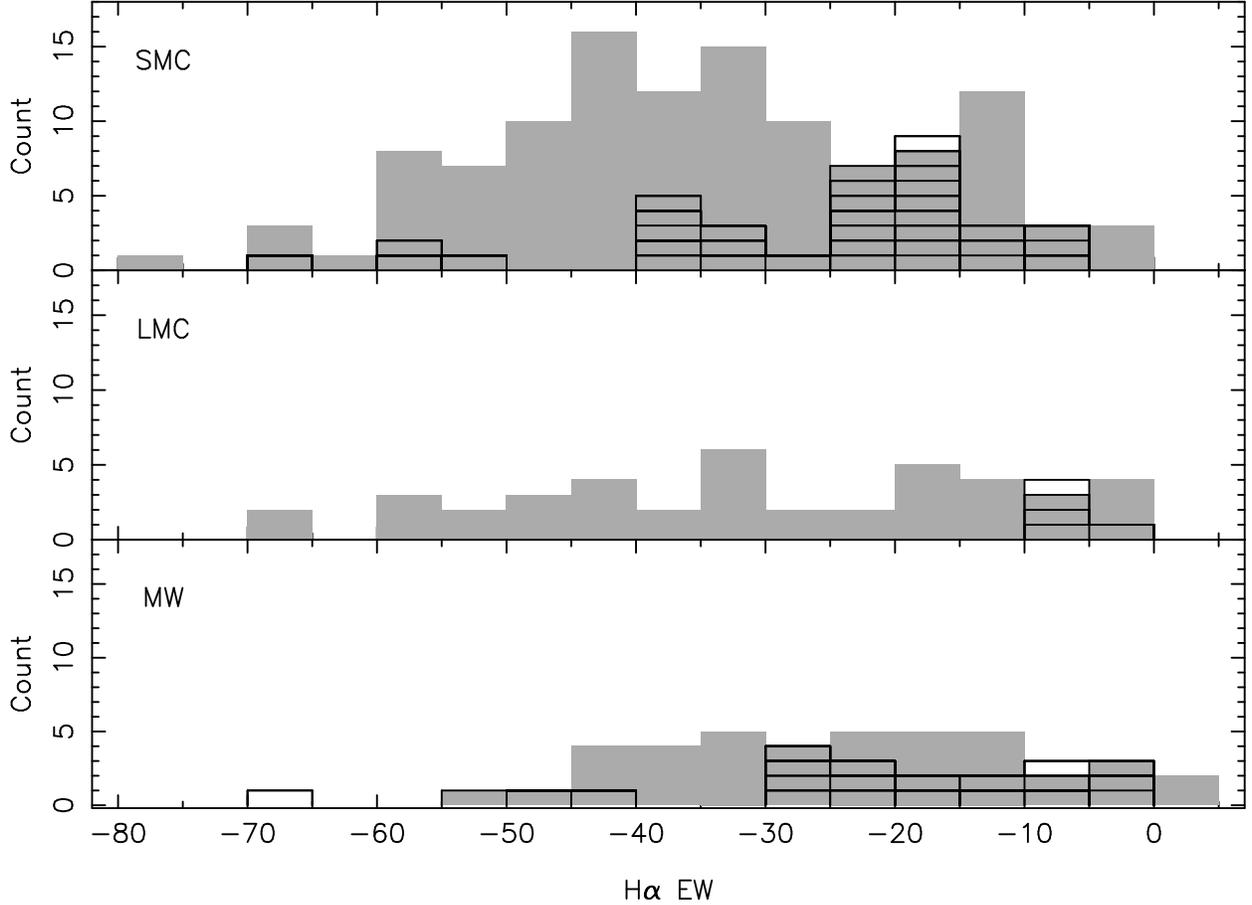}}
\caption{Distribution of \Ha equivalent widths for Be-XRBs (open histograms) and Be stars (solid histograms). {\it SMC (top panel):}
The Be-XRBs data are taken from the present work supplemented by that
of Coe \etal (2005), while for the Be stars we used data from the work
of Martayan \etal (2007a; Be stars in NGC330). {\it LMC (middle
panel):} The \Ha EW measurements for Be-XRBs are taken from Liu \etal
(2005) and references therein, and for Be stars from the work of
Martayan \etal (2006a; NGC2004 Be stars). {\it Milky Way (bottom
panel):} The values for the Be-XRBs are taken from Reig (2007), and
Liu \etal (2006; and references therein), and for the Be stars from
the works of Ashok \etal (1984), Fabregat \& Reglero (1990), and Dachs
\etal (1986, 1992). For the SMC, only sources with spectral types B3 or earlier were used, while for the LMC and the Milky Way the spectral-type limit used above is B2.}\label{EWdistr}
\end{figure}

\begin{figure}
\begin{center}
\rotatebox{270}{\includegraphics[width=12cm]{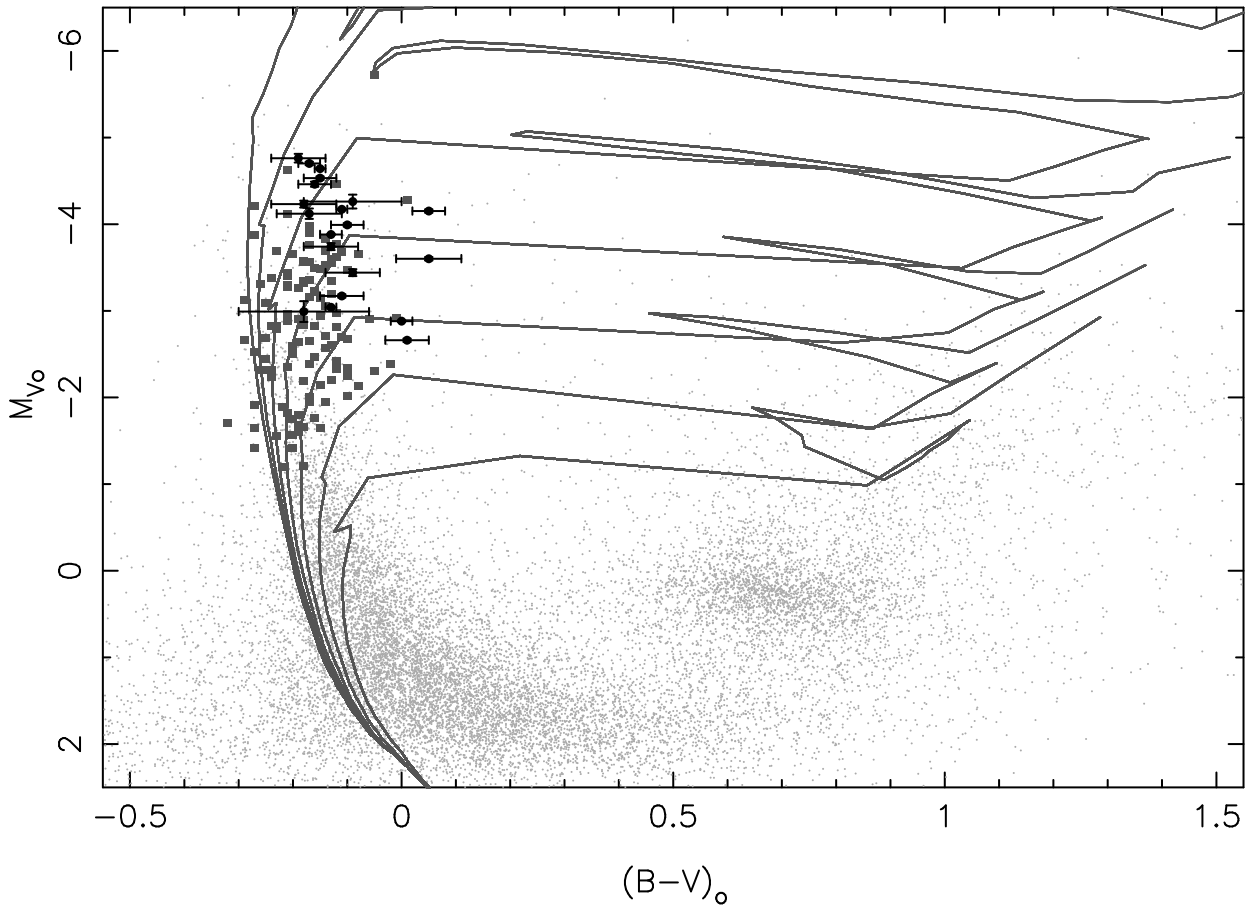}}
\caption{${\rm M_{{V}_{o}}}$ vs. ${\rm (B-V)_{o}}$ CMD of the optical counterparts (black circles) of the X-ray sources with 2dF spectra. All data used in this plot are corrected for extinction (using E(B-V)=0.09 and Av=0.29 for the SMC, following Antoniou \etal 2009). With gray squares the NGC330 Be stars from Martayan \etal (2007b) are also shown, while with small gray dots we present the OGLE-II stars that lie in our \chandra field 4 and have $<0.2$ mag errors in the V and B band. The main sequence, the red giant branch, and the red clump loci are clearly seen. Overlaid are the isochrones (solid
lines) from Geneva database
(Lejeune \& Schaerer 2001) for ages of 8.7 Myr, 15.5 Myr, 27.5 Myr, 49.0
Myr, 87.1 Myr, 154.9 Myr and 275.4 Myr (from top to bottom).}\label{CMD2dF}
\end{center}
\end{figure}

\begin{figure}
\centering
\rotatebox{270}{\includegraphics[width=12cm]{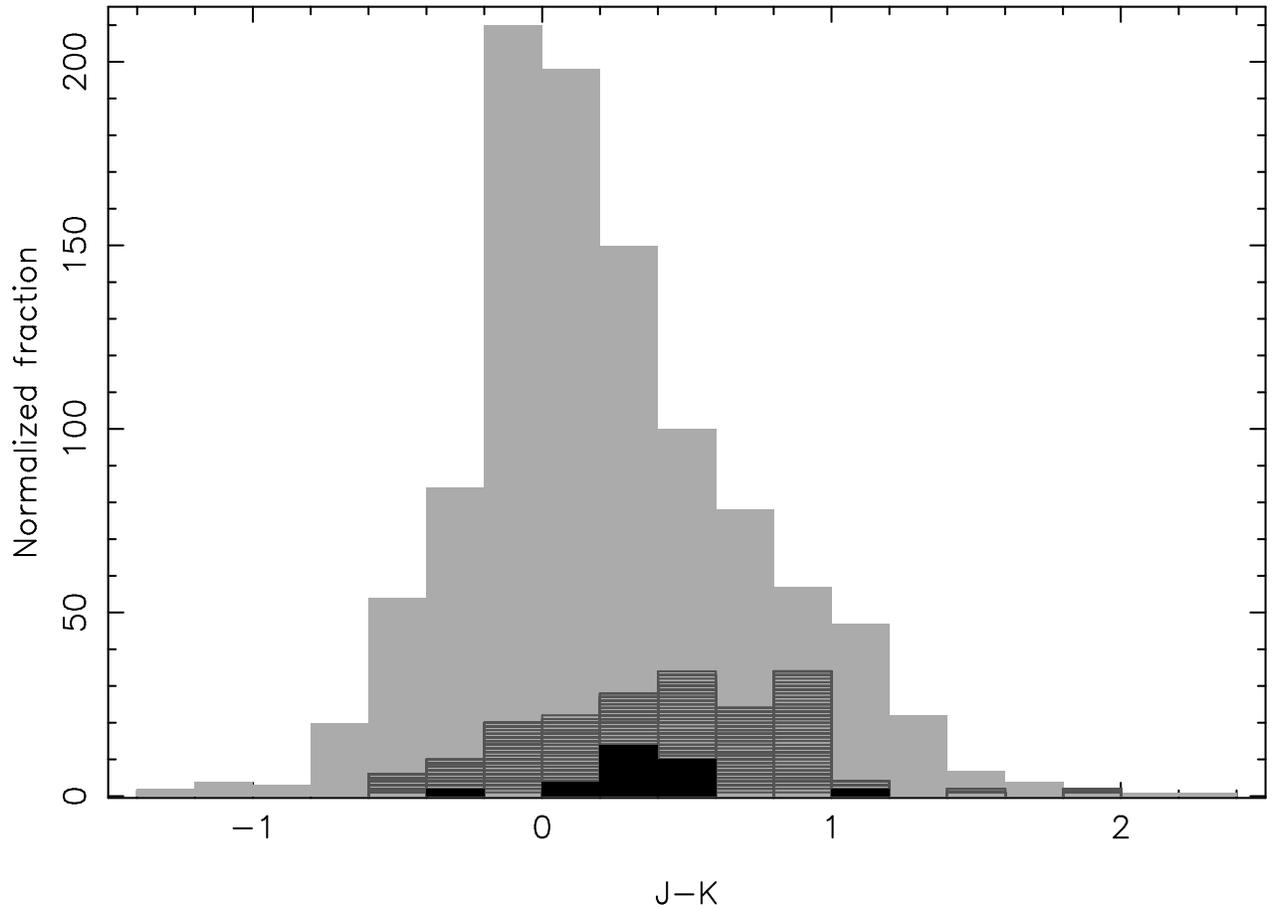}}
\caption{Histogram of J-K color values for the Be-XRBs of the present
work (shown in black; filled histogram). Be stars of luminosity class
III-V from the 2dF
spectroscopic survey of the SMC (Evans \etal 2004) are shown in dark
gray (hatched histogram), while all B-type stars without emission
(i.e. excluding
the Be stars sample) from the same survey are shown in light gray (filled
histogram). The number of sources in the Be-XRB and Be stars sample is
normalized by a factor of 2.}\label{JKo}
\end{figure}

\begin{figure}
\centering
\rotatebox{270}{\includegraphics[width=6.0cm]{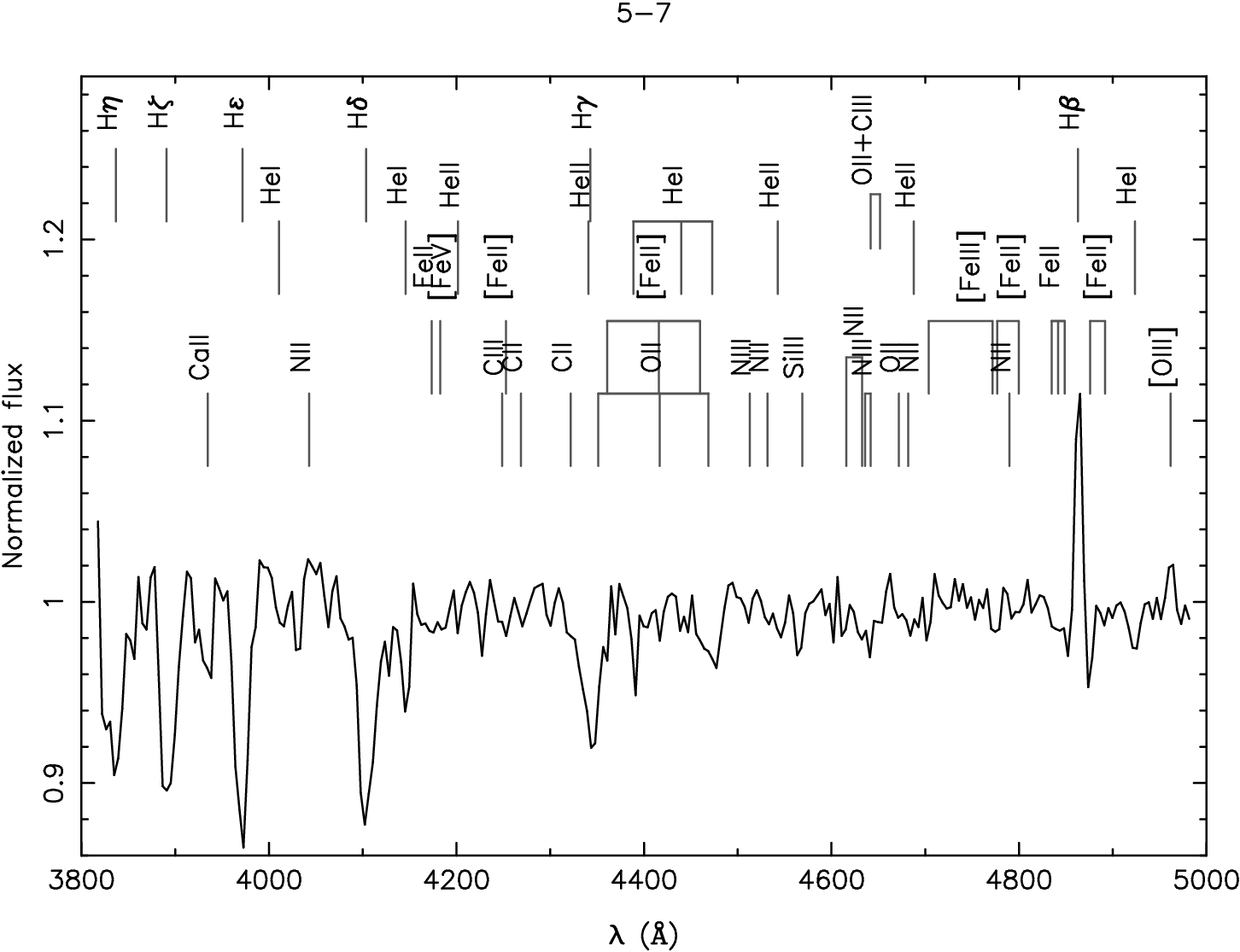}}
\vspace{0.5cm}
\rotatebox{270}{\includegraphics[width=6.0cm]{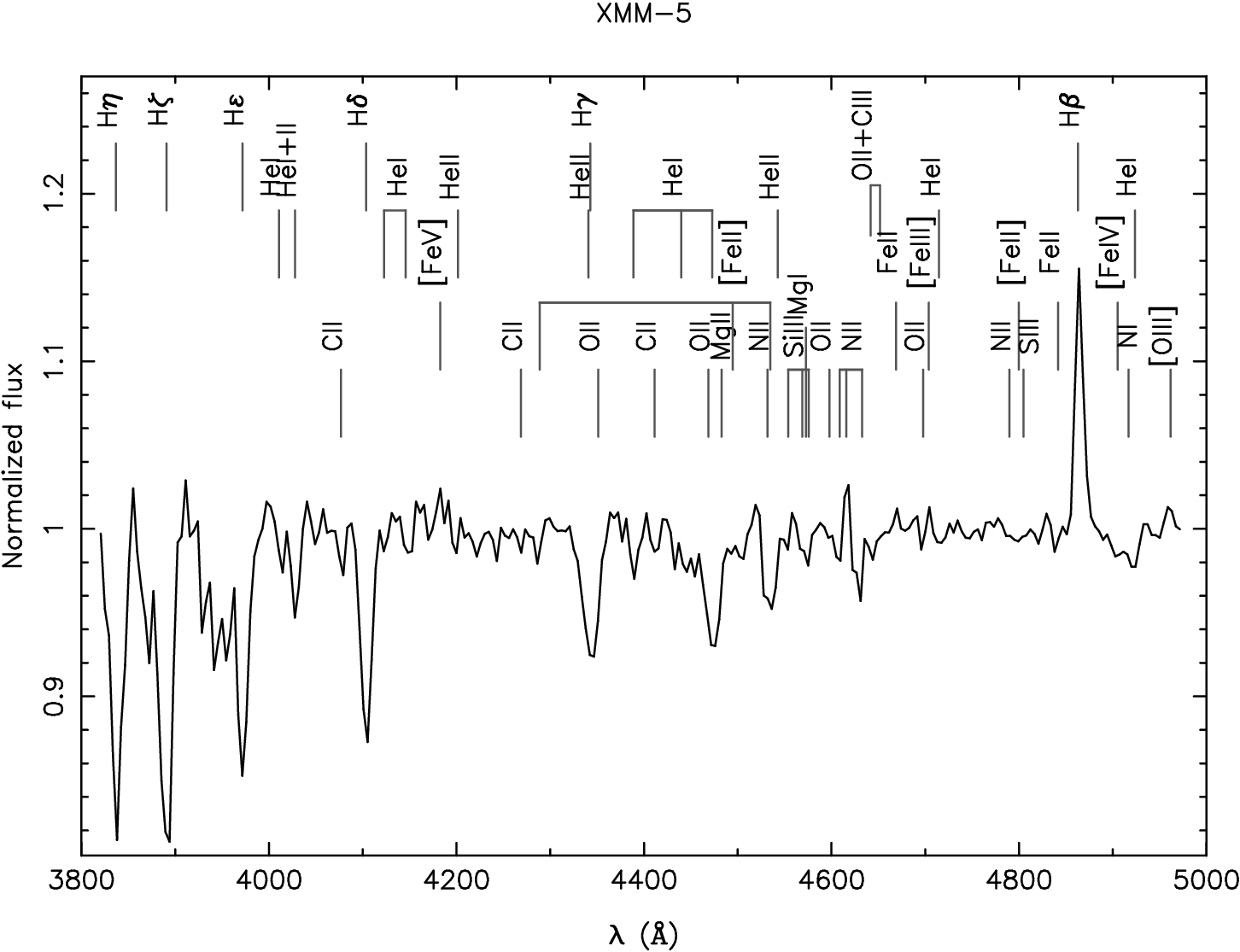}}
\vspace{0.5cm}
\rotatebox{270}{\includegraphics[width=6.0cm]{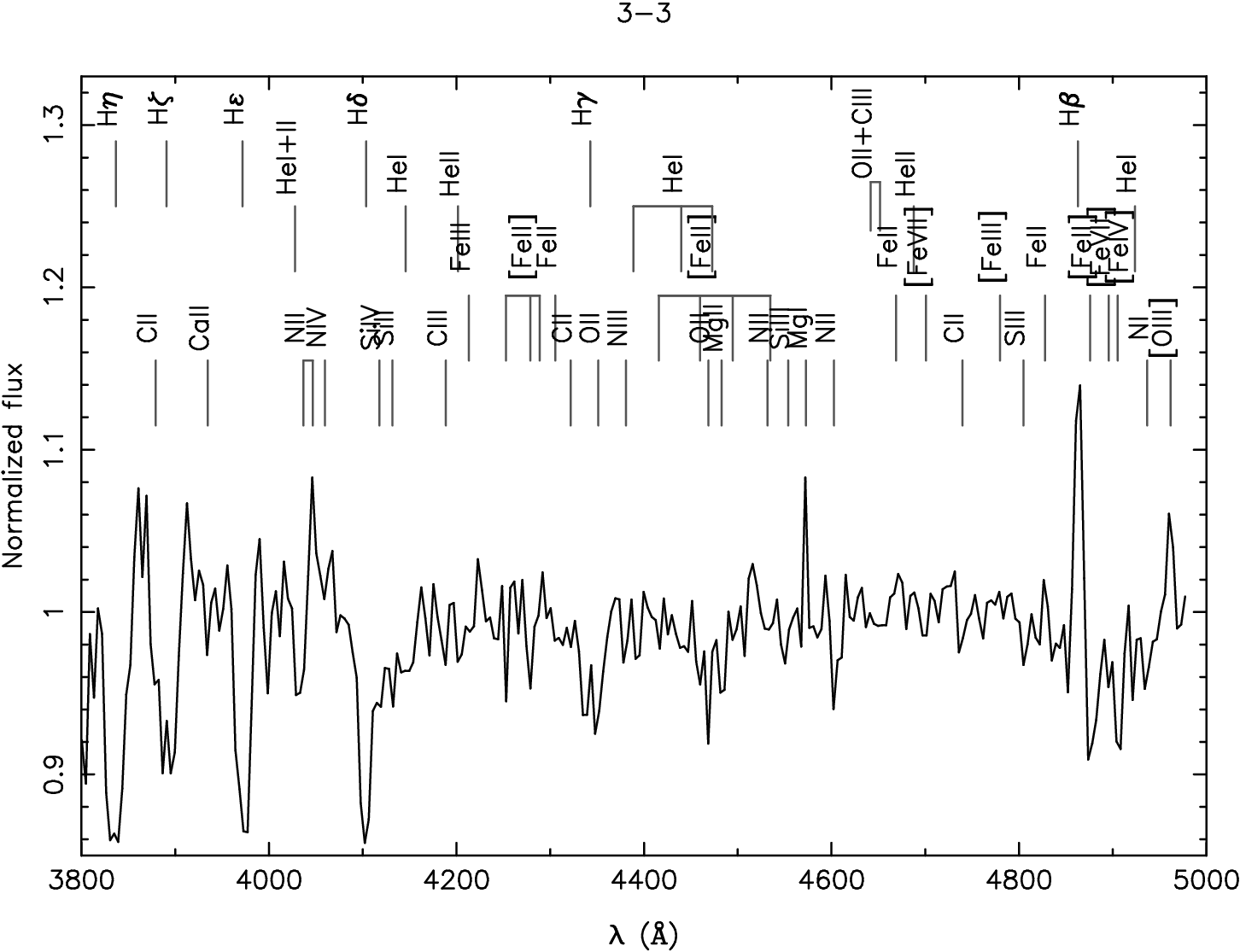}}
\rotatebox{270}{\includegraphics[width=6.0cm]{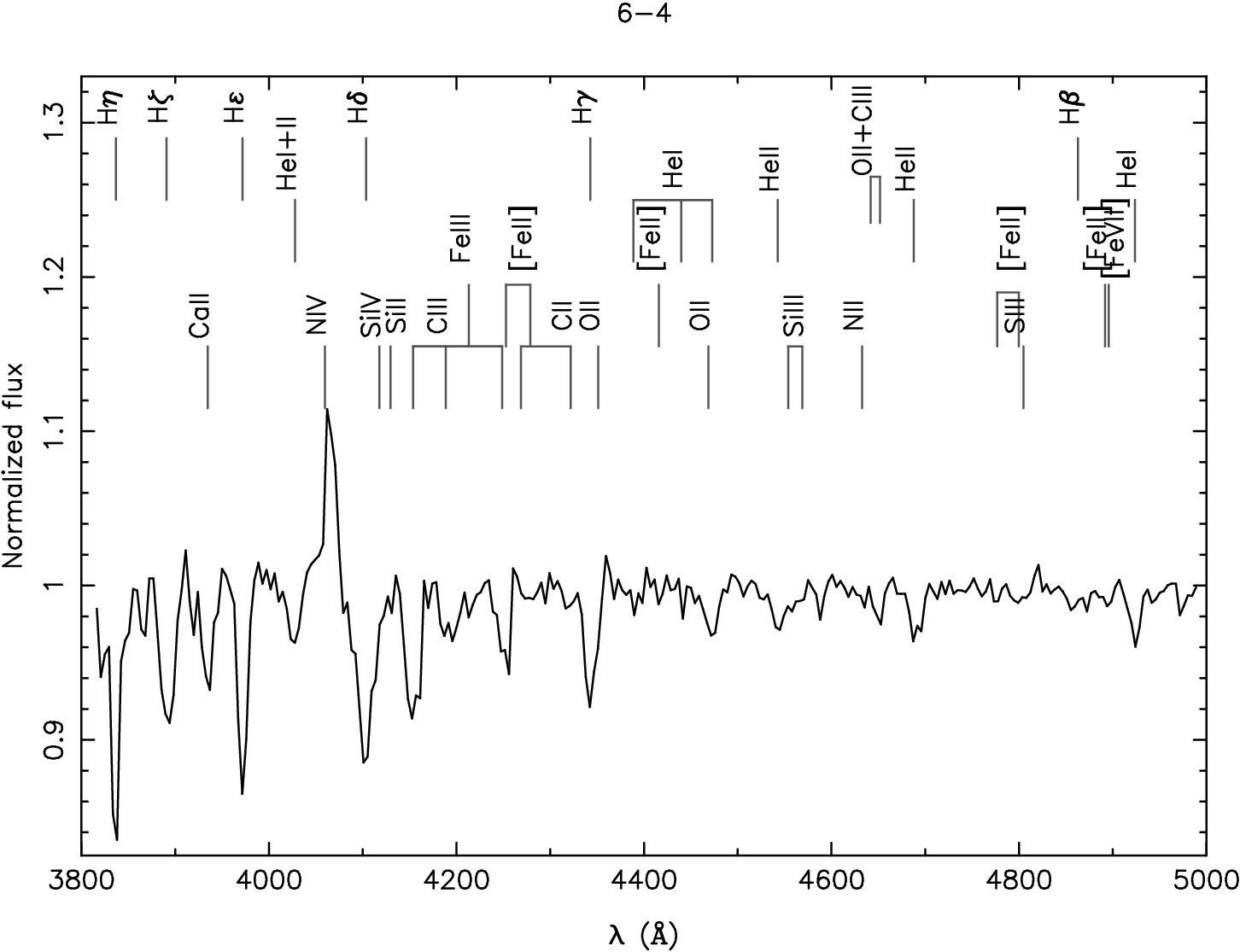}}
\vspace{0.5cm}
\rotatebox{270}{\includegraphics[width=6.0cm]{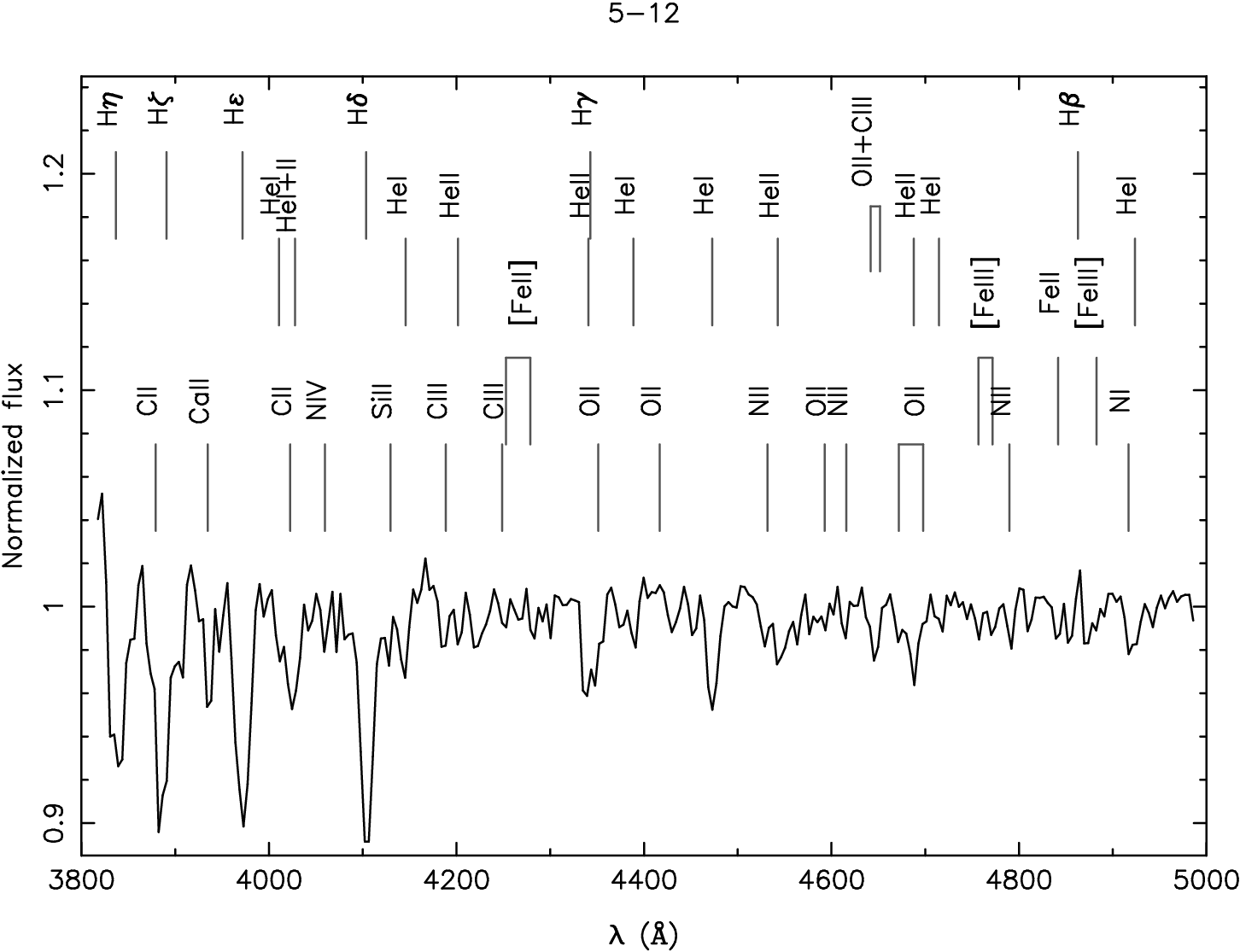}}
\rotatebox{270}{\includegraphics[width=6.0cm]{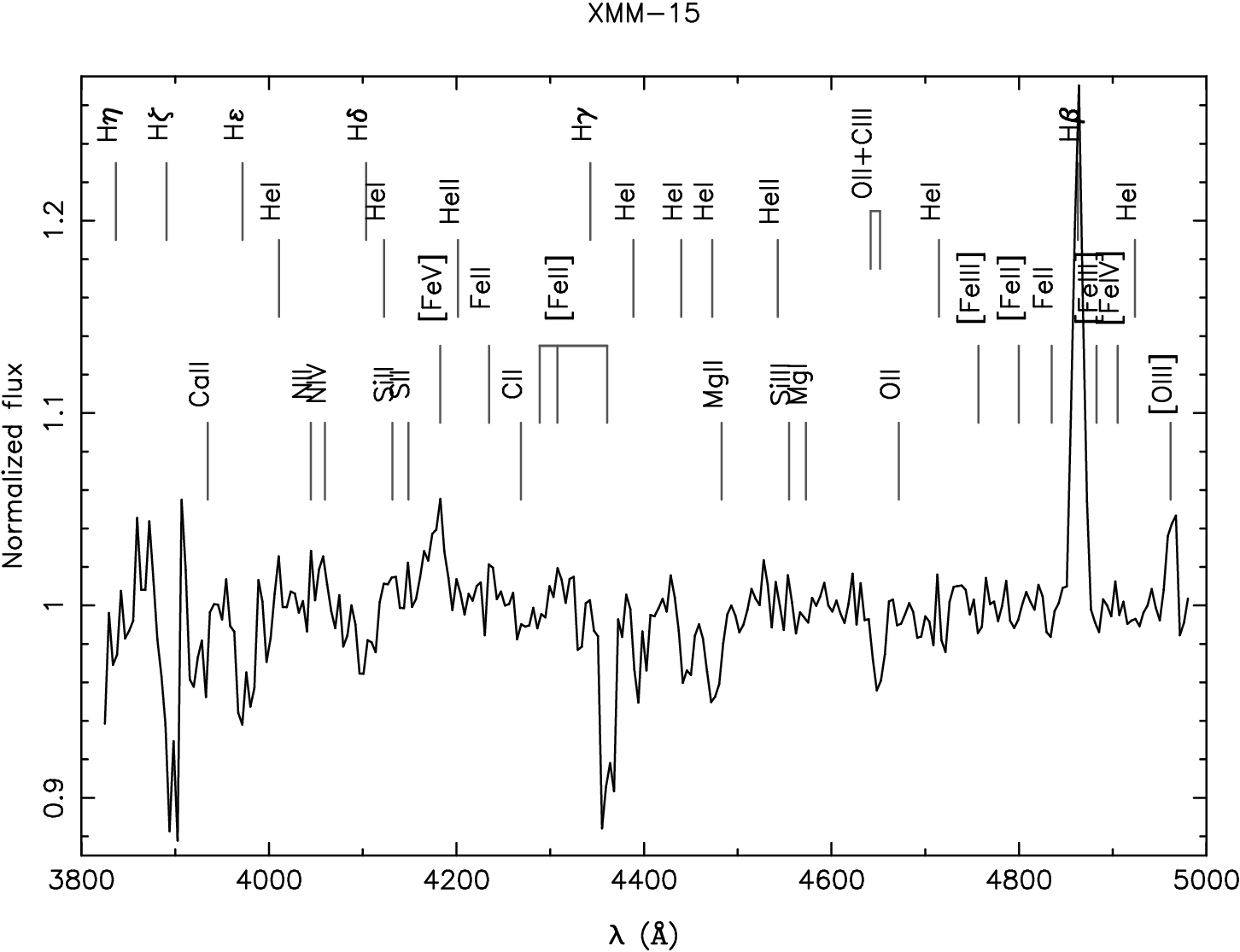}}
\caption{Sample normalized spectra for sources 5-7, XMM-5, 3-3, 6-4,
5-12, and XMM-15 in the 3800\AA-5000\AA\ wavelength range. Characteristic spectral features are marked.}\label{normspectra}
\end{figure}

\begin{figure}
\centering
\rotatebox{270}{\includegraphics[width=12cm]{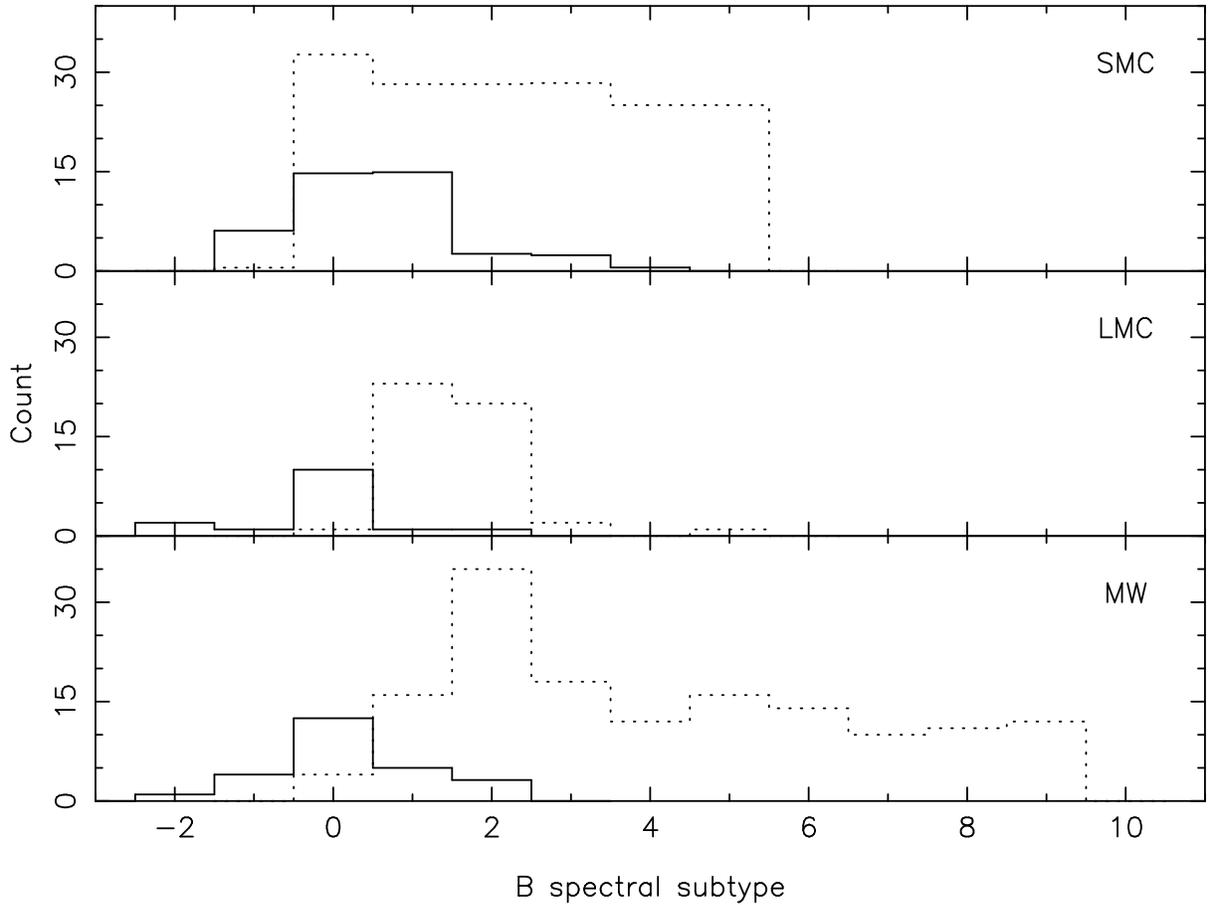}}
\caption{Comparison of the B spectral subtype distributions of Be-XRBs (solid histograms) to Be stars (dashed histograms) in the SMC (top panel), the LMC (middle panel), and the Milky Way (bottom panel). Negative spectral subtypes correspond to O-type stars.}\label{Bspecsubtdistr}
\end{figure}

\begin{figure}
\centering
\rotatebox{270}{\includegraphics[width=12cm]{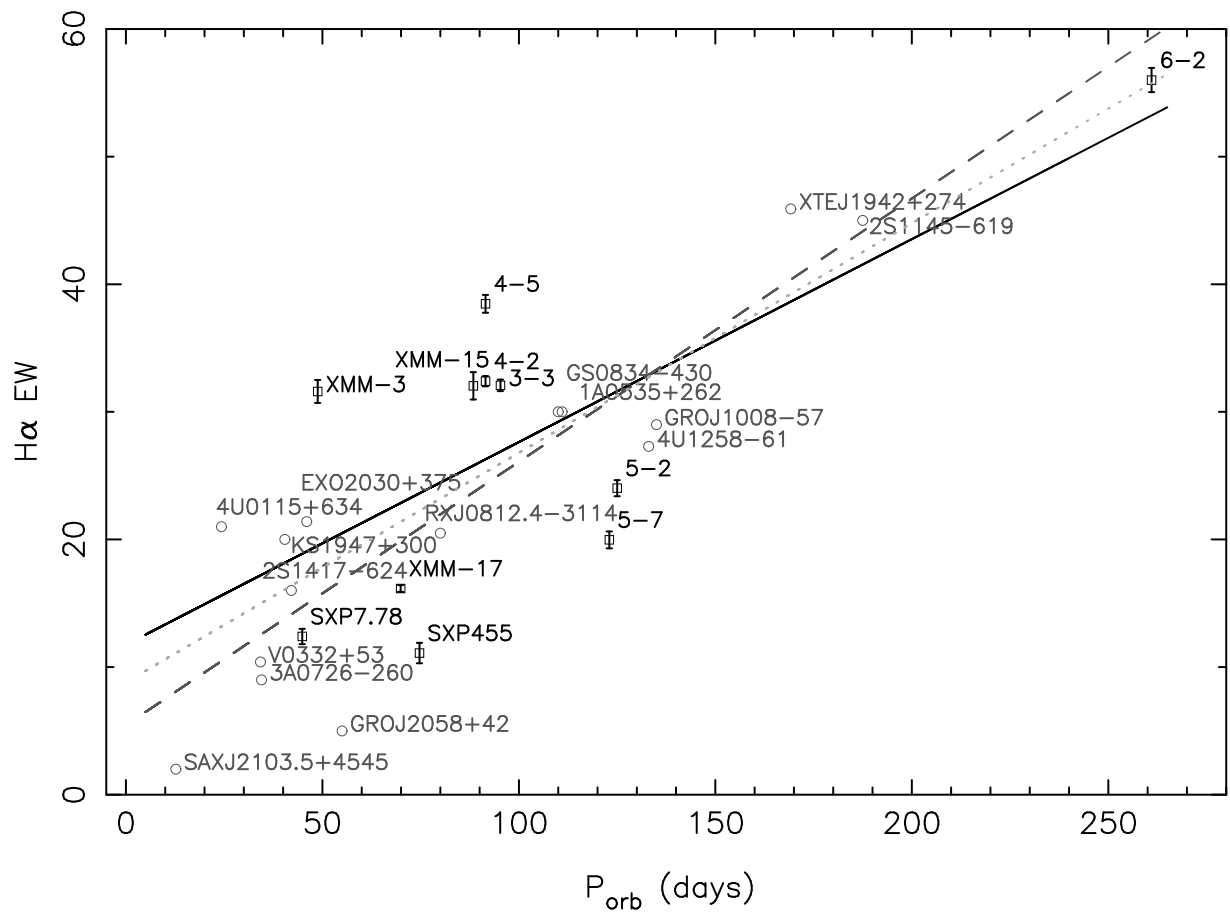}}
\caption{\Ha EW vs. orbital period (in days) for SMC (squares) and Galactic (circles) Be-XRBs. For the SMC sample, we used values from the present work and that of Liu \etal (2005) and Coe \etal (2005).
The Galactic sample (and source ID) is taken from Reig (2007). The linear regression for the SMC data sample is shown with a solid line, for the Galaxy with a dashed line and for the combined sample with a dotted line.}\label{EWPorb}
\end{figure}

\begin{figure}
\centering
\rotatebox{270}{\includegraphics[width=12cm]{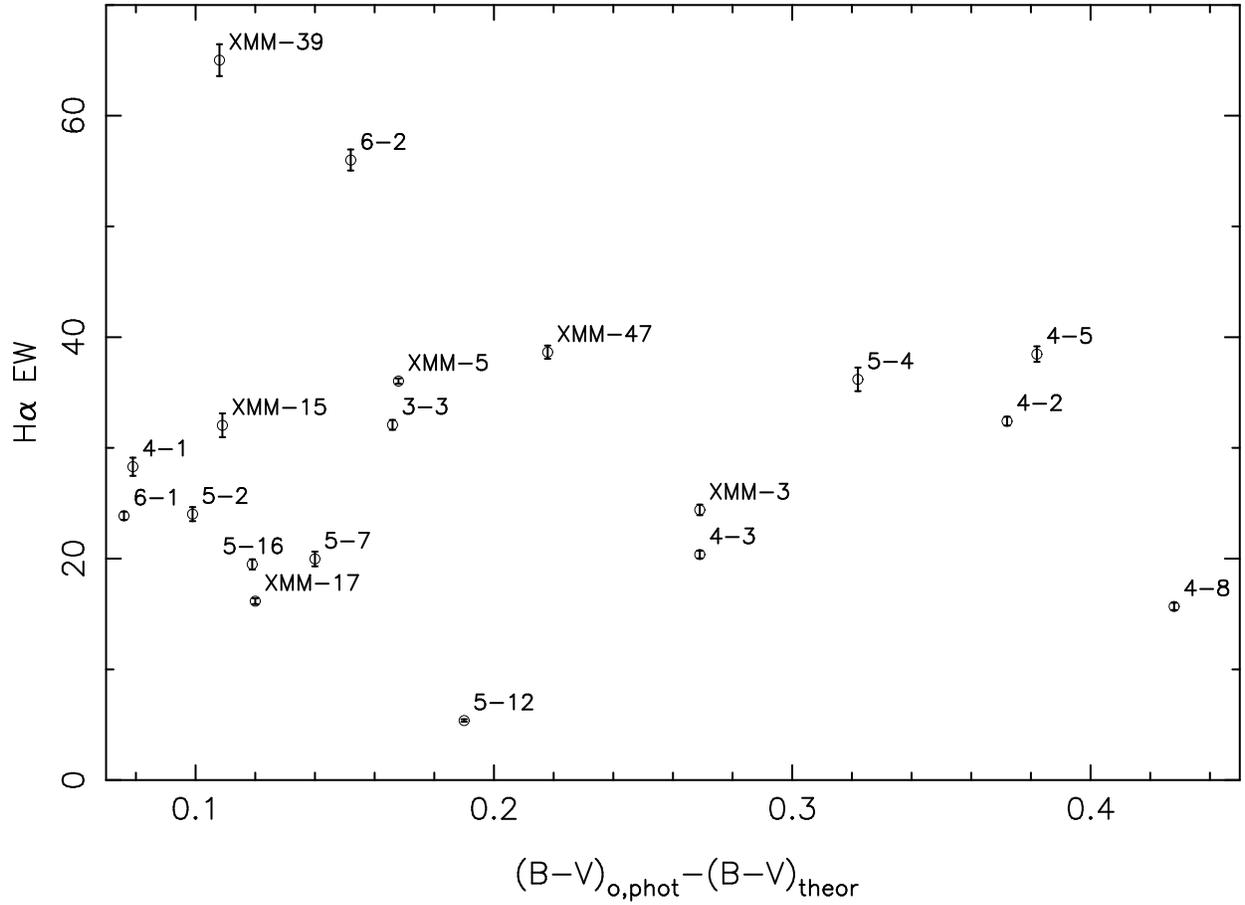}}
\caption{\Ha EW vs. the intrinsic reddening caused by the decretion
disk around Be stars, which is defined as the difference in the B-V color derived from observations (photometry) and theory (using the Geneva isochrones for the SMC metallicity; Z=0.004).}\label{LocalReddening}
\end{figure}

\end{document}